\documentclass[12pt]{iopart_footnotes}

\usepackage{etoolbox}
\newtoggle{postreferee}
\newtoggle{devmode}

\usepackage{iopams}

\expandafter\let\csname equation*\endcsname=\relax
\expandafter\let\csname endequation*\endcsname=\relax

\usepackage[authoryear]{natbib}

\newcommand\eprint[1]{\href{https://arXiv.org/abs/#1}{{\tt [arXiv:#1]}}}
 
\newcommand{\projecttitle}{Does relativistic cosmology software handle emergent volume evolution?}
\newcommand{\projectversion}{cc5ca58}
\newcommand{\projectzenodoid}{zenodo.6794222}
\newcommand{\projectzenodohref}{\href{https://zenodo.org/record/6794222}{zenodo.6794222}}
\newcommand{\projectzenodohrefShowURL}{\href{https://zenodo.org/record/6794222}{https://zenodo.org/record/6794222}}
\newcommand{\projectzenodofilesbase}{https://zenodo.org/record/6794222/files}
\newcommand{\projectgitrepository}{\url{https://codeberg.org/boud/gevcurvtest}}
\newcommand{\projectgitrepositoryPlainURL}{https://codeberg.org/boud/gevcurvtest}
\newcommand{\projectgitrepositoryarchived}{\href{https://archive.softwareheritage.org/swh:1:rev:5a062b564f33ad137dd244b86ee5b758bc26b808\%3Borigin=https://codeberg.org/boud/gevcurvtest}{swh:1:rev:5a062b564f33ad137dd244b86ee5b758bc26b808}}
\newcommand{\projectgitrepositoryarchivedPlainURL}{https://archive.softwareheritage.org/swh:1:rev:5a062b564f33ad137dd244b86ee5b758bc26b808\%3Borigin=https://codeberg.org/boud/gevcurvtest}

\newcommand{\projectgevpatchfileSWHhref}{\href{https://archive.softwareheritage.org/swh:1:cnt:043550246bb1c9c26664a0dbebedb334693033b1\%3Borigin=https://codeberg.org/boud/gevcurvtest\%3Bvisit=swh:1:snp:bab903fdadb2439d1555d1600c9c5b303970cbb1\%3Banchor=swh:1:rev:56849c3e096ba77ed6083c0c730ea4498d04b67d\%3Bpath=/reproduce/software/patches/20210915\%5Fphi\%5Fsource.patch}{https://archive.softwareheritage.org/}\\\href{https://archive.softwareheritage.org/swh:1:cnt:043550246bb1c9c26664a0dbebedb334693033b1\%3Borigin=https://codeberg.org/boud/gevcurvtest\%3Bvisit=swh:1:snp:bab903fdadb2439d1555d1600c9c5b303970cbb1\%3Banchor=swh:1:rev:56849c3e096ba77ed6083c0c730ea4498d04b67d\%3Bpath=/reproduce/software/patches/20210915\%5Fphi\%5Fsource.patch}{swh:1:cnt:043550246bb1c9c26664a0dbebedb334693033b1}}

\newcommand{\mpgraficname}{{\sc mpgrafic}}

\newcommand{\dtfename}{{\sc dtfe}}

\newcommand{\inhomogname}{{\sc inhomog}}
\newcommand{\inhomogversion}{0.1.10-1da3bed}
\newcommand{\ramsesscalavname}{{\sc ramses-scalav}}

\newcommand{\cosmdistname}{{\sc cosmdist}}
\newcommand{\cosmdistversion}{0.3.12}
\newcommand{\gevolutionname}{{\sc gevolution}}
\newcommand{\gevolutionversion}{1.2-0404c0b}

\newcommand{\Lboxvalue}{40}

\newcommand{\NcrootInhomogHTwEightvalue}{128}
\newcommand{\NcrootGevolutionHTwEightvalue}{128}
\newcommand{\InitPerturbInhomogPlusvalue}{0.001}
\newcommand{\InitPerturbInhomogMinusvalue}{-0.001}

\newcommand{\OmegaMEdSvalue}{1.0}

\newcommand{\OmegaLEdSvalue}{0.0}

\newcommand{\HubbleEdSvalue}{37.7}

\newcommand{\OmegaMLCDMvalue}{0.3089}

\newcommand{\OmegaLLCDMvalue}{0.6911}

\newcommand{\HubbleLCDMvalue}{67.74}

\newcommand{\ramsesparamlevelmaxEdSThirtTwoname}{{\tt levelmax\_EdS\_ThirtTwo}}
\newcommand{\ramsesparamlevelmaxEdSThirtTwovalue}{10}

\newcommand{\ramsesparamlevelmaxEdSHTwEightname}{{\tt levelmax\_EdS\_HTwEight}}
\newcommand{\ramsesparamlevelmaxEdSHTwEightvalue}{10}

\newcommand{\InitPerturbGevnPlusvalue}{0.001}

\newcommand{\gevnparamZeromodeamplitudeEdSPlusThirtTwovalue}{0.001}

\newcommand{\gevnparamZeromodeamplitudeLCDMPlusThirtTwovalue}{0.001}

\newcommand{\gevnparamZeromodeamplitudeEdSMinusThirtTwovalue}{-0.001}

\newcommand{\gevnparamZeromodeamplitudeLCDMMinusThirtTwovalue}{-0.001}

\newcommand{\gevnparamZeromodeamplitudeEdSPlusHTwEightvalue}{0.001}

\newcommand{\gevnparamZeromodeamplitudeLCDMPlusHTwEightvalue}{0.001}

\newcommand{\gevnparamZeromodeamplitudeEdSMinusHTwEightvalue}{-0.001}

\newcommand{\gevnparamZeromodeamplitudeLCDMMinusHTwEightvalue}{-0.001}
 \newcommand{\InhomogAccuracyPercentEdSzdzzouThirtTwovalue}{-0.0015}
\newcommand{\InhomogAccuracyPercentEdSmzdzzouThirtTwovalue}{-0.0017}
\newcommand{\InhomogAccuracyPercentLCDMzdzzouThirtTwovalue}{-0.0012}
\newcommand{\InhomogAccuracyPercentLCDMmzdzzouThirtTwovalue}{-0.0011}
\newcommand{\InhomogAccuracyPercentEdSzdzzouSixtFourvalue}{-0.0031}
\newcommand{\InhomogAccuracyPercentEdSmzdzzouSixtFourvalue}{-0.0033}
\newcommand{\InhomogAccuracyPercentLCDMzdzzouSixtFourvalue}{-0.0021}
\newcommand{\InhomogAccuracyPercentLCDMmzdzzouSixtFourvalue}{-0.0020}
\newcommand{\InhomogAccuracyPercentEdSzdzzouHTwEightvalue}{-0.0058}
\newcommand{\InhomogAccuracyPercentEdSmzdzzouHTwEightvalue}{-0.0059}
\newcommand{\InhomogAccuracyPercentLCDMzdzzouHTwEightvalue}{-0.0033}
\newcommand{\InhomogAccuracyPercentLCDMmzdzzouHTwEightvalue}{-0.0034}
\newcommand{\InhomogExpectedAratioEdSzdzzouThirtTwovalue}{1.00517}
\newcommand{\InhomogPerturbAratioEdSzdzzouThirtTwovalue}{1.00516}
\newcommand{\InhomogExpectedAratioEdSmzdzzouThirtTwovalue}{0.99480}
\newcommand{\InhomogPerturbAratioEdSmzdzzouThirtTwovalue}{0.99479}
\newcommand{\InhomogExpectedAratioLCDMzdzzouThirtTwovalue}{1.00423}
\newcommand{\InhomogPerturbAratioLCDMzdzzouThirtTwovalue}{1.00422}
\newcommand{\InhomogExpectedAratioLCDMmzdzzouThirtTwovalue}{0.99576}
\newcommand{\InhomogPerturbAratioLCDMmzdzzouThirtTwovalue}{0.99575}

\newcommand{\InhomogExpectedAratioEdSzdzzouHTwEightvalue}{1.00971}
\newcommand{\InhomogPerturbAratioEdSzdzzouHTwEightvalue}{1.00965}
\newcommand{\InhomogExpectedAratioEdSmzdzzouHTwEightvalue}{0.99021}
\newcommand{\InhomogPerturbAratioEdSmzdzzouHTwEightvalue}{0.99015}
\newcommand{\InhomogExpectedAratioLCDMzdzzouHTwEightvalue}{1.00739}
\newcommand{\InhomogPerturbAratioLCDMzdzzouHTwEightvalue}{1.00735}
\newcommand{\InhomogExpectedAratioLCDMmzdzzouHTwEightvalue}{0.99257}
\newcommand{\InhomogPerturbAratioLCDMmzdzzouHTwEightvalue}{0.99253}
\newcommand{\GevolutionAccuracyPercentMaxEdSzdzzouThirtTwovalue}{0.013}

\newcommand{\GevolutionAccuracyPercentMaxEdSmzdzzouThirtTwovalue}{0.013}

\newcommand{\GevolutionAccuracyPercentMaxLCDMzdzzouThirtTwovalue}{0.013}

\newcommand{\GevolutionAccuracyPercentMaxLCDMmzdzzouThirtTwovalue}{0.013}

\newcommand{\GevolutionAccuracyPercentMaxEdSzdzzouSixtFourvalue}{0.013}

\newcommand{\GevolutionAccuracyPercentMaxEdSmzdzzouSixtFourvalue}{0.013}

\newcommand{\GevolutionAccuracyPercentMaxLCDMzdzzouSixtFourvalue}{0.013}

\newcommand{\GevolutionAccuracyPercentMaxLCDMmzdzzouSixtFourvalue}{0.013}

\newcommand{\GevolutionAccuracyPercentMaxEdSzdzzouHTwEightvalue}{0.012}

\newcommand{\GevolutionAccuracyPercentMaxEdSmzdzzouHTwEightvalue}{0.012}

\newcommand{\GevolutionAccuracyPercentMaxLCDMzdzzouHTwEightvalue}{0.013}

\newcommand{\GevolutionAccuracyPercentMaxLCDMmzdzzouHTwEightvalue}{0.013}

\newcommand{\GevolutionExpectedAratioFinalEdSzdzzouThirtTwovalue}{1.00000}
\newcommand{\GevolutionPhiAratioFinalEdSzdzzouThirtTwovalue}{1.00000}
\newcommand{\GevolutionExpectedAratioFinalEdSmzdzzouThirtTwovalue}{1.00000}
\newcommand{\GevolutionPhiAratioFinalEdSmzdzzouThirtTwovalue}{1.00000}
\newcommand{\GevolutionExpectedAratioFinalLCDMzdzzouThirtTwovalue}{1.00000}
\newcommand{\GevolutionPhiAratioFinalLCDMzdzzouThirtTwovalue}{1.00000}
\newcommand{\GevolutionExpectedAratioFinalLCDMmzdzzouThirtTwovalue}{1.00000}
\newcommand{\GevolutionPhiAratioFinalLCDMmzdzzouThirtTwovalue}{1.00000}

\newcommand{\GevolutionExpectedAratioFinalEdSzdzzouHTwEightvalue}{1.00000}
\newcommand{\GevolutionPhiAratioFinalEdSzdzzouHTwEightvalue}{1.00000}
\newcommand{\GevolutionExpectedAratioFinalEdSmzdzzouHTwEightvalue}{1.00000}
\newcommand{\GevolutionPhiAratioFinalEdSmzdzzouHTwEightvalue}{1.00000}
\newcommand{\GevolutionExpectedAratioFinalLCDMzdzzouHTwEightvalue}{1.00000}
\newcommand{\GevolutionPhiAratioFinalLCDMzdzzouHTwEightvalue}{1.00000}
\newcommand{\GevolutionExpectedAratioFinalLCDMmzdzzouHTwEightvalue}{1.00000}
\newcommand{\GevolutionPhiAratioFinalLCDMmzdzzouHTwEightvalue}{1.00000}

\newcommand{\InhomogEdszZzoHtweightOmegaMzRefvalue}{1.000}

\newcommand{\InhomogEdszZzoHtweightOmegaLamzRefvalue}{0.000}

\newcommand{\InhomogEdszZzoHtweightOmegaKzRefvalue}{0.000}

\newcommand{\InhomogEdszZzoHtweightHzRefvalue}{37.700}

\newcommand{\InhomogEdszZzoHtweightAInitRefvalue}{0.033}

\newcommand{\InhomogEdszZzoHtweightOmegaMzEffvalue}{0.952}

\newcommand{\InhomogEdszZzoHtweightOmegaLamzEffvalue}{0.000}

\newcommand{\InhomogEdszZzoHtweightOmegaKzEffvalue}{0.048}

\newcommand{\InhomogEdszZzoHtweightADotZEffvalue}{38.620}

\newcommand{\InhomogEdsmzZzoHtweightOmegaMzRefvalue}{1.000}

\newcommand{\InhomogEdsmzZzoHtweightOmegaLamzRefvalue}{0.000}

\newcommand{\InhomogEdsmzZzoHtweightOmegaKzRefvalue}{0.000}

\newcommand{\InhomogEdsmzZzoHtweightHzRefvalue}{37.700}

\newcommand{\InhomogEdsmzZzoHtweightAInitRefvalue}{0.033}

\newcommand{\InhomogEdsmzZzoHtweightOmegaMzEffvalue}{1.053}

\newcommand{\InhomogEdsmzZzoHtweightOmegaLamzEffvalue}{0.000}

\newcommand{\InhomogEdsmzZzoHtweightOmegaKzEffvalue}{-0.053}

\newcommand{\InhomogEdsmzZzoHtweightADotZEffvalue}{36.757}

\newcommand{\InhomogLcdmzZzoHtweightOmegaMzRefvalue}{0.309}

\newcommand{\InhomogLcdmzZzoHtweightOmegaLamzRefvalue}{0.691}

\newcommand{\InhomogLcdmzZzoHtweightOmegaKzRefvalue}{0.000}

\newcommand{\InhomogLcdmzZzoHtweightHzRefvalue}{67.740}

\newcommand{\InhomogLcdmzZzoHtweightAInitRefvalue}{0.034}

\newcommand{\InhomogLcdmzZzoHtweightOmegaMzEffvalue}{0.304}

\newcommand{\InhomogLcdmzZzoHtweightOmegaLamzEffvalue}{0.681}

\newcommand{\InhomogLcdmzZzoHtweightOmegaKzEffvalue}{0.015}

\newcommand{\InhomogLcdmzZzoHtweightADotZEffvalue}{68.244}

\newcommand{\InhomogLcdmmzZzoHtweightOmegaMzRefvalue}{0.309}

\newcommand{\InhomogLcdmmzZzoHtweightOmegaLamzRefvalue}{0.691}

\newcommand{\InhomogLcdmmzZzoHtweightOmegaKzRefvalue}{0.000}

\newcommand{\InhomogLcdmmzZzoHtweightHzRefvalue}{67.740}

\newcommand{\InhomogLcdmmzZzoHtweightAInitRefvalue}{0.034}

\newcommand{\InhomogLcdmmzZzoHtweightOmegaMzEffvalue}{0.314}

\newcommand{\InhomogLcdmmzZzoHtweightOmegaLamzEffvalue}{0.702}

\newcommand{\InhomogLcdmmzZzoHtweightOmegaKzEffvalue}{-0.015}

\newcommand{\InhomogLcdmmzZzoHtweightADotZEffvalue}{67.232}

\newcommand{\InhomogEdszZzoThirttworesolutionAInitRefvalue}{0.060}

\newcommand{\InhomogEdszZzoSixtfourresolutionAInitRefvalue}{0.044}

\newcommand{\InhomogEdszZzoHtweightresolutionAInitRefvalue}{0.033}

\newcommand{\GevolutionEdszZzoHtweightOmegaMzRefvalue}{1.000}

\newcommand{\GevolutionEdszZzoHtweightOmegaLamzRefvalue}{0.000}

\newcommand{\GevolutionEdszZzoHtweightOmegaKzRefvalue}{0.000}

\newcommand{\GevolutionEdszZzoHtweightHzRefvalue}{37.700}

\newcommand{\GevolutionEdszZzoHtweightPhiInitvalue}{0.001}

\newcommand{\GevolutionEdszZzoHtweightAInitRefvalue}{0.010}

\newcommand{\GevolutionEdszZzoHtweightOmegaMzEffvalue}{1.000}

\newcommand{\GevolutionEdszZzoHtweightOmegaLamzEffvalue}{0.000}

\newcommand{\GevolutionEdszZzoHtweightOmegaKzEffvalue}{0.000}

\newcommand{\GevolutionEdszZzoHtweightADotZEffvalue}{37.700}

\newcommand{\GevolutionEdsmzZzoHtweightOmegaMzRefvalue}{1.000}

\newcommand{\GevolutionEdsmzZzoHtweightOmegaLamzRefvalue}{0.000}

\newcommand{\GevolutionEdsmzZzoHtweightOmegaKzRefvalue}{0.000}

\newcommand{\GevolutionEdsmzZzoHtweightHzRefvalue}{37.700}

\newcommand{\GevolutionEdsmzZzoHtweightPhiInitvalue}{-0.001}

\newcommand{\GevolutionEdsmzZzoHtweightAInitRefvalue}{0.010}

\newcommand{\GevolutionEdsmzZzoHtweightOmegaMzEffvalue}{1.000}

\newcommand{\GevolutionEdsmzZzoHtweightOmegaLamzEffvalue}{0.000}

\newcommand{\GevolutionEdsmzZzoHtweightOmegaKzEffvalue}{0.000}

\newcommand{\GevolutionEdsmzZzoHtweightADotZEffvalue}{37.700}

\newcommand{\GevolutionLcdmzZzoHtweightOmegaMzRefvalue}{0.309}

\newcommand{\GevolutionLcdmzZzoHtweightOmegaLamzRefvalue}{0.691}

\newcommand{\GevolutionLcdmzZzoHtweightOmegaKzRefvalue}{0.000}

\newcommand{\GevolutionLcdmzZzoHtweightHzRefvalue}{67.740}

\newcommand{\GevolutionLcdmzZzoHtweightPhiInitvalue}{0.001}

\newcommand{\GevolutionLcdmzZzoHtweightAInitRefvalue}{0.010}

\newcommand{\GevolutionLcdmzZzoHtweightOmegaMzEffvalue}{0.309}

\newcommand{\GevolutionLcdmzZzoHtweightOmegaLamzEffvalue}{0.691}

\newcommand{\GevolutionLcdmzZzoHtweightOmegaKzEffvalue}{-0.000}

\newcommand{\GevolutionLcdmzZzoHtweightADotZEffvalue}{67.740}

\newcommand{\GevolutionLcdmmzZzoHtweightOmegaMzRefvalue}{0.309}

\newcommand{\GevolutionLcdmmzZzoHtweightOmegaLamzRefvalue}{0.691}

\newcommand{\GevolutionLcdmmzZzoHtweightOmegaKzRefvalue}{0.000}

\newcommand{\GevolutionLcdmmzZzoHtweightHzRefvalue}{67.740}

\newcommand{\GevolutionLcdmmzZzoHtweightPhiInitvalue}{-0.001}

\newcommand{\GevolutionLcdmmzZzoHtweightAInitRefvalue}{0.010}

\newcommand{\GevolutionLcdmmzZzoHtweightOmegaMzEffvalue}{0.309}

\newcommand{\GevolutionLcdmmzZzoHtweightOmegaLamzEffvalue}{0.691}

\newcommand{\GevolutionLcdmmzZzoHtweightOmegaKzEffvalue}{0.000}

\newcommand{\GevolutionLcdmmzZzoHtweightADotZEffvalue}{67.740}

  \newcommand{\machinearchitecture}{x86\_64}
\newcommand{\machinebyteorder}{Little Endian}

 \togglefalse{postreferee}   \togglefalse{devmode}    

\usepackage{amsmath}

\usepackage{csquotes}

\iftoggle{postreferee}{
\newcommand\postrefereechanges[1]{{\bf \color{myred} \large #1}}
        \newcommand\postrefereestart{ \bf \color{myred} }
        \newcommand\postrefereestop{ \rm \color{black} }
        \usepackage{color}  \definecolor{myred}{rgb}{0.7,0.0,0.2}
        }{
\newcommand\postrefereechanges[1]{#1}
        \newcommand\postrefereestart{  }
        \newcommand\postrefereestop{ }
        }

\usepackage{hyperref}

\usepackage{graphicx}

\usepackage[font={footnotesize}]{caption}

\usepackage{xcolor}

\usepackage[many]{tcolorbox}

\usepackage{tipa}

\newcommand{\mktab}[1]{\textcolor{black!30!white}{\_\_\_TAB\_\_\_}}

\usepackage{tikz}
\usetikzlibrary{graphs}
\usetikzlibrary{external}
\usetikzlibrary{positioning}
\tikzsetexternalprefix{tikz/}

\usepackage{pgfplots}
\pgfplotsset{compat=newest}
\usepgfplotslibrary{groupplots}
\pgfplotsset{
  axis line style={thick},
  tick style={semithick},
  tick label style = {font=\footnotesize},
  every axis label = {font=\footnotesize},
  legend style = {font=\footnotesize},
  label style = {font=\footnotesize}
  }

\tikzset{node-terminal/.style={
  rectangle,
  very thick,
  draw=blue!50,
  text centered,
  top color=white,
  minimum size=6mm,
  text width=2.1cm,
  rounded corners=3mm,
  bottom color=blue!20,
  font=\ttfamily}}

\tikzset{node-nonterminal/.style={
  rectangle,
  very thick,
  text centered,
  top color=white,
  text width=2.1cm,
  minimum size=6mm,
  draw=green!50!black!50,
  bottom color=green!80!black!50,
  font=\ttfamily}}

\tikzset{node-nonterminal-thin/.style={
  rectangle,
  thick,
  text centered,
  top color=white,
  text width=2cm,
  minimum size=2mm,
  draw=green!50!black!50,
  bottom color=green!80!black!50,
  font=\ttfamily\scriptsize}}

\tikzset{node-makefile/.style={
  thick,
  rectangle,
  anchor=south,
  minimum width=2.6cm,
  minimum height=5cm,
  draw=green!50!black!50,
  fill=black!10!green!12!white,
}}

\tikzset{node-point/.style={
  circle,
  black!50,
  inner sep=0pt,
  minimum size=0pt,
  fill=white}}

\tikzset{ bbox/.style={
  rectangle,
  minimum width=2.5cm,
  rounded corners=2mm,
  very thick,draw=blue!50,
  top color=white,
  bottom color=blue!20 } }

\tikzset{ rbox/.style={
    rectangle,
    dotted,
    minimum width=2.5cm,
    rounded corners=2mm,
    very thick,draw=red!50!black!50,
    top color=white,
    bottom color=red!50!black!20 } }

\tikzset{ gbox/.style={
    rectangle,
    minimum width=2.5cm,
    very thick,
    draw=green!50!black!50,
    top color=white,
    bottom color=green!50!black!20 } }

\tikzset{ dirbox/.style={
    thick,
    rectangle,
    anchor=north,
    text centered,
    font=\ttfamily,
    minimum width=15cm,
    minimum height=7.5cm,
    draw=brown!50!black!50,
    fill=brown!10!white }}

\providecommand\cqg{Classical \& Quantum Gravity}

\providecommand\cqg{CQG}

\providecommand\jcap{JCAP}

\providecommand\PRL{Physical Review Letters}
\providecommand\prd{Physical Review D}

\providecommand\APPBPS{Acta Physica Polon.\/ B Proc.\/ Suppl.\/}

\providecommand\aap{A\&A}

\providecommand\apjs{ApJS}
\providecommand\mnras{MNRAS}

\providecommand\pnas{Proceedings of the National Academy of Sciences}
\providecommand\pnas{Proc.~Nat.~Acad.~Sci.}

\providecommand\annrevnucpartphys{Annual Review of Nuclear and Particle Science}
\providecommand\annrevnucpartphys{Ann.~Rev.~Nucl.~Part.~Sci.}

\providecommand\CiSE{Comp.~in~Sci.~Eng.}
\providecommand\nat{Nature}

\usepackage{pstricks}

\usepackage{etoolbox} 

\providecommand\SSS{Sect.~}
\providecommand\SSSS{Sections~}
\providecommand\diffd{\mathrm{d}}

\newcommand{\CQ}{{\cal Q}}
\newcommand{\CD}{{\cal D}}

\newcommand{\CP}{{\cal P}}
\newcommand{\CR}{{\cal R}}
\newcommand{\CV}{{\cal V}}

\newcommand{\scalaverageD}[1]{\left\langle #1 \right\rangle_\CD}

\newcommand\invI{{\mathrm{I}}}
\newcommand\invII{{\mathrm{II}}}
\newcommand\invIII{{\mathrm{III}}}

\newcommand\initavinvI{{\initial{\invI}}}
\newcommand\initavinvII{{\initial{\invII}}}
\newcommand\initavinvIII{{\initial{\invIII}}}

\newcommand{\initial}[1]{{{#1}_{\mathbf i}}}
\newcommand{\nrmlstn}[1]{{{#1}_{\mathrm n}}} \newcommand{\currepoch}[1]{{{#1}_{\mathrm 0}}} \newcommand{\refmodel}[1]{\mbox{${#1}^{\mathrm r}$}} \newcommand{\effmodel}[1]{\mbox{${#1}^{\mathrm e}$}}  \newcommand{\expected}[1]{{{#1}_{\mathrm{FLRW}}}} \newcommand{\dotPhi}{\dot{\Phi\rule{0ex}{1.7ex}}} 

\newcommand\LATfieldtwo{{\sc latfield2}}

\newcommand\Omm{\Omega_{\mathrm{m}}}\newcommand\Ommzero{\Omega_{\mathrm{m0}}}  \newcommand\OmLam{\Omega_{\Lambda}} \newcommand\OmLamzero{\Omega_{\Lambda0}} \newcommand\OmLamzeroref{\Omega_{\Lambda0}^{\mathrm{ref}}}

\newcommand\Omk{\Omega_{\mathrm{k}}}

\providecommand\eeuler{\mathrm{e}}

\newcommand\mS{\mathbb{S}}
\newcommand\mE{\mathbb{E}}

\newcommand\mT{\mathbb{T}}
 \iftoggle{devmode}{

\newcommand{\InhomogAccuracyPercentEdSmzdzzouHTwEightvalue}{\InhomogAccuracyPercentEdSmzdzzouThirtTwovalue}
\newcommand{\InhomogAccuracyPercentEdSzdzzouHTwEightvalue}{\InhomogAccuracyPercentEdSzdzzouThirtTwovalue}
\newcommand{\InhomogAccuracyPercentLCDMmzdzzouHTwEightvalue}{\InhomogAccuracyPercentLCDMmzdzzouThirtTwovalue}
\newcommand{\InhomogAccuracyPercentLCDMzdzzouHTwEightvalue}{\InhomogAccuracyPercentLCDMzdzzouThirtTwovalue}
\newcommand{\InhomogEdsmzZzoHtweightADotZEffvalue}{\InhomogEdsmzZzoThirttwoADotZEffvalue}
\newcommand{\InhomogEdsmzZzoHtweightAInitRefvalue}{\InhomogEdsmzZzoThirttwoAInitRefvalue}
\newcommand{\InhomogEdsmzZzoHtweightHzRefvalue}{\InhomogEdsmzZzoThirttwoHzRefvalue}
\newcommand{\InhomogEdsmzZzoHtweightOmegaKzEffvalue}{\InhomogEdsmzZzoThirttwoOmegaKzEffvalue}
\newcommand{\InhomogEdsmzZzoHtweightOmegaKzRefvalue}{\InhomogEdsmzZzoThirttwoOmegaKzRefvalue}
\newcommand{\InhomogEdsmzZzoHtweightOmegaLamzEffvalue}{\InhomogEdsmzZzoThirttwoOmegaLamzEffvalue}
\newcommand{\InhomogEdsmzZzoHtweightOmegaLamzRefvalue}{\InhomogEdsmzZzoThirttwoOmegaLamzRefvalue}
\newcommand{\InhomogEdsmzZzoHtweightOmegaMzEffvalue}{\InhomogEdsmzZzoThirttwoOmegaMzEffvalue}
\newcommand{\InhomogEdsmzZzoHtweightOmegaMzRefvalue}{\InhomogEdsmzZzoThirttwoOmegaMzRefvalue}
\newcommand{\InhomogEdszZzoHtweightADotZEffvalue}{\InhomogEdszZzoThirttwoADotZEffvalue}
\newcommand{\InhomogEdszZzoHtweightAInitRefvalue}{\InhomogEdszZzoThirttwoAInitRefvalue}
\newcommand{\InhomogEdszZzoHtweightHzRefvalue}{\InhomogEdszZzoThirttwoHzRefvalue}
\newcommand{\InhomogEdszZzoHtweightOmegaKzEffvalue}{\InhomogEdszZzoThirttwoOmegaKzEffvalue}
\newcommand{\InhomogEdszZzoHtweightOmegaKzRefvalue}{\InhomogEdszZzoThirttwoOmegaKzRefvalue}
\newcommand{\InhomogEdszZzoHtweightOmegaLamzEffvalue}{\InhomogEdszZzoThirttwoOmegaLamzEffvalue}
\newcommand{\InhomogEdszZzoHtweightOmegaLamzRefvalue}{\InhomogEdszZzoThirttwoOmegaLamzRefvalue}
\newcommand{\InhomogEdszZzoHtweightOmegaMzEffvalue}{\InhomogEdszZzoThirttwoOmegaMzEffvalue}
\newcommand{\InhomogEdszZzoHtweightOmegaMzRefvalue}{\InhomogEdszZzoThirttwoOmegaMzRefvalue}
\newcommand{\InhomogExpectedAratioEdSmzdzzouHTwEightvalue}{\InhomogExpectedAratioEdSmzdzzouThirtTwovalue}
\newcommand{\InhomogExpectedAratioEdSzdzzouHTwEightvalue}{\InhomogExpectedAratioEdSzdzzouThirtTwovalue}
\newcommand{\InhomogExpectedAratioLCDMmzdzzouHTwEightvalue}{\InhomogExpectedAratioLCDMmzdzzouThirtTwovalue}
\newcommand{\InhomogExpectedAratioLCDMzdzzouHTwEightvalue}{\InhomogExpectedAratioLCDMzdzzouThirtTwovalue}
\newcommand{\InhomogLcdmmzZzoHtweightADotZEffvalue}{\InhomogLcdmmzZzoThirttwoADotZEffvalue}
\newcommand{\InhomogLcdmmzZzoHtweightAInitRefvalue}{\InhomogLcdmmzZzoThirttwoAInitRefvalue}
\newcommand{\InhomogLcdmmzZzoHtweightHzRefvalue}{\InhomogLcdmmzZzoThirttwoHzRefvalue}
\newcommand{\InhomogLcdmmzZzoHtweightOmegaKzEffvalue}{\InhomogLcdmmzZzoThirttwoOmegaKzEffvalue}
\newcommand{\InhomogLcdmmzZzoHtweightOmegaKzRefvalue}{\InhomogLcdmmzZzoThirttwoOmegaKzRefvalue}
\newcommand{\InhomogLcdmmzZzoHtweightOmegaLamzEffvalue}{\InhomogLcdmmzZzoThirttwoOmegaLamzEffvalue}
\newcommand{\InhomogLcdmmzZzoHtweightOmegaLamzRefvalue}{\InhomogLcdmmzZzoThirttwoOmegaLamzRefvalue}
\newcommand{\InhomogLcdmmzZzoHtweightOmegaMzEffvalue}{\InhomogLcdmmzZzoThirttwoOmegaMzEffvalue}
\newcommand{\InhomogLcdmmzZzoHtweightOmegaMzRefvalue}{\InhomogLcdmmzZzoThirttwoOmegaMzRefvalue}
\newcommand{\InhomogLcdmzZzoHtweightADotZEffvalue}{\InhomogLcdmzZzoThirttwoADotZEffvalue}
\newcommand{\InhomogLcdmzZzoHtweightAInitRefvalue}{\InhomogLcdmzZzoThirttwoAInitRefvalue}
\newcommand{\InhomogLcdmzZzoHtweightHzRefvalue}{\InhomogLcdmzZzoThirttwoHzRefvalue}
\newcommand{\InhomogLcdmzZzoHtweightOmegaKzEffvalue}{\InhomogLcdmzZzoThirttwoOmegaKzEffvalue}
\newcommand{\InhomogLcdmzZzoHtweightOmegaKzRefvalue}{\InhomogLcdmzZzoThirttwoOmegaKzRefvalue}
\newcommand{\InhomogLcdmzZzoHtweightOmegaLamzEffvalue}{\InhomogLcdmzZzoThirttwoOmegaLamzEffvalue}
\newcommand{\InhomogLcdmzZzoHtweightOmegaLamzRefvalue}{\InhomogLcdmzZzoThirttwoOmegaLamzRefvalue}
\newcommand{\InhomogLcdmzZzoHtweightOmegaMzEffvalue}{\InhomogLcdmzZzoThirttwoOmegaMzEffvalue}
\newcommand{\InhomogLcdmzZzoHtweightOmegaMzRefvalue}{\InhomogLcdmzZzoThirttwoOmegaMzRefvalue}
\newcommand{\InhomogPerturbAratioEdSmzdzzouHTwEightvalue}{\InhomogPerturbAratioEdSmzdzzouThirtTwovalue}
\newcommand{\InhomogPerturbAratioEdSzdzzouHTwEightvalue}{\InhomogPerturbAratioEdSzdzzouThirtTwovalue}
\newcommand{\InhomogPerturbAratioLCDMmzdzzouHTwEightvalue}{\InhomogPerturbAratioLCDMmzdzzouThirtTwovalue}
\newcommand{\InhomogPerturbAratioLCDMzdzzouHTwEightvalue}{\InhomogPerturbAratioLCDMzdzzouThirtTwovalue}
\newcommand{\NcrootInhomogHTwEightvalue}{\NcrootInhomogThirtTwovalue}
\newcommand{\ramsesparamlevelmaxEdSHTwEightname}{\ramsesparamlevelmaxEdSThirtTwoname}
\newcommand{\ramsesparamlevelmaxEdSHTwEightvalue}{\ramsesparamlevelmaxEdSThirtTwovalue}

\newcommand{\InhomogAccuracyPercentEdSmzdzzouSixtFourvalue}{\InhomogAccuracyPercentEdSmzdzzouThirtTwovalue}
\newcommand{\InhomogAccuracyPercentEdSzdzzouSixtFourvalue}{\InhomogAccuracyPercentEdSzdzzouThirtTwovalue}
\newcommand{\InhomogAccuracyPercentLCDMmzdzzouSixtFourvalue}{\InhomogAccuracyPercentLCDMmzdzzouThirtTwovalue}
\newcommand{\InhomogAccuracyPercentLCDMzdzzouSixtFourvalue}{\InhomogAccuracyPercentLCDMzdzzouThirtTwovalue}

\newcommand{\InhomogEdszZzoHtweightresolutionAInitRefvalue}{\InhomogEdszZzoThirttworesolutionAInitRefvalue}
\newcommand{\InhomogEdszZzoSixtfourresolutionAInitRefvalue}{\InhomogEdszZzoThirttworesolutionAInitRefvalue}

\newcommand{\GevolutionAccuracyPercentMaxEdSmzdzzouHTwEightvalue}{\GevolutionAccuracyPercentMaxEdSmzdzzouThirtTwovalue}
\newcommand{\GevolutionAccuracyPercentMaxEdSzdzzouHTwEightvalue}{\GevolutionAccuracyPercentMaxEdSzdzzouThirtTwovalue}
\newcommand{\GevolutionAccuracyPercentMaxLCDMmzdzzouHTwEightvalue}{\GevolutionAccuracyPercentMaxLCDMmzdzzouThirtTwovalue}
\newcommand{\GevolutionAccuracyPercentMaxLCDMzdzzouHTwEightvalue}{\GevolutionAccuracyPercentMaxLCDMzdzzouThirtTwovalue}
\newcommand{\GevolutionEdsmzZzoHtweightADotZEffvalue}{\GevolutionEdsmzZzoThirttwoADotZEffvalue}
\newcommand{\GevolutionEdsmzZzoHtweightAInitRefvalue}{\GevolutionEdsmzZzoThirttwoAInitRefvalue}
\newcommand{\GevolutionEdsmzZzoHtweightHzRefvalue}{\GevolutionEdsmzZzoThirttwoHzRefvalue}
\newcommand{\GevolutionEdsmzZzoHtweightOmegaKzEffvalue}{\GevolutionEdsmzZzoThirttwoOmegaKzEffvalue}
\newcommand{\GevolutionEdsmzZzoHtweightOmegaKzRefvalue}{\GevolutionEdsmzZzoThirttwoOmegaKzRefvalue}
\newcommand{\GevolutionEdsmzZzoHtweightOmegaLamzEffvalue}{\GevolutionEdsmzZzoThirttwoOmegaLamzEffvalue}
\newcommand{\GevolutionEdsmzZzoHtweightOmegaLamzRefvalue}{\GevolutionEdsmzZzoThirttwoOmegaLamzRefvalue}
\newcommand{\GevolutionEdsmzZzoHtweightOmegaMzEffvalue}{\GevolutionEdsmzZzoThirttwoOmegaMzEffvalue}
\newcommand{\GevolutionEdsmzZzoHtweightOmegaMzRefvalue}{\GevolutionEdsmzZzoThirttwoOmegaMzRefvalue}
\newcommand{\GevolutionEdsmzZzoHtweightPhiInitvalue}{\GevolutionEdsmzZzoThirttwoPhiInitvalue}
\newcommand{\GevolutionEdszZzoHtweightADotZEffvalue}{\GevolutionEdszZzoThirttwoADotZEffvalue}
\newcommand{\GevolutionEdszZzoHtweightAInitRefvalue}{\GevolutionEdszZzoThirttwoAInitRefvalue}
\newcommand{\GevolutionEdszZzoHtweightHzRefvalue}{\GevolutionEdszZzoThirttwoHzRefvalue}
\newcommand{\GevolutionEdszZzoHtweightOmegaKzEffvalue}{\GevolutionEdszZzoThirttwoOmegaKzEffvalue}
\newcommand{\GevolutionEdszZzoHtweightOmegaKzRefvalue}{\GevolutionEdszZzoThirttwoOmegaKzRefvalue}
\newcommand{\GevolutionEdszZzoHtweightOmegaLamzEffvalue}{\GevolutionEdszZzoThirttwoOmegaLamzEffvalue}
\newcommand{\GevolutionEdszZzoHtweightOmegaLamzRefvalue}{\GevolutionEdszZzoThirttwoOmegaLamzRefvalue}
\newcommand{\GevolutionEdszZzoHtweightOmegaMzEffvalue}{\GevolutionEdszZzoThirttwoOmegaMzEffvalue}
\newcommand{\GevolutionEdszZzoHtweightOmegaMzRefvalue}{\GevolutionEdszZzoThirttwoOmegaMzRefvalue}
\newcommand{\GevolutionEdszZzoHtweightPhiInitvalue}{\GevolutionEdszZzoThirttwoPhiInitvalue}
\newcommand{\GevolutionExpectedAratioFinalEdSmzdzzouHTwEightvalue}{\GevolutionExpectedAratioFinalEdSmzdzzouThirtTwovalue}
\newcommand{\GevolutionExpectedAratioFinalEdSzdzzouHTwEightvalue}{\GevolutionExpectedAratioFinalEdSzdzzouThirtTwovalue}
\newcommand{\GevolutionExpectedAratioFinalLCDMmzdzzouHTwEightvalue}{\GevolutionExpectedAratioFinalLCDMmzdzzouThirtTwovalue}
\newcommand{\GevolutionExpectedAratioFinalLCDMzdzzouHTwEightvalue}{\GevolutionExpectedAratioFinalLCDMzdzzouThirtTwovalue}
\newcommand{\GevolutionLcdmmzZzoHtweightADotZEffvalue}{\GevolutionLcdmmzZzoThirttwoADotZEffvalue}
\newcommand{\GevolutionLcdmmzZzoHtweightAInitRefvalue}{\GevolutionLcdmmzZzoThirttwoAInitRefvalue}
\newcommand{\GevolutionLcdmmzZzoHtweightHzRefvalue}{\GevolutionLcdmmzZzoThirttwoHzRefvalue}
\newcommand{\GevolutionLcdmmzZzoHtweightOmegaKzEffvalue}{\GevolutionLcdmmzZzoThirttwoOmegaKzEffvalue}
\newcommand{\GevolutionLcdmmzZzoHtweightOmegaKzRefvalue}{\GevolutionLcdmmzZzoThirttwoOmegaKzRefvalue}
\newcommand{\GevolutionLcdmmzZzoHtweightOmegaLamzEffvalue}{\GevolutionLcdmmzZzoThirttwoOmegaLamzEffvalue}
\newcommand{\GevolutionLcdmmzZzoHtweightOmegaLamzRefvalue}{\GevolutionLcdmmzZzoThirttwoOmegaLamzRefvalue}
\newcommand{\GevolutionLcdmmzZzoHtweightOmegaMzEffvalue}{\GevolutionLcdmmzZzoThirttwoOmegaMzEffvalue}
\newcommand{\GevolutionLcdmmzZzoHtweightOmegaMzRefvalue}{\GevolutionLcdmmzZzoThirttwoOmegaMzRefvalue}
\newcommand{\GevolutionLcdmmzZzoHtweightPhiInitvalue}{\GevolutionLcdmmzZzoThirttwoPhiInitvalue}
\newcommand{\GevolutionLcdmzZzoHtweightADotZEffvalue}{\GevolutionLcdmzZzoThirttwoADotZEffvalue}
\newcommand{\GevolutionLcdmzZzoHtweightAInitRefvalue}{\GevolutionLcdmzZzoThirttwoAInitRefvalue}
\newcommand{\GevolutionLcdmzZzoHtweightHzRefvalue}{\GevolutionLcdmzZzoThirttwoHzRefvalue}
\newcommand{\GevolutionLcdmzZzoHtweightOmegaKzEffvalue}{\GevolutionLcdmzZzoThirttwoOmegaKzEffvalue}
\newcommand{\GevolutionLcdmzZzoHtweightOmegaKzRefvalue}{\GevolutionLcdmzZzoThirttwoOmegaKzRefvalue}
\newcommand{\GevolutionLcdmzZzoHtweightOmegaLamzEffvalue}{\GevolutionLcdmzZzoThirttwoOmegaLamzEffvalue}
\newcommand{\GevolutionLcdmzZzoHtweightOmegaLamzRefvalue}{\GevolutionLcdmzZzoThirttwoOmegaLamzRefvalue}
\newcommand{\GevolutionLcdmzZzoHtweightOmegaMzEffvalue}{\GevolutionLcdmzZzoThirttwoOmegaMzEffvalue}
\newcommand{\GevolutionLcdmzZzoHtweightOmegaMzRefvalue}{\GevolutionLcdmzZzoThirttwoOmegaMzRefvalue}
\newcommand{\GevolutionLcdmzZzoHtweightPhiInitvalue}{\GevolutionLcdmzZzoThirttwoPhiInitvalue}
\newcommand{\GevolutionPhiAratioFinalEdSmzdzzouHTwEightvalue}{\GevolutionPhiAratioFinalEdSmzdzzouThirtTwovalue}
\newcommand{\GevolutionPhiAratioFinalEdSzdzzouHTwEightvalue}{\GevolutionPhiAratioFinalEdSzdzzouThirtTwovalue}
\newcommand{\GevolutionPhiAratioFinalLCDMmzdzzouHTwEightvalue}{\GevolutionPhiAratioFinalLCDMmzdzzouThirtTwovalue}
\newcommand{\GevolutionPhiAratioFinalLCDMzdzzouHTwEightvalue}{\GevolutionPhiAratioFinalLCDMzdzzouThirtTwovalue}
\newcommand{\NcrootGevolutionHTwEightvalue}{\NcrootGevolutionThirtTwovalue}

\newcommand{\gevnparamZeromodeamplitudeEdSPlusHTwEightvalue}{\gevnparamZeromodeamplitudeEdSPlusThirtTwovalue}
\newcommand{\gevnparamZeromodeamplitudeEdSMinusHTwEightvalue}{\gevnparamZeromodeamplitudeEdSMinusThirtTwovalue}
\newcommand{\gevnparamZeromodeamplitudeLCDMPlusHTwEightvalue}{\gevnparamZeromodeamplitudeLCDMPlusThirtTwovalue}
\newcommand{\gevnparamZeromodeamplitudeLCDMMinusHTwEightvalue}{\gevnparamZeromodeamplitudeLCDMMinusThirtTwovalue}

\newcommand{\GevolutionAccuracyPercentMaxEdSmzdzzouSixtFourvalue}{\GevolutionAccuracyPercentMaxEdSmzdzzouThirtTwovalue}
\newcommand{\GevolutionAccuracyPercentMaxEdSzdzzouSixtFourvalue}{\GevolutionAccuracyPercentMaxEdSzdzzouThirtTwovalue}
\newcommand{\GevolutionAccuracyPercentMaxLCDMmzdzzouSixtFourvalue}{\GevolutionAccuracyPercentMaxLCDMmzdzzouThirtTwovalue}
\newcommand{\GevolutionAccuracyPercentMaxLCDMzdzzouSixtFourvalue}{\GevolutionAccuracyPercentMaxLCDMzdzzouThirtTwovalue}

}{
}

\newtoggle{showDetailedDerivations}
\togglefalse{showDetailedDerivations} 

\begin{document}\title[Relativistic cosmology and volume evolution]{\projecttitle}

\author[Borkowska \& Roukema]{Justyna Borkowska$^{1}$ \&
         Boudewijn F. Roukema$^{1,2}$}
\address{$^1$ Institute of Astronomy, Faculty of Physics,
           Astronomy and Informatics, Nicolaus Copernicus
           University, Grudziadzka 5, 87-100 Toru\'n, Poland
         \\ $^2$ Univ Lyon, Ens de Lyon, Univ Lyon1, CNRS, Centre de
           Recherche Astrophysique de Lyon UMR5574, F--69007, Lyon,
           France}
\ead{jborkowska astro.uni.torun.pl}

\vspace{10pt}
\begin{indented}
\item[]Accepted \ldots Received \ldots; in original form \ldots
\end{indented}

\begin{abstract}
{Several software packages for relativistic cosmological simulations that do not fully implement the Einstein equation have recently been developed.
    Two of the free-licensed ones are {\inhomogname} and {\gevolutionname}.
    A key question is whether globally emergent volume evolution that is faster than that of a Friedmannian reference model results from the averaged effects of structure formation.}
{Checking that emergent volume evolution is correctly modelled by the packages is thus needed.}
{We numerically replace the software's default random realisation of initial seed fluctuations by a fluctuation of spatially constant amplitude in a simulation's initial conditions.
    The average volume evolution of the perturbed model should follow that of a Friedmannian expansion history that corresponds to the original Friedmannian reference solution modified by the insertion of the spatially constant perturbation.
        We derive the equations that convert from the perturbed reference solution to the effective solution.}
{We find that {\inhomogname} allows emergent volume evolution correctly at first order through to the current epoch.
    For initial conditions with a resolution of $N=\NcrootInhomogHTwEightvalue^3$ particles and an initial non-zero extrinsic curvature invariant $\initial{\invI} = \InitPerturbInhomogPlusvalue$, {\inhomogname} matches an exact Friedmannian solution to \postrefereechanges{$\InhomogAccuracyPercentEdSzdzzouHTwEightvalue$\% (Einstein--de~Sitter, EdS) or $\InhomogAccuracyPercentLCDMzdzzouHTwEightvalue$\%} ($\Lambda$CDM).
We find that {\gevolutionname} models the decaying mode to fair accuracy, and excludes the growing mode by construction.
    For $N=\NcrootGevolutionHTwEightvalue^3$ and an initial scalar potential $\Phi = \InitPerturbGevnPlusvalue$, {\gevolutionname} is accurate for the decaying mode to \postrefereechanges{$\GevolutionAccuracyPercentMaxEdSzdzzouHTwEightvalue$\% (EdS) or $\GevolutionAccuracyPercentMaxLCDMzdzzouHTwEightvalue$\%} ($\Lambda$CDM).
  }
{We conclude that this special case of an exact non-linear solution for a perturbed Friedmannian model provides a robust calibration for relativistic cosmological simulations.}
\end{abstract}


\section{Introduction} \label{s-intro}

We introduce a method of investigating whether the recently-developed free-software--licensed packages for relativistic cosmological simulations are able to model global volume evolution that may emerge as the averaged effect of inhomogeneous non-linear spatial (3-Ricci) curvature evolution.
We use simulation packages that are both licensed as free software and distributed publicly, so the whole of the scientific community is able to test the software, modify it, introduce changes to improve scientific accuracy, add functionality or correct bugs, and freely redistribute these modified versions.

We apply our method to two fast codes that take shortcuts from full solutions of the Einstein equation.
These are {\gevolutionname} \citep{AdamekDDK15,AdamekDDK16code}\footnote{Full history (v1.0, v1.1, v1.2): \url{https://codeberg.org/boud/gevolution}}, which uses second-order expansions within the Poisson gauge and has become quite popular; and {\inhomogname} \citep{Roukema17silvir,RO19flatness}\footnote{\url{https://codeberg.org/boud/inhomog}}, which uses the relativistic Zel'dovich approximation to carry out very fast calculations that in terms of standard linear perturbation theory of structure formation are effectively non-perturbative calculations \citep{BKS00,BuchRZA1,BuchRZA2}.
The {\inhomogname} package is based on the relativistic Zel'dovich approximation (RZA) for the evolution of the kinematical backreaction $\CQ_\CD$ \citep{BuchRZA1,BuchRZA2}, i.e. the $\CQ_\CD$ Zel'dovich approximation (QZA) \citep{Roukema17silvir}.
This approach uses a flow-orthogonal foliation of spacetime and a Cartan co-frame formalism.
This allows for an approach that is effectively \enquote*{non-linear} in the context of linear perturbation theory.
For example, {\inhomogname} and the QZA can be used to derive the curvature and density \enquote*{$\Omega$} parameter dependence on the expansion rate during dark-matter halo collapse without needing restrictive simplifications such as Birkhoff's law or spherical symmetry \citep{RO19flatness,VignBuch19,Ostrowski19}.
While {\gevolutionname} is designed within a general-relativistic framework -- a Poisson gauge line element -- it linearises some expressions derived within the Poisson gauge and truncates others to second order.
The ability of {\inhomogname} and {\gevolutionname} to incorporate deviations from a flat Friedmann--Lema{\^i}tre--Robertson--Walker (FLRW) reference model will be investigated in this work.

Several packages that, in principle, solve the Einstein equation fully include the following.
The Einstein Toolkit (ET) approach of \citet{BentivegnaBruni15} and \citet{Macpherson17}'s {\sc FLRWSolver}\footnote{\url{https://github.com/hayleyjm/FLRWSolver_public}} uses the BSSN formalism \postrefereechanges{(named after the authors: \citealt{BaumgarteShapiro99,ShibataNakamura95})}.
In terms of the formulae justifying their calculations, the ET calculations aim to be fully relativistic, without truncating Maclaurin expansions.
Another free-licensed package for relativistic cosmological simulations using the BSSN formalism is {\sc cosmograph} \citep{GiblinMS17}\footnote{\url{https://github.com/cwru-pat/cosmograph}}.
The {\sc gramses} package \citep{Barrera2019} is not yet public, but should be distributable \postrefereechanges{(as per copyleft licensing rules) with a CECILL-compatible\footnote{\url{https://cecill.info/index.en.html}}} licence if it is derived from the {\sc ramses} code \citep{Teyssier02}, since the latter is licensed under CECILL.

Accurate modelling of non-linear 3-Ricci (spatial) curvature is a key challenge that all of these software packages need to meet if they are to be used to test the hypothesis that structure formation leads to recently emerging negative mean curvature that physically explains \enquote*{dark energy} \citep{Rasanen06negemergcurv,
  NambuTanimoto05,KaiNambu07, Rasanen08peakmodel,
  Larena09template,Chiesa14Larenatest, WiegBuch10,BuchRas12,
  Wiltshire09timescape,DuleyWilt13,NazerW15CMB, ROB13,
  Barbosa16viscos, BolejkoC10SwissSzek, LRasSzybka13,
  Roukema17silvir, Sussman15LTBpertGIC,Chirinos16LTBav,
  Bolejko17negcurv,Bolejko17styro,
  Kras81kevolving,Kras82kevolving,Kras83kevolving,Stichel16,Stichel18,
  Coley10scalars,KasparSvitek14Cartan,RaczDobos16}.

We present a method of testing whether global volume evolution beyond that of an FLRW reference model, which could emerge from the inhomogeneous evolution of 3-Ricci curvature, is accurately modelled.
We introduce a numerically uniform perturbation on a flat FLRW reference model.
Depending on a particular code's structure and design, this should yield the global scale factor evolution of a curved or flat (depending on the perturbation) FLRW model, if the code accurately models the dynamical role of curvature and allows the emergence of global volume evolution beyond that of the reference model.
This calibration test will not work on all codes: a simulation that checks for global mathematical consistency in every successive spatial hypersection of the cosmological pseudo-manifold would detect that the topology and curvature become inconsistent in this case.
However, neither {\inhomogname} nor {\gevolutionname} claim to check consistency between global spatial topology and curvature on the fly.
They only establish this consistency in the initial conditions.
Thus, the calibration test proposed here should test emergent volume evolution without causing the code to fail due to geometry--topology constraints.

We present our method in \SSS\ref{s-algebra-generic}.
We rewrite the standard FLRW relations in a convenient form in \SSS\ref{s-algebra-FLRW-rewritten} and we present the generic way of relating the effective model to the reference model, given a perturbation, in \SSS\ref{s-algebra-ref-to-eff}.
Time matching between a pair of original and expected FLRW models is discussed in \SSS{\ref{s-time-matching}.
The relations for the FLRW scale factor solutions for the perturbed models for the specific codes are given for {\inhomogname} (\SSS\ref{s-algebra-inhomog}) and {\gevolutionname} (\SSS\ref{s-algebra-gev}).
The software details of inserting the perturbations in the codes are briefly described in \SSS\ref{s-method-inhomog} for {\inhomogname} and in \SSS\ref{s-method-gev} for {\gevolutionname}.
Results are presented in \SSS\ref{s-results}.
Appendices include a derivation of the Hamiltonian constraint (\ref{app-phiprime}) and the Raychaudhuri equation (\ref{app-Raychaudhuri}) for the {\gevolutionname} case, and the relations between the EdS reference and effective solutions illustrating decaying and growing modes (\ref{app-growing-decaying-modes}).
We discuss in \SSS\ref{s-discuss} and conclude in \SSS\ref{s-conclu}.

This package aims to be fully reproducible \citep{Akhlaghi2020maneage}, with a source package at \postrefereechanges{{\projectzenodoid}}\footnote{{\projectzenodohrefShowURL}} and live\footnote{\projectgitrepository} and archived\footnote{\projectgitrepositoryarchived} {\sc git} repositories.
This paper was produced with git commit \postrefereechanges{{\projectversion}} of the source package, which was configured, compiled and run on a {\machinebyteorder} {\machinearchitecture} architecture.

\section{Method} \label{s-method}

To test the accuracy of the packages' modelling of emergent volume evolution, we perform the simulations with and without the addition of a spatially uniform perturbation to the initial conditions of the simulation.
\postrefereechanges{We call the model without (or prior to) the introduction of a perturbation the \enquote*{reference} model, while the \enquote*{effective} model (equivalently, the \enquote*{emergent} model) is the reference model with the addition of a weak, uniform perturbation.
  We additionally use the term \enquote*{expected} model to refer to the analytically expected effective model, as opposed to the numerically simulated effective model.
  The analytical calculations, presented in \SSS\ref{s-algebra-generic}, represent the ideal, expected behaviour of the code.}
As stated above, the expected behaviour of a code that handles the perturbation correctly is to switch from the FLRW reference model to an effective model whose scale factor evolution is that of a new FLRW $a(t)$ solution.
This is, in principle, a very simple test.
However, we are not aware of any cosmological simulation codes that, as cosmological time increases, replace their initial FLRW reference model $a(t)$ solution by a new FLRW $a(t)$ solution in a large (or global) spatial domain.
The approximate methods of the codes tested in this paper are not guaranteed by construction to pass this calibration test.
The degree to which these codes fail the test for a global perturbation should provide a guideline to the degree to which they may be inaccurate on more local domains.

The generic method of doing this testing is explained via a thought experiment in \SSS\ref{s-thought-exper} and given algebraically in \SSSS\ref{s-norm-Hubble}, \ref{s-algebra-FLRW-rewritten}, \ref{s-algebra-ref-to-eff} and \ref{s-time-matching}.
The specific elements of testing {\inhomogname} are given algebraically in \SSS\ref{s-algebra-inhomog} and in terms of software and calculations in \SSS\ref{s-method-inhomog}.
Similarly, the specific elements of testing {\gevolutionname} are described in \SSS\ref{s-algebra-gev} (algebra; including the choice of perturbation mode -- a solution of the Hamiltonian and Raychaudhuri equations) and in \SSS\ref{s-method-gev} (software and calculations).
In each case, we start with an FLRW reference model defined by current-epoch values of the cosmological parameters.
The expected result is that the average volume evolution of the model should follow the evolution of an \enquote*{emergent} Friedmannian model with current-epoch values of the cosmological parameters that are different to those of the reference model.
\postrefereechanges{We consider two FLRW models: an Einstein--de Sitter universe (EdS) model and a model with a non-zero cosmological constant $\Lambda$ and cold dark matter particles ($\Lambda$CDM).
Together, EdS and $\Lambda$CDM span a reasonable range of flat FLRW models.
Both are of special interest to cosmology: EdS models are often used as a background model in works considering the possibility of dark energy being an effect of structure formation (e.g. \citealt{ROB13}; for a wider list of papers, see \SSS\ref{s-intro}), while $\Lambda$CDM (with an undetermined topology) is the current standard model in cosmology.}

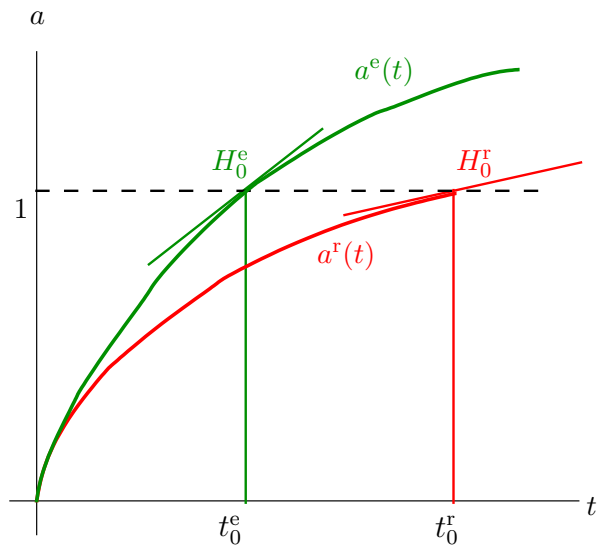
\begin{figure}
\centering
  \begingroup\makeatletter \ifx\SetFigFont\undefined \gdef\SetFigFont#1#2#3#4#5{\reset@font\fontsize{#1}{#2pt}\fontfamily{#3}\fontseries{#4}\fontshape{#5}\selectfont}\fi
\endgroup \begin{pspicture}(-0.0000,-0.0000)(7.7322,7.3893)
\psline[linewidth=0.0159,linecolor=black]{-}(0.3895,0.2688)(0.3895,6.7098)
\psline[linewidth=0.0159,linecolor=black]{-}(0.0254,0.7260)(7.6052,0.7260)
\newrgbcolor{greeni}{0.0000 0.5600 0.0000}\psline[linewidth=0.0318,linecolor=greeni]{-}(3.1686,4.8768)(3.1686,0.6858)
\psline[linewidth=0.0318,linecolor=red]{-}(5.9309,4.8768)(5.9309,0.6858)
\psline[linewidth=0.0318,linecolor=greeni]{-}(1.8690,3.8629)(4.1974,5.6832)
\psline[linewidth=0.0318,linecolor=red]{-}(4.4662,4.5254)(7.6412,5.2324)
\psline[linewidth=0.0476,linejoin=2,linecolor=red]{-}(0.3873,0.7070)(0.3873,0.7091)(0.3895,0.7154)(0.3895,0.7260)
	(0.3916,0.7408)(0.3958,0.7641)(0.4001,0.7916)(0.4064,0.8276)
	(0.4149,0.8657)(0.4233,0.9102)(0.4339,0.9588)(0.4466,1.0118)
	(0.4614,1.0668)(0.4784,1.1261)(0.4974,1.1853)(0.5207,1.2510)
	(0.5461,1.3166)(0.5778,1.3885)(0.6117,1.4626)(0.6519,1.5431)
	(0.7006,1.6277)(0.7535,1.7187)(0.8128,1.8140)(0.8805,1.9135)
	(0.9462,2.0045)(1.0118,2.0913)(1.0732,2.1696)(1.1303,2.2394)
	(1.1790,2.3008)(1.2213,2.3516)(1.2573,2.3940)(1.2869,2.4257)
	(1.3102,2.4532)(1.3271,2.4723)(1.3420,2.4892)(1.3547,2.5019)
	(1.3674,2.5125)(1.3801,2.5252)(1.3949,2.5379)(1.4118,2.5527)
	(1.4351,2.5717)(1.4647,2.5972)(1.5007,2.6310)(1.5473,2.6691)
	(1.6044,2.7199)(1.6743,2.7771)(1.7526,2.8427)(1.8436,2.9189)
	(1.9452,2.9993)(2.0532,3.0840)(2.1484,3.1559)(2.2437,3.2258)
	(2.3326,3.2914)(2.4151,3.3507)(2.4871,3.4036)(2.5527,3.4502)
	(2.6077,3.4883)(2.6543,3.5221)(2.6924,3.5496)(2.7220,3.5729)
	(2.7474,3.5899)(2.7665,3.6047)(2.7834,3.6174)(2.7961,3.6280)
	(2.8088,3.6386)(2.8215,3.6470)(2.8363,3.6576)(2.8533,3.6682)
	(2.8744,3.6830)(2.9041,3.6999)(2.9400,3.7211)(2.9845,3.7444)
	(3.0395,3.7740)(3.1051,3.8100)(3.1835,3.8481)(3.2745,3.8947)
	(3.3761,3.9455)(3.4883,3.9984)(3.6110,4.0576)(3.7401,4.1148)
	(3.8629,4.1677)(3.9857,4.2185)(4.1063,4.2672)(4.2228,4.3138)
	(4.3371,4.3561)(4.4450,4.3963)(4.5508,4.4323)(4.6524,4.4662)
	(4.7498,4.4979)(4.8450,4.5275)(4.9361,4.5551)(5.0271,4.5826)
	(5.1139,4.6080)(5.1985,4.6313)(5.2811,4.6524)(5.3615,4.6736)
	(5.4398,4.6948)(5.5139,4.7117)(5.5838,4.7308)(5.6494,4.7456)
	(5.7108,4.7604)(5.7658,4.7731)(5.8145,4.7837)(5.8547,4.7942)
	(5.8907,4.8027)(5.9182,4.8070)(5.9394,4.8133)(5.9542,4.8154)
	(5.9648,4.8175)(5.9690,4.8197)(5.9711,4.8197)
\psline[linewidth=0.0318,linestyle=dashed,dash=0.1905 0.1905,linejoin=2,linecolor=black]{-}(0.3704,4.8514)(7.0570,4.8514)
\psline[linewidth=0.0476,linejoin=2,linecolor=greeni]{-}(0.3873,0.7070)(0.3873,0.7091)(0.3873,0.7154)(0.3895,0.7260)
	(0.3916,0.7429)(0.3958,0.7641)(0.4001,0.7916)(0.4043,0.8255)
	(0.4106,0.8636)(0.4191,0.9038)(0.4297,0.9504)(0.4403,0.9991)
	(0.4551,1.0520)(0.4720,1.1091)(0.4932,1.1726)(0.5186,1.2425)
	(0.5461,1.3208)(0.5821,1.4055)(0.6223,1.5007)(0.6689,1.6023)
	(0.7112,1.6891)(0.7514,1.7716)(0.7895,1.8479)(0.8234,1.9135)
	(0.8509,1.9706)(0.8742,2.0172)(0.8932,2.0532)(0.9059,2.0807)
	(0.9165,2.0997)(0.9229,2.1145)(0.9292,2.1273)(0.9335,2.1378)
	(0.9377,2.1484)(0.9462,2.1611)(0.9546,2.1781)(0.9694,2.2013)
	(0.9885,2.2331)(1.0160,2.2733)(1.0499,2.3283)(1.0943,2.3940)
	(1.1472,2.4723)(1.2086,2.5654)(1.2806,2.6691)(1.3589,2.7813)
	(1.4245,2.8723)(1.4901,2.9612)(1.5515,3.0480)(1.6087,3.1263)
	(1.6616,3.1983)(1.7082,3.2618)(1.7505,3.3189)(1.7844,3.3676)
	(1.8140,3.4078)(1.8373,3.4438)(1.8584,3.4734)(1.8754,3.4989)
	(1.8902,3.5221)(1.9029,3.5412)(1.9135,3.5602)(1.9262,3.5793)
	(1.9389,3.5983)(1.9558,3.6216)(1.9748,3.6470)(1.9981,3.6766)
	(2.0257,3.7105)(2.0595,3.7529)(2.1019,3.8015)(2.1505,3.8566)
	(2.2077,3.9201)(2.2733,3.9920)(2.3474,4.0704)(2.4299,4.1550)
	(2.5167,4.2418)(2.6077,4.3328)(2.7072,4.4281)(2.8025,4.5170)
	(2.8935,4.5995)(2.9739,4.6715)(3.0459,4.7350)(3.1094,4.7879)
	(3.1602,4.8323)(3.2046,4.8704)(3.2406,4.8980)(3.2681,4.9212)
	(3.2914,4.9382)(3.3083,4.9530)(3.3232,4.9636)(3.3380,4.9721)
	(3.3528,4.9826)(3.3676,4.9911)(3.3867,5.0038)(3.4100,5.0186)
	(3.4396,5.0377)(3.4777,5.0631)(3.5242,5.0927)(3.5793,5.1287)
	(3.6470,5.1731)(3.7253,5.2239)(3.8142,5.2790)(3.9116,5.3425)
	(4.0195,5.4081)(4.1296,5.4758)(4.2418,5.5414)(4.3498,5.6028)
	(4.4492,5.6578)(4.5403,5.7087)(4.6228,5.7510)(4.6926,5.7870)
	(4.7540,5.8166)(4.8048,5.8420)(4.8472,5.8611)(4.8810,5.8780)
	(4.9107,5.8907)(4.9339,5.8991)(4.9551,5.9076)(4.9763,5.9140)
	(4.9953,5.9203)(5.0144,5.9267)(5.0377,5.9330)(5.0652,5.9436)
	(5.0991,5.9563)(5.1393,5.9690)(5.1858,5.9881)(5.2409,6.0092)
	(5.3065,6.0346)(5.3806,6.0643)(5.4631,6.0960)(5.5541,6.1320)
	(5.6473,6.1680)(5.7446,6.2039)(5.8632,6.2463)(5.9732,6.2844)
	(6.0748,6.3161)(6.1637,6.3458)(6.2463,6.3669)(6.3182,6.3881)
	(6.3839,6.4029)(6.4431,6.4156)(6.4982,6.4262)(6.5490,6.4368)
	(6.5955,6.4431)(6.6379,6.4495)(6.6760,6.4537)(6.7077,6.4558)
	(6.7374,6.4601)(6.7606,6.4622)(6.7797,6.4622)(6.7945,6.4643)
	(6.8030,6.4643)(6.8072,6.4643)(6.8093,6.4643)
\begingroup\SetFigFont{11}{13.2}{\familydefault}{\mddefault}{\updefault}\rput[lb](0.2984,7.0760){$a$}\endgroup \begingroup\SetFigFont{11}{13.2}{\familydefault}{\mddefault}{\updefault}\rput[lb](7.6983,0.5419){$t$}\endgroup \begingroup\SetFigFont{11}{13.2}{\familydefault}{\mddefault}{\updefault}\rput[lb](0.0868,4.4873){1}\endgroup \begingroup\SetFigFont{12}{14.4}{\familydefault}{\mddefault}{\updefault}\rput[lb](2.8088,0.1566){$t_0^{\mathrm e}$}\endgroup \begingroup\SetFigFont{12}{14.4}{\familydefault}{\mddefault}{\updefault}\rput[lb](5.6875,0.1461){$t_0^{\mathrm r}$}\endgroup \begingroup\SetFigFont{11}{13.2}{\familydefault}{\mddefault}{\updefault}\rput[lb](2.7136,5.0250){\textcolor{greeni}{$H^{\mathrm e}_0$}}\endgroup \begingroup\SetFigFont{11}{13.2}{\familydefault}{\mddefault}{\updefault}\rput[lb](5.9648,5.0313){\textcolor{red}{$H^{\mathrm{r}}_0$}}\endgroup \begingroup\SetFigFont{11}{13.2}{\familydefault}{\mddefault}{\updefault}\rput[lb](4.1169,3.8015){\textcolor{red}{$a^{\mathrm{r}}(t)$}}\endgroup \begingroup\SetFigFont{11}{13.2}{\familydefault}{\mddefault}{\updefault}\rput[lb](4.6164,6.2526){\textcolor{greeni}{$a^{\mathrm e}(t)$}}\endgroup \end{pspicture}
   \caption{Schematic figure showing the relation between scale factor evolution in the biverse model (usually called the \enquote*{background} model plus perturbations), in which the two models are referred to here as a pair of \enquote*{reference} (superscript $\refmodel{{}}$) and \enquote*{effective} (superscript $\effmodel{{}}$) models, respectively.
    The illustration assumes that the effective scale factor evolution, $\effmodel{a}(t)$, deviates significantly from that of the reference model, $\refmodel{a}(t)$.
    The two models are matched {\em ab initio}, in contrast to the typical observational matching of models, which requires models to match at the current epoch.
    Thus, the epochs at which the respective scale factors reach unity differ (in the illustration, $\effmodel{t}_0 < \refmodel{t}_0$).
    If the effective model is the $\Lambda$CDM model fit to observations (i.e., $\effmodel{t}_0 \approx 13.8$~Gyr) and the reference model is an EdS model, and the two are matched {\em ab initio}, then there are two different Hubble constants: $\effmodel{H}_0 := \effmodel{H}(\effmodel{t}_0) > \refmodel{H}_0 := \refmodel{H}(\refmodel{t}_0)$.
    Moreover, the reference model has $\refmodel{H}(\effmodel{t}_0) < \effmodel{H}_0$.
    In this paper, by construction, the expected behaviour of the effective model is an FLRW solution.
    \label{f-biverse-model}}
\end{figure}

\subsection{Thought experiment} \label{s-thought-exper}

This can be thought of in more detail via the following thought experiment, more complex than our simplified case.
Let us suppose that a time slice, either in the flow-orthogonal foliation of {\inhomogname} or the Poisson-gauge coordinate system of {\gevolutionname}, is big in volume, and homogeneous except for a \postrefereechanges{compact (finite volume)} domain $\CP$.
We assume that $\CP$ is small compared to the global volume $\CV$, but big compared to the volume scale $\CD$ of numerical gravitational averaging, i.e. $|\CD| \ll |\CP| \ll |\CV|$, where $|\cdot|$ gives the volume of a spatial domain.
We also assume that $\CP$ is itself numerically homogeneous and isotropic after averaging at the scale $\CD$, while the expansion rate or gravitational potential of $\CP$ is offset from that of the global part of the time slice.
If the simulation ignores the internal $\CD$-scale averaging and applies the Einstein equation at large scales, then the domain $\CP$ should undergo scale factor evolution $a(t)$ that can be approximated by an FLRW solution different to that of the reference model, apart from effects at the boundary between $\CP$ and its complement that may gradually propagate to the interior of $\CP$.

To avoid these boundary effects and provide an unambiguous test method, we choose $\CP$ to be the global spatial section rather than just a small part of it, i.e. $\CP \equiv \CV$.
Moreover, we remove all numerical sources of spatial inhomogeneity.
Spatial inhomogeneity remains an implicit effect represented inside of the $\CD$ scale (if the code allows that), as we consider averaging to have already been performed inside the $\CD$ scale.
Thus, we expect FLRW-like effective $a(t)$ behaviour after the perturbation has been applied, without any additional complications.
This provides us with a robust calibration test.

The algebra and calculus for this test are, in principle, simple.
However, the intuition is less simple, because of the habits of thinking in terms of the standard model.
The standard model is mathematically a biverse model: a reference pseudo-manifold and a perturbed pseudo-manifold mathematically co-exist as parallel universes, related to one another at early times by a diffeomorphism.
The reference pseudo-manifold is usually called a \enquote*{background} model.
However, in order to have the property that positive and negative perturbations statistically cancel, it is usually assumed that averaging each spatial section (time slice) of the perturbed pseudo-manifold at successive cosmological times yields an alternative definition of the background that exactly matches the original reference model.
Thus, the usual assumption is that the effective scale factor evolution is assumed to be identical to the reference model scale factor evolution.
This makes the term \enquote*{background} ambiguous when we are considering relativistic cosmology, in which the expansion--structure decoupling hypothesis is dropped.
Here, the primary question of interest is how the emergent average behaviour differs from the reference model.
Since we wish to avoid ambiguity between the reference model and the average of the perturbed model, we avoid the ambiguous term \enquote*{background}.

\subsection{Normalisation and Hubble constants} \label{s-norm-Hubble}

For both software packages, our goal is to study the emergence of super- or sub-Friedmannian volume evolution.
This requires caution when normalising FLRW models.
Normalisation usually refers to setting the scale factor to unity at the \enquote*{current epoch}.
However, two different FLRW models that are normalised this way will, in general, have different current ages, so that at a given lookback time corresponding to an early epoch, the two models will disagree in scale factor evolution.
In particular, a great enough lookback time in one model will correspond to a pre-big-bang epoch in the other.
For a simulation that starts at an early time and evolves forwards in time with both a reference FLRW model and effective super-Friedmannian expansion, a different observational normalisation is needed.
This was first shown independently by \citet{RaczDobos16} and \citet{RMBO16Hbg1}.
These papers used an EdS reference model, with the aim of approximately matching $\Lambda$CDM behaviour (a proxy for observations) in the effective model.
In this case, the initial scale factor growth in the reference and effective models is nearly identical as a function of a common cosmological time parameter, but later evolution of the scale factor in the effective model is super-Friedmannian.
The effective model reaches a scale factor of unity after about 13.8~Gyr, while the reference model takes several more Gyr to reach a scale factor of unity.
Thus, the constant values of the cosmological parameters used to parametrise the two models must differ.
As found by \citet{RaczDobos16} and \citet{RMBO16Hbg1}, and with the usual subscript ${}_0$ for the epoch of the scale factor reaching unity, an observationally realistic EdS reference model matched this way has $H_0 = 37.7$~km/s/Mpc, while the effective $\Lambda$CDM model has the usual $H_0 \approx 70$~km/s/Mpc.
The reference and effective models reach unity scale factor after about 17.3~Gyr and 13.8~Gyr, respectively.

Figure~\ref{f-biverse-model} illustrates this schematically.
The reference $\refmodel{{}}$ and effective~$\effmodel{{}}$ models are matched {\em ab initio}, requiring $\refmodel{a}(t) \approx \effmodel{a}(t)$ at early times, where the time parameter $t$ is identical for the two models.
Using this notation, the matching \citep{RaczDobos16,RMBO16Hbg1} between a reference EdS model and an effective $\Lambda$CDM model yields an effective model that reaches $\effmodel{H}(\effmodel{t}_0) \approx \effmodel{H}(13.8~\mathrm{Gyr}) \approx 70$~km/s/Mpc for a reference EdS model with $\refmodel{H}_0 \approx \refmodel{H}(17.3~\mathrm{Gyr}) \approx 37.7$~km/s/Mpc.

This multiplicity of Hubble constants is contrary to the usual intuition of \enquote*{the} Hubble constant.
Confusion can arise if the different constants are not clearly distinguished.
Thus, in a similar spirit to \citet{RMBO16Hbg1}, the following subsections present the physical reasoning and further notation.
These make it possible to compare the scale factor evolution in the reference FLRW model to the emergent FLRW scale factor evolution that would be expected if the computer code were accurate.
We ignore the contributions of radiation and neutrinos, but our test calibration could easily be extended to include them.

\subsection{\postrefereechanges{Software pipeline}}

\postrefereechanges{We would like the reader to be able to straightforwardly reproduce \citep{RougierHinsen2017Repdefn} our results, i.e., check the method and obtain identical results using precisely identified versions of high-level software, lower level software libraries, compilers and input values.
We provide the full calculation pipeline for this paper using the Maneage template (\citealt{Akhlaghi2020maneage}; see also \citealt{Akhlaghi19IAU335,Infantesainz2020,PeperRouk2021}) which aims to provide a solution to the \enquote*{reproducibility crisis} in science \citep{Peng15,Baker16,Fanelli18}.
Using the free-licensed source package, available as a version of record at {\projectzenodohref} and in \href{\projectgitrepositoryPlainURL}{live} and \href{\projectgitrepositoryarchivedPlainURL}{archived} {\sc git} repositories, the aim is that the reader should be able to produce the results of this paper on any Unix-like operating system.
Git commit hashes are used to identify contributing software sources and the full package itself.
In particular, git commit {\projectversion} of the source package was used to produce this paper.}

\postrefereechanges{For each code, we test both an EdS reference model, in which emergent volume growth may match that of the $\Lambda$CDM observational proxy, and a $\Lambda$CDM reference model itself.
While the amount of extra volume growth is not intended to be matched to observations in this paper, since the aim is to test the accuracy of the codes against the known FLRW solution, it is still preferable to use realistic values of cosmological density parameters $\Omega$ and Hubble constants $H_0$ ($\Omega_{X0}$ for the cosmological parameter of type $X$, and $H_0$, with the \enquote*{0} subscript indicating values when $a(t)=1$).
Thus, the EdS reference model constants are $\currepoch{\refmodel{\Omm}} = \OmegaMEdSvalue$, $\currepoch{\refmodel{\OmLam}} = \OmegaLEdSvalue$, and $\currepoch{\refmodel{H}} = \HubbleEdSvalue$~km/s/Mpc, as explained below (\SSS\ref{s-method}; \citealt{RaczDobos16,RMBO16Hbg1}).
The $\Lambda$CDM reference model constants are $\currepoch{\refmodel{\Omm}} = \OmegaMLCDMvalue$, $\currepoch{\refmodel{\OmLam}} = \OmegaLLCDMvalue$, and $\currepoch{\refmodel{H}} = \HubbleLCDMvalue$~km/s/Mpc.}

\subsection{Emergent FLRW model parameters: expected behaviour} \label{s-algebra-generic}

The generic method of parametrisation of a perturbed simulation is common to the two codes considered here (and, we expect, to other codes).
We define the following terminology.

\postrefereechanges{We need to calculate the cosmological parameter constants $\Omega_{X0}$ for the effective FLRW $a(t)$ model.}
The effective model and its constants are determined by the reference model modified by the addition of a uniform perturbation.
This kind of perturbation is referred to by \citet{AdamekDDK16code} as a \enquote*{homogeneous mode}.
In \citet{AdamekDDK16code}'s terminology, the emergent (effective) model represents \enquote*{absorbing the homogeneous mode}, and, in principle, could be used to replace the original reference model.
The replacement of the reference model by the effective model would enable the simulation to continue with the homogeneous mode removed (we briefly return to this in \SSS\ref{s-discuss-gevolution}).
The \enquote*{absorption} procedure is more likely to be necessary in the case of {\gevolutionname} rather than {\inhomogname}.

We first present the common steps shared by the two codes (\SSSS\ref{s-algebra-FLRW-rewritten}, \ref{s-algebra-ref-to-eff}, \ref{s-time-matching}), followed by the steps specific to the two codes in \SSSS\ref{s-algebra-inhomog} and \ref{s-algebra-gev}.

\subsubsection{Standard FLRW relations rewritten} \label{s-algebra-FLRW-rewritten}

We first write the standard FLRW relations in a general form, allowing the cosmological parameters (constants) to be evaluated at an arbitrary epoch (not necessarily the current epoch).
These relations represent, respectively, comoving conservation of rest-frame mass, normalisation by the expansion rate squared, constancy of dark energy modelled as a cosmological constant, and the Raychaudhuri equation:
\begin{align}
  \Omm &= \nrmlstn{\Omm} \,\nrmlstn{H}^2
    {a}^{-1} {\dot{a}}^{-2} {\nrmlstn{a}}^{3}
\nonumber\\
  \Omk &= \nrmlstn{\Omk} \,\nrmlstn{H}^2 \left(\frac{\dot{a}}{\nrmlstn{a}}\right)^{-2}
  \nonumber\\
  \OmLam &= \nrmlstn{\OmLam} \,\nrmlstn{H}^2
  \left(\frac{a}{\dot{a}}\right)^{2}
\nonumber\\
  \dot{a} &= \nrmlstn{\dot{a}}
     \sqrt{ \nrmlstn{\Omm} \left(\frac{a}{\nrmlstn{a}}\right)^{-1} +
            \nrmlstn{\Omk} +
            \nrmlstn{\OmLam} \left(\frac{a}{\nrmlstn{a}}\right)^{2} }
     \,.
  \label{e-FLRW-generalised}
\end{align}
The subscript $\nrmlstn{{}}$ indicates an epoch chosen for normalisation; the time-dependent cosmological parameters are $\Omm$, the non-relativistic matter density parameter, $\OmLam$, the dark energy parameter, and $\Omk$, the curvature parameter.
The derivative of the scale factor $a$ with respect to cosmological time $t$ is denoted $\dot{a}$.
The Hubble parameter is $H := \dot{a}/a$.

In the case of {\gevolutionname}, which uses conformal time in its internal calculations and output, conversions between conformal time $\tau$ and proper time $t$, via $\diffd t = a \diffd \tau$, are needed.
A convenient relation is ${\cal H} \equiv \dot{a}$, where ${\cal H}$ is \citet{AdamekDDK16code}'s conformal Hubble parameter.
We do not consider the $\dot{a} < 0$ case, which would require the negative square root for $\dot{a}$ in the expression for $\dot{a}$ in Eq.~\eqref{e-FLRW-generalised}.

As stated above, we retain the standard conventions of using the subscript $\currepoch{{}}$ for the epoch at which the scale factor attains unity, so that $\currepoch{a} :=1$.
The cosmological parameter values at an early epoch $\initial{t}$ are denoted by the subscript $\initial{{}}$.
Thus, we can rewrite the relations from Eq.~\eqref{e-FLRW-generalised} for the reference model (where the superscript $\refmodel{{}}$ is used to differentiate reference model variables from those of the effective model):
\begin{align}
  \initial{\refmodel{\Omm}} &= \currepoch{\refmodel{\Omm}} \,{{\currepoch{\refmodel{H}}}^{2}} {\left({\initial{\refmodel{a}}}\right)^{-1}} {\left({\,\initial{\refmodel{\dot{a}}}}\right)^{-2}}
  \nonumber\\
  \initial{\refmodel{\Omk}} &= \currepoch{\refmodel{\Omk}} \,{{\currepoch{\refmodel{H}}}^{2}} {\left({\,\initial{{\refmodel{\dot{a}}}}}\right)^{-2}}
  \nonumber\\
  \initial{\refmodel{\OmLam}} &= \currepoch{\refmodel{\OmLam}} \,{{\currepoch{\refmodel{H}}}^{2}} {{\initial{\refmodel{a}}}^{2}} {\left({\,\initial{\refmodel{\dot{a}}}}\right)^{-2}}
  \nonumber\\
  \initial{\refmodel{\dot{a}}} &= \currepoch{\refmodel{H}}
     \sqrt{ \currepoch{\refmodel{\Omm}} {{\initial{\refmodel{a}}}^{-1}} +
            \currepoch{\refmodel{\Omk}} +
            \currepoch{\refmodel{\OmLam}} {{\initial{\refmodel{a}}}^{2}} }
     \,.
  \label{e-FLRW-initial-ref}
\end{align}

\subsubsection{Reference FLRW to emergent FLRW conversion: common steps} \label{s-algebra-ref-to-eff}
Steps to convert from the reference FLRW $a(t)$ solution to the effective FLRW solution that are common to the two codes include the following.

We assume by default that the mass of the spatial section in an FLRW model with a $\mT^3$ (3-torus) spatial section, which is the case with both {\inhomogname} and {\gevolutionname}, is conserved, i.e.,
\begin{align}
  M = \Omm H^2 a^3 \,,
  \label{e-massconservation}
\end{align}
and allow a corrective factor depending on the assumptions of the modelling approach.

We assume the same cosmological constant $\Lambda$ for the reference and emergent models, since we do not consider dark energy models with more exotic equations of state.
This yields the conversion
\begin{align}
  \initial{\effmodel{\OmLam}} &:= \refmodel{\OmLam}
                     \left( \frac{\initial{\effmodel{a}}}{\initial{\refmodel{a}}} \right)^2
                     \left( \frac{\initial{\refmodel{\dot{a}}}}{\initial{\effmodel{\dot{a}}}} \right)^2
                            \label{e-OmegaLami-eff} \,.
\end{align}

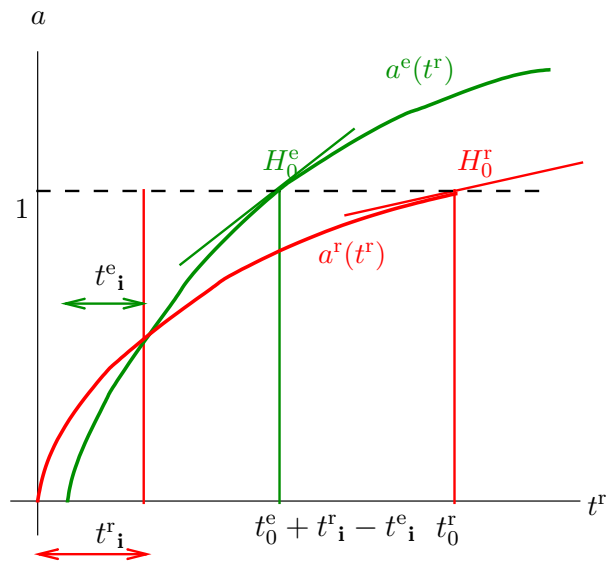
\begin{figure}
\centering
  \begingroup\makeatletter \ifx\SetFigFont\undefined \gdef\SetFigFont#1#2#3#4#5{\reset@font\fontsize{#1}{#2pt}\fontfamily{#3}\fontseries{#4}\fontshape{#5}\selectfont}\fi
\endgroup \begin{pspicture}(-0.0000,-0.0000)(7.7322,7.4803)
\psline[linewidth=0.0159,linecolor=black]{-}(0.3895,0.3598)(0.3895,6.8009)
\psline[linewidth=0.0159,linecolor=black]{-}(0.0254,0.8170)(7.6052,0.8170)
\psline[linewidth=0.0318,linecolor=red]{-}(5.9309,4.9678)(5.9309,0.7768)
\psline[linewidth=0.0318,linecolor=red]{-}(4.4662,4.6164)(7.6412,5.3234)
\newrgbcolor{greeni}{0.0000 0.5600 0.0000}\psline[linewidth=0.0318,linecolor=greeni]{-}(2.2754,3.9539)(4.6037,5.7743)
\psline[linewidth=0.0318,linecolor=greeni]{-}(3.6047,4.9678)(3.6047,0.7768)
\psline[linewidth=0.0318,linecolor=red]{-}(1.7992,4.9678)(1.7992,0.7768)
\psline[linewidth=0.0318,linecolor=greeni]{-}(0.7599,3.4438)(1.7865,3.4438)
\psline[linewidth=0.0318,linecolor=greeni]{c-c}(1.0139,3.3803)(0.7599,3.4438)(1.0139,3.5073)
\psline[linewidth=0.0318,linecolor=greeni]{c-c}(1.5325,3.5073)(1.7865,3.4438)(1.5325,3.3803)
\psline[linewidth=0.0318,linecolor=red]{-}(0.4043,0.1101)(1.8373,0.1101)
\psline[linewidth=0.0318,linecolor=red]{c-c}(0.6583,0.0466)(0.4043,0.1101)(0.6583,0.1736)
\psline[linewidth=0.0318,linecolor=red]{c-c}(1.5833,0.1736)(1.8373,0.1101)(1.5833,0.0466)
\psline[linewidth=0.0318,linestyle=dashed,dash=0.1905 0.1905,linejoin=2,linecolor=black]{-}(0.3704,4.9424)(7.0570,4.9424)
\psline[linewidth=0.0476,linejoin=2,linecolor=greeni]{-}(0.7853,0.7980)(0.7853,0.8001)(0.7853,0.8065)(0.7874,0.8170)
	(0.7895,0.8340)(0.7937,0.8551)(0.7980,0.8826)(0.8022,0.9165)
	(0.8086,0.9546)(0.8170,0.9948)(0.8276,1.0414)(0.8382,1.0901)
	(0.8530,1.1430)(0.8700,1.2002)(0.8911,1.2636)(0.9165,1.3335)
	(0.9440,1.4118)(0.9800,1.4965)(1.0202,1.5917)(1.0668,1.6933)
	(1.1091,1.7801)(1.1494,1.8627)(1.1875,1.9389)(1.2213,2.0045)
	(1.2488,2.0616)(1.2721,2.1082)(1.2912,2.1442)(1.3039,2.1717)
	(1.3144,2.1907)(1.3208,2.2056)(1.3271,2.2183)(1.3314,2.2288)
	(1.3356,2.2394)(1.3441,2.2521)(1.3525,2.2691)(1.3674,2.2923)
	(1.3864,2.3241)(1.4139,2.3643)(1.4478,2.4194)(1.4923,2.4850)
	(1.5452,2.5633)(1.6065,2.6564)(1.6785,2.7601)(1.7568,2.8723)
	(1.8225,2.9633)(1.8881,3.0522)(1.9494,3.1390)(2.0066,3.2173)
	(2.0595,3.2893)(2.1061,3.3528)(2.1484,3.4100)(2.1823,3.4586)
	(2.2119,3.4989)(2.2352,3.5348)(2.2564,3.5645)(2.2733,3.5899)
	(2.2881,3.6132)(2.3008,3.6322)(2.3114,3.6513)(2.3241,3.6703)
	(2.3368,3.6893)(2.3537,3.7126)(2.3728,3.7380)(2.3961,3.7677)
	(2.4236,3.8015)(2.4575,3.8439)(2.4998,3.8925)(2.5485,3.9476)
	(2.6056,4.0111)(2.6712,4.0831)(2.7453,4.1614)(2.8279,4.2460)
	(2.9146,4.3328)(3.0057,4.4238)(3.1051,4.5191)(3.2004,4.6080)
	(3.2914,4.6905)(3.3718,4.7625)(3.4438,4.8260)(3.5073,4.8789)
	(3.5581,4.9234)(3.6026,4.9615)(3.6386,4.9890)(3.6661,5.0123)
	(3.6893,5.0292)(3.7063,5.0440)(3.7211,5.0546)(3.7359,5.0631)
	(3.7507,5.0736)(3.7655,5.0821)(3.7846,5.0948)(3.8079,5.1096)
	(3.8375,5.1287)(3.8756,5.1541)(3.9222,5.1837)(3.9772,5.2197)
	(4.0450,5.2641)(4.1233,5.3149)(4.2122,5.3700)(4.3095,5.4335)
	(4.4175,5.4991)(4.5275,5.5668)(4.6397,5.6324)(4.7477,5.6938)
	(4.8472,5.7489)(4.9382,5.7997)(5.0207,5.8420)(5.0906,5.8780)
	(5.1520,5.9076)(5.2028,5.9330)(5.2451,5.9521)(5.2790,5.9690)
	(5.3086,5.9817)(5.3319,5.9902)(5.3530,5.9986)(5.3742,6.0050)
	(5.3933,6.0113)(5.4123,6.0177)(5.4356,6.0240)(5.4631,6.0346)
	(5.4970,6.0473)(5.5372,6.0600)(5.5838,6.0791)(5.6388,6.1002)
	(5.7044,6.1256)(5.7785,6.1553)(5.8611,6.1870)(5.9521,6.2230)
	(6.0452,6.2590)(6.1426,6.2950)(6.2611,6.3373)(6.3712,6.3754)
	(6.4728,6.4071)(6.5617,6.4368)(6.6442,6.4580)(6.7162,6.4791)
	(6.7818,6.4939)(6.8411,6.5066)(6.8961,6.5172)(6.9469,6.5278)
	(6.9935,6.5341)(7.0358,6.5405)(7.0739,6.5447)(7.1056,6.5469)
	(7.1353,6.5511)(7.1586,6.5532)(7.1776,6.5532)(7.1924,6.5553)
	(7.2009,6.5553)(7.2051,6.5553)(7.2073,6.5553)
\psline[linewidth=0.0476,linejoin=2,linecolor=red]{-}(0.3873,0.7980)(0.3873,0.8001)(0.3895,0.8065)(0.3895,0.8170)
	(0.3916,0.8319)(0.3958,0.8551)(0.4001,0.8826)(0.4064,0.9186)
	(0.4149,0.9567)(0.4233,1.0012)(0.4339,1.0499)(0.4466,1.1028)
	(0.4614,1.1578)(0.4784,1.2171)(0.4974,1.2763)(0.5207,1.3420)
	(0.5461,1.4076)(0.5778,1.4796)(0.6117,1.5536)(0.6519,1.6341)
	(0.7006,1.7187)(0.7535,1.8098)(0.8128,1.9050)(0.8805,2.0045)
	(0.9462,2.0955)(1.0118,2.1823)(1.0732,2.2606)(1.1303,2.3304)
	(1.1790,2.3918)(1.2213,2.4426)(1.2573,2.4850)(1.2869,2.5167)
	(1.3102,2.5442)(1.3271,2.5633)(1.3420,2.5802)(1.3547,2.5929)
	(1.3674,2.6035)(1.3801,2.6162)(1.3949,2.6289)(1.4118,2.6437)
	(1.4351,2.6628)(1.4647,2.6882)(1.5007,2.7220)(1.5473,2.7601)
	(1.6044,2.8109)(1.6743,2.8681)(1.7526,2.9337)(1.8436,3.0099)
	(1.9452,3.0903)(2.0532,3.1750)(2.1484,3.2470)(2.2437,3.3168)
	(2.3326,3.3824)(2.4151,3.4417)(2.4871,3.4946)(2.5527,3.5412)
	(2.6077,3.5793)(2.6543,3.6132)(2.6924,3.6407)(2.7220,3.6639)
	(2.7474,3.6809)(2.7665,3.6957)(2.7834,3.7084)(2.7961,3.7190)
	(2.8088,3.7296)(2.8215,3.7380)(2.8363,3.7486)(2.8533,3.7592)
	(2.8744,3.7740)(2.9041,3.7910)(2.9400,3.8121)(2.9845,3.8354)
	(3.0395,3.8650)(3.1051,3.9010)(3.1835,3.9391)(3.2745,3.9857)
	(3.3761,4.0365)(3.4883,4.0894)(3.6110,4.1487)(3.7401,4.2058)
	(3.8629,4.2587)(3.9857,4.3095)(4.1063,4.3582)(4.2228,4.4048)
	(4.3371,4.4471)(4.4450,4.4873)(4.5508,4.5233)(4.6524,4.5572)
	(4.7498,4.5889)(4.8450,4.6186)(4.9361,4.6461)(5.0271,4.6736)
	(5.1139,4.6990)(5.1985,4.7223)(5.2811,4.7435)(5.3615,4.7646)
	(5.4398,4.7858)(5.5139,4.8027)(5.5838,4.8218)(5.6494,4.8366)
	(5.7108,4.8514)(5.7658,4.8641)(5.8145,4.8747)(5.8547,4.8853)
	(5.8907,4.8937)(5.9182,4.8980)(5.9394,4.9043)(5.9542,4.9064)
	(5.9648,4.9085)(5.9690,4.9107)(5.9711,4.9107)
\begingroup\SetFigFont{11}{13.2}{\familydefault}{\mddefault}{\updefault}\rput[lb](0.2984,7.1670){$a$}\endgroup \begingroup\SetFigFont{11}{13.2}{\familydefault}{\mddefault}{\updefault}\rput[lb](7.6983,0.6329){$\refmodel{t}$}\endgroup \begingroup\SetFigFont{11}{13.2}{\familydefault}{\mddefault}{\updefault}\rput[lb](0.0868,4.5784){1}\endgroup \begingroup\SetFigFont{12}{14.4}{\familydefault}{\mddefault}{\updefault}\rput[lb](5.6875,0.2371){$t_0^{\mathrm r}$}\endgroup \begingroup\SetFigFont{11}{13.2}{\familydefault}{\mddefault}{\updefault}\rput[lb](5.9648,5.1223){\textcolor{red}{$H^{\mathrm{r}}_0$}}\endgroup \begingroup\SetFigFont{11}{13.2}{\familydefault}{\mddefault}{\updefault}\rput[lb](3.3761,5.1160){\textcolor{greeni}{$H^{\mathrm e}_0$}}\endgroup \begingroup\SetFigFont{12}{14.4}{\familydefault}{\mddefault}{\updefault}\rput[lb](3.2745,0.2477){$t_0^{\mathrm e} +\initial{\refmodel{t}} - \initial{\effmodel{t}}$}\endgroup \begingroup\SetFigFont{11}{13.2}{\familydefault}{\mddefault}{\updefault}\rput[lb](5.0123,6.3437){\textcolor{greeni}{$a^{\mathrm e}(\refmodel{t})$}}\endgroup \begingroup\SetFigFont{11}{13.2}{\familydefault}{\mddefault}{\updefault}\rput[lb](4.1169,3.8925){\textcolor{red}{$a^{\mathrm{r}}(\refmodel{t})$}}\endgroup \begingroup\SetFigFont{12}{14.4}{\familydefault}{\mddefault}{\updefault}\rput[lb](1.1472,3.6237){$\initial{\effmodel{t}}$}\endgroup \begingroup\SetFigFont{12}{14.4}{\familydefault}{\mddefault}{\updefault}\rput[lb](1.1578,0.1926){$\initial{\refmodel{t}}$}\endgroup \end{pspicture}
   \caption{As for Fig.~\ref{f-biverse-model}, but taking into account that the \enquote*{initial} epoch at which the perturbation is added to the reference model is $\initial{\refmodel{t}} > 0$ in the reference model, not $t=0$.
    The FLRW parametrisation of the expected effective model needs to take into account the difference between the two FLRW time parametrisations (see \SSS\protect\ref{s-time-matching}).
    \label{f-timeshift}}
\end{figure}

We set the curvature parameter of the emergent model to satisfy the Hamiltonian constraint:
\begin{align}
  \initial{\effmodel{\Omk}} &:= 1 -  \initial{\effmodel{\Omm}} - \initial{\effmodel{\OmLam}}
  \,.
    \label{e-Omegaki-eff}
\end{align}

Provided that we have expressions for $\initial{\effmodel{a}}$, $\initial{\effmodel{\Omm}}$, and $\initial{\effmodel{\dot{a}}}$, we can now use the standard FLRW relations of Eq.~\eqref{e-FLRW-generalised} to write
\begin{align}
  \currepoch{\effmodel{\Omm}} &= \initial{\effmodel{\Omm}}
  \,\initial{\effmodel{H}}^2
            \left({\currepoch{\effmodel{a}}}\right)^{-1}
            \left({\currepoch{\effmodel{\dot{a}}}}\right)^{-2}
            {\initial{\effmodel{a}}}^{3}
\nonumber\\
  \currepoch{\effmodel{\Omk}} &= \initial{\effmodel{\Omk}} \,\initial{\effmodel{H}}^2 \left(\frac{\currepoch{\effmodel{\dot{a}}}}{\initial{\effmodel{a}}}\right)^{-2}
  \nonumber\\
  \currepoch{\effmodel{\OmLam}} &= \initial{\effmodel{\OmLam}} \,\initial{\effmodel{H}}^2
\left(\frac{\currepoch{\effmodel{a}}}{\currepoch{\effmodel{\dot{a}}}}\right)^{2}
  \nonumber\\
  \currepoch{\effmodel{\dot{a}}} &= \initial{\effmodel{\dot{a}}}
     \sqrt{ \initial{\effmodel{\Omm}} \left(\frac{\currepoch{\effmodel{a}}}{\initial{\effmodel{a}}}\right)^{-1} +
            \initial{\effmodel{\Omk}} +
            \initial{\effmodel{\OmLam}} \left(\frac{\currepoch{\effmodel{a}}}{\initial{\effmodel{a}}}\right)^{2} }
     \,.
  \label{e-FLRW-effective-currepoch}
\end{align}
As stated above, we adopt $\currepoch{\effmodel{a}} = 1$, so the cosmological time at the epoch at which this occurs is $\currepoch{\effmodel{t}}$, which, in general, is not the same as the epoch $\currepoch{\refmodel{t}} \approx 13.8$~Gyr in the reference model (when $\currepoch{\refmodel{a}} = 1$).

Three of the variables needed to successfully construct an emergent FLRW model using the above equations are $\initial{\effmodel{a}}$, $\initial{\effmodel{\dot{a}}}$ and $\initial{\effmodel{\Omm}}$.
Because of the differences in the analytical justifications and corresponding code structure, the expressions for these two variables differ between {\inhomogname} and {\gevolutionname}.
These expressions are given in the respective sections (\SSS\ref{s-algebra-inhomog} and \ref{s-algebra-gev}).

\subsubsection{FLRW time matching} \label{s-time-matching}

In the method presented here, our intervention in the reference model, which is expected to switch it to an effective model, is carried out at an epoch that is early, but is later than the initial singularity.
We expect the expansion behaviour of the effective model to correspond to an exact FLRW $a(t)$ solution.
However, when we switch between the reference FLRW $a(t)$ model and the effective one, the universe ages $\initial{\refmodel{t}}$ and $\initial{\effmodel{t}}$, as calculated from their corresponding FLRW parameter values at the corresponding reference and effective scale factors, will, in general, differ at the switching epoch.
In other words, the scheme in Fig.~\ref{f-biverse-model} is an analytical oversimplification; a numerical simulation does not start at the initial singularity.
This is illustrated in Fig.~\ref{f-timeshift}.
Functionally, we can write
\begin{align}
  \initial{\refmodel{t}} &=
  t_{\mathrm{FLRW}} \left(\initial{\refmodel{a}},
           \refmodel{\currepoch{\Omm}},
           \refmodel{\currepoch{\OmLam}},
           \refmodel{\currepoch{H}} \right)
  \\
  \initial{\effmodel{t}} &=
  t_{\mathrm{FLRW}} \left(\initial{\effmodel{a}},
           \effmodel{\currepoch{\Omm}},
           \effmodel{\currepoch{\OmLam}},
           \effmodel{\currepoch{H}} \right) \,.
\end{align}
We match the two models at the switching epoch by setting
\begin{align}
  \effmodel{t} - \initial{\effmodel{t}} &= \refmodel{t} - \initial{\refmodel{t}}  \,.
  \label{e-match-t}
\end{align}
At the switching epoch itself,
$\effmodel{t} - \initial{\effmodel{t}} = 0 =\refmodel{t} - \initial{\refmodel{t}}\,,$
even though, in general, $\initial{\refmodel{t}} \ne \initial{\effmodel{t}}$.
The effective model, including Eq.~\eqref{e-match-t}, is only meaningful for $\effmodel{t} \ge \initial{\effmodel{t}}$.
The effective model universe age $\effmodel{t}$ does not represent the integral of proper time since the initial singularity; it only represents the time parameter of the FLRW $a(t)$ solution that, together with the parameters $\effmodel{\{ \currepoch{\Omm}, \currepoch{\OmLam}, \currepoch{H}\}}$, yields the expected values of $\effmodel{a}$.

For the purposes of FLRW conversions, which we calculate using {\cosmdistname}-{\cosmdistversion}\footnote{\url{https://codeberg.org/boud/cosmdist}}, we match reference and effective model \enquote*{redshifts} with scale factors in the usual way:
\begin{align}
  \refmodel{a}(t) & =: \left(1+\refmodel{z}\right)^{-1} \\
  \effmodel{a}(t) & =: \left(1+\effmodel{z}\right)^{-1} \,.
\end{align}

\subsubsection{Emergent FLRW model parameters: {\inhomogname}} \label{s-algebra-inhomog}

\postrefereechanges{In this work, as described below in \SSS\ref{s-method-inhomog}, {\inhomogname} uses a power spectrum of Gaussian initial conditions to generate $N$-body simulation initial conditions, infer initial invariants of the extrinsic curvature in spatial subdomains $\CD$, and use the relativistic Zel'dovich approximation (RZA) to calculate the evolution of the kinematical backreaction $\CQ_\CD$ \citep{BuchRZA1,BuchRZA2}, i.e. the $\CQ_\CD$ Zel'dovich approximation (QZA) \citep{Roukema17silvir}.
  The evolution of $\CQ_\CD$ yields the evolution of effective scale factors $a_{\CD}(t)$ and density parameters in each domain $\CD$.}
\postrefereechanges{The intervention in {\inhomogname} is to replace the value of the extrinsic curvature invariant $\initavinvI$ in each domain $\CD$ by a single small non-zero global value} (\SSS\ref{s-method-inhomog}).
As in \citet[][{\SSS}VI.A]{BuchRZA2}, \citet[][eq.~(20)]{Roukema17silvir}, this should not affect the scale factor at the epoch of intervention, but should affect the expansion rate, giving the relations \postrefereechanges{between the reference model and the emergent} model of
\begin{align}
  \initial{\effmodel{a}} &= \initial{\refmodel{a}}
  \label{e-ai-eff-inhomog}
  \\
  \initial{\effmodel{\dot{a}}} &=
  \initial{\refmodel{\dot{a}}} \left(1 + \frac{\initavinvI}{3} \right)
  \,.
  \label{e-adoti-eff-inhomog}
\end{align}

Mass conservation (Eq.~\eqref{e-massconservation}), corrected to satisfy the relativistic Zel'dovich approximation initial conditions (\citealt{BuchRZA2}, eq.~(92); \citealt{Roukema17silvir}, eq.~(26)) gives
\begin{align}
  \initial{\effmodel{\Omm}} &= \initial{\refmodel{\Omm}}
  \left(1 + \frac{\initavinvI}{3} \right)^{-2}
  \left(1 - \initavinvI\right) \,.
  \label{e-Omegami-eff-inhomog}
\end{align}

Here, at the instant of inserting the perturbation in the {\inhomogname} simulation, we have implicitly made a choice of projection between curved and flat space.
In the case of the {\gevolutionname} simulation (see \SSS\ref{s-method-projection} below), this projection is implicit at every time step that evaluates quantities on a mesh.
The scalar averaging approach normally does not impose any particular projection to flat space.

Using Eqs~\eqref{e-FLRW-initial-ref} for the reference model, Eqs~\eqref{e-ai-eff-inhomog} and \eqref{e-adoti-eff-inhomog} to obtain $\initial{\effmodel{a}}$ and $\initial{\effmodel{\dot{a}}}$, Eqs~\eqref{e-Omegami-eff-inhomog}, \eqref{e-OmegaLami-eff}, and \eqref{e-Omegaki-eff} for the effective initial $\Omega$ parameters, and Eq.~\eqref{e-FLRW-effective-currepoch} for the effective cosmological parameters at $\currepoch{\effmodel{t}}$ , we obtain successive conversions from
$\refmodel{\{\currepoch{a}, \currepoch{\dot{a}}, \currepoch{\Omm}, \currepoch{\OmLam}, \currepoch{\Omk}, \currepoch{H}\}}$ to
$\refmodel{\{\initial{a}, \initial{\dot{a}}, \initial{\Omm}, \initial{\OmLam}, \initial{\Omk}, \initial{H}\}}$ to
$\effmodel{\{\initial{a}, \initial{\dot{a}}, \initial{\Omm}, \initial{\OmLam}, \initial{\Omk}, \initial{H}\}}$ to
$\effmodel{\{\currepoch{a}, \currepoch{\dot{a}}, \currepoch{\Omm}, \currepoch{\OmLam}, \currepoch{\Omk}, \currepoch{H}\}}$.

\subsubsection{Emergent FLRW model parameters: {\gevolutionname}} \label{s-algebra-gev}

The emergence of an effective scale factor different to that of the reference model in {\gevolutionname} is described in \citet[][\SSS5.3]{AdamekDDK16code} as the emergence of homogeneous modes.

\postrefereechanges{The} {\gevolutionname} package works with a three-dimensional grid of particles and data fields, scattered across a lattice created by a C++ library {\LATfieldtwo} \citep{Daverio16latfield2}.
The biverse nature of the model is built as in standard perturbation theory, as stated above, in the Poisson gauge.
In the Poisson gauge, there is an unperturbed background, which we refer to here as the reference model to avoid ambiguity, and in parallel there is a nearly identical, but perturbed, universe.
However, the aim of {\gevolutionname} is to evolve the perturbed model in a background-free way, i.e., independently of the reference \postrefereechanges{model, using the reference model only to set the initial conditions of the simulation.}
Whether or not this aim is achieved is a question \postrefereechanges{that} the current work partly investigates.

\postrefereechanges{The} scale factor evolution of the reference model, $\refmodel{a}(t)$ (or $\refmodel{a}(\tau)$), is \postrefereechanges{a} solution \postrefereechanges{from the FLRW family}.
\postrefereechanges{In the code, it} is evaluated \postrefereechanges{independently} of the evolution of structure in the simulation.
If deviations from $\refmodel{a}$ are small, i.e. if $\vert \effmodel{a} - \refmodel{a}\vert \ll 1 $ at all times, then the Maclaurin expansions used to justify approximations in the code should remain accurate.
On the contrary, if there is significant non-Friedmannian effective volume \postrefereechanges{evolution -- as expected, for example,} for the structure-induced emergent dark energy \postrefereechanges{hypothesis --} then the use of low-order Maclaurin expansions may become inaccurate.

\paragraph{Choice of projection} \label{s-method-projection}

\postrefereechanges{An implicit assumption commonly made by numerical relativity codes -- including the case of {\gevolutionname} -- is that the line element contains all the information that is necessary to infer any relevant geometrical properties.
  Analytically, this is correct.
  However, numerically, some of the geometrical information is lost when performing a projection from a patch of a curved space to a corresponding patch of a flat space, and discretising to a mesh.
  The choice of projection necessarily implies a selection of which properties are conserved in the projection and which are distorted; this selection affects the mesh discretisation.
  For example, any two-dimensional sky projection from the 2-sphere $\mS^2$ into the flat 2-plane $\mE^2$ can conserve either areas (via the Jacobian), angles, or geodesics, but not all simultaneously.
  Two examples of non-conservation of properties are the following.
  The 2-sphere $\mS^2$ cannot be tiled exactly by a Cartesian mesh of isometric quadrilaterals; though it can be tiled exactly in a non-Cartesian mesh by six isometric squares, with straight equal-length edges and corner angles of 120$^\circ$ (not 90$^\circ$; the corner angles are \enquote*{distorted} if $\mS^2$ is projected to the ordinary cube).
  A projection of $\mS^2$ to $\mE^2$ followed by a mesh using parallels of latitude and meridians would lead to the poles being surrounded by small triangles on $\mS^2$, appearing as rectangles in $\mE^2$, with edges along parallels of latitude that are geodesic on $\mE^2$ but non-geodesic on $\mS^2$; geodesics on $\mS^2$ would in general be non-geodesic on $\mE^2$.
  
  For more generic cases, such as a perturbed 3-manifold modelling a spatial slice of the Universe, the projection and discretisation will affect the numerical treatment of continuous fields in curved space.
  Thus, most numerical relativity simulations necessarily make a projection and associated discretisation, where the discretisation is itself associated with the numerical resolution of the simulation.
  In principle, the inaccuracies introduced by the implicit choices of projection and discretisation can be investigated numerically by carrying out simulations of successively higher resolution and checking for numerical convergence.}

\postrefereechanges{Here,} we are interested in the particular case of {\gevolutionname} and {\LATfieldtwo}.
The assumption of {\LATfieldtwo} is that changes in cell shapes are not needed for the physical problem of interest; statistical combinations of pointwise estimates (\enquote*{finite differences}), \postrefereechanges{are} assumed to be sufficient to encode the spatially distributed geometry.
Thus, for {\gevolutionname} to handle curvature on the scale of discretisation into a mesh, prior to using a Fourier transform, a choice of projection has to be made.
\postrefereechanges{As stated above, this will result in an approximation that cannot conserve all the geometric characteristics of a cell.}
The particle-to-mesh interpolation \citep[][app.~B]{AdamekDDK16code} uses a combination of regular-cubical--pyramidal cloud-in-cell and nearest-grid-point averaging in coordinate space, i.e. the computationally simplest projection is adopted.
In the (deprojected) curved 3-manifold itself, the cubical pyramids are, in general, irregular.
For the Hamiltonian constraint, discrete averages based on sampling that appears to be uniform in flat space, rather than being intrinsically uniform on the curved 3-manifold, are used \citet[][app.~C, e.g. (C.1), (C.3)]{AdamekDDK16code}.

\postrefereechanges{In} our work, we wish to minimise changes to the internal structure of each code.
In this context, we assume that the volume of a \enquote*{cubical} cell in the perturbed model is given by the cube of the side length of the cell, where the side length is given by the (perturbed) line element in the Poisson gauge.
A more detailed model for future calibration tests could perform sub-cell discretisation uniformly in the curved space rather than in the flat space.
The choice of cubing the Poisson-gauge line-element side length translates to an assumption that {\gevolutionname}'s approximation of curvature only affects the dynamics in the domain, not the geometry on sub-cell scales that is implicitly represented by the particle distribution.
Alternative choices could be based on constant curvature models of the cells surrounding a point.
However, none of these alternative choices of geometrical projection appear to be included in {\gevolutionname}, so we do not attempt to model them.

\paragraph{Reference FLRW to emergent FLRW conversion} \label{s-algebra-ref-to-eff-old}

\postrefereechanges{The Poisson-gauge line element -- used by default in the following form in v1.1 or later of {\gevolutionname} \citep[][eq.~(6)]{Adamek2017radiation}\footnote{For example, {\tt git diff 0906a1e..fe756e7 gevolution.hpp} in the \href{https://codeberg.org/boud/gevolution}{full history}.} -- is}
\begin{align}
  \diffd s^2 &= a^2 \left[-{\eeuler}^{2\Psi} \diffd \tau^2 -
    2B_i \diffd x^i \diffd \tau +
    \delta_{ij} \eeuler^{-2\Phi} \diffd x^i \diffd x^j + h_{ij} \diffd x^i \diffd x^j \right]
  \,,
  \label{e-line-element}
\end{align}
where $\tau$ is conformal time in the reference model \postrefereechanges{(defined by $\diffd \tau = a^{-1} \diffd t$)}, Roman indices $i,j$ are spatial, giving $x^i$ as the spatial coordinates.
\postrefereechanges{The parameters that define the model are $\Psi$, $\Phi$, $B_i$ and $h_{ij}$, where $\Psi$ and $\Phi$ are scalar functions, and  $B_i$ and $h_{ij}$ are vector and tensor modes, respectively.
We use the Kronecker $\delta_{ij}$ ($\delta_{ij} = 1$ when $i=j$, $\delta_{ij} = 0$ otherwise; similarly for $\delta^{ij}$ and $\delta^i_j$ below).
For this work, as we are modelling a perturbation effectively corresponding to the homogeneous mode, we set $\chi := \Phi - \Psi$ to zero, as in \citet[][\SSS5.3]{AdamekDDK16code} for the homogeneous mode.}
Although the vector and tensor modes appear in the line element, for the purposes of this test their homogeneous modes have been set to zero, so they do not contribute to the emergent FLRW $a(t)$ model that is expected.
We write $\dot{\ } := \diffd/\diffd t$ and ${\ }' := \diffd/\diffd \tau$, so that
\begin{align}
  \frac{\diffd\Phi}{\diffd t} =: \dotPhi = \frac{\Phi'}{\refmodel{a}} := \frac{\diffd \Phi}{\refmodel{a} \diffd\tau} \,.
  \label{e-phi-t-tau-relation}
\end{align}

\postrefereechanges{Continuing with biverse terminology, the reference model has the line element of Eq.~\eqref{e-line-element} with $\Psi \equiv \Phi \equiv 0$, while the perturbed model only has the constraint $\Psi \equiv \Phi$ (as stated above).
At time $\initial{t}$, we now want to convert the set of model-defining parameters $\refmodel{\{\initial{a}, \initial{\dot{a}}, \initial{\Omm}, \initial{\OmLam}, \initial{\Omk}, \initial{H}\}}$ of the reference model to the set $\effmodel{\{\initial{a}, \initial{\dot{a}}, \initial{\Omm}, \initial{\OmLam}, \initial{\Omk}, \initial{H}\}}$ of an FLRW effective model.
The latter should correspond to the reference model perturbed by the spatially constant potential $\Phi$.}

According to the \postrefereechanges{perturbed metric} (Eq.~\eqref{e-line-element}), distance scales proportionately to $\eeuler^{-\Phi}$, and the scale factor is linearly proportional to distance.
We compare the reference and perturbed model at the same epoch $\initial{t},$ giving
\begin{align}
  \initial{\effmodel{a}} &:= \effmodel{a}(\initial{t}) = \initial{\refmodel{a}} \, \eeuler^{-\Phi} \,.
  \label{e-ai-eff-gev}
\end{align}

\postrefereechanges{To relate the density parameters, we assume that the total mass content of a given volume of the Universe is the same for the reference and emergent models, i.e. $\effmodel{M} = \refmodel{M}$, and use the relations $a \propto \eeuler^{-\Phi}$ and $H := \dot{a} / a$.
We calculate the perturbed volume by assuming that it can be described as the cube of the side length described by the perturbed metric (as described in \SSS\ref{s-method-projection}, strictly valid only in flat space).
Equation~\eqref{e-massconservation} yields}
\begin{align}
  \initial{\effmodel{\Omm}} &:= \initial{\refmodel{\Omm}}
       \,\eeuler^{\Phi}
       \left( \frac{\initial{\refmodel{\dot{a}}}}{\initial{\effmodel{\dot{a}}}} \right)^2
       \,.
       \label{e-Omegami-eff-gev}
\end{align}

\begin{figure}
  \centering
  \includegraphics[width=0.6\columnwidth]{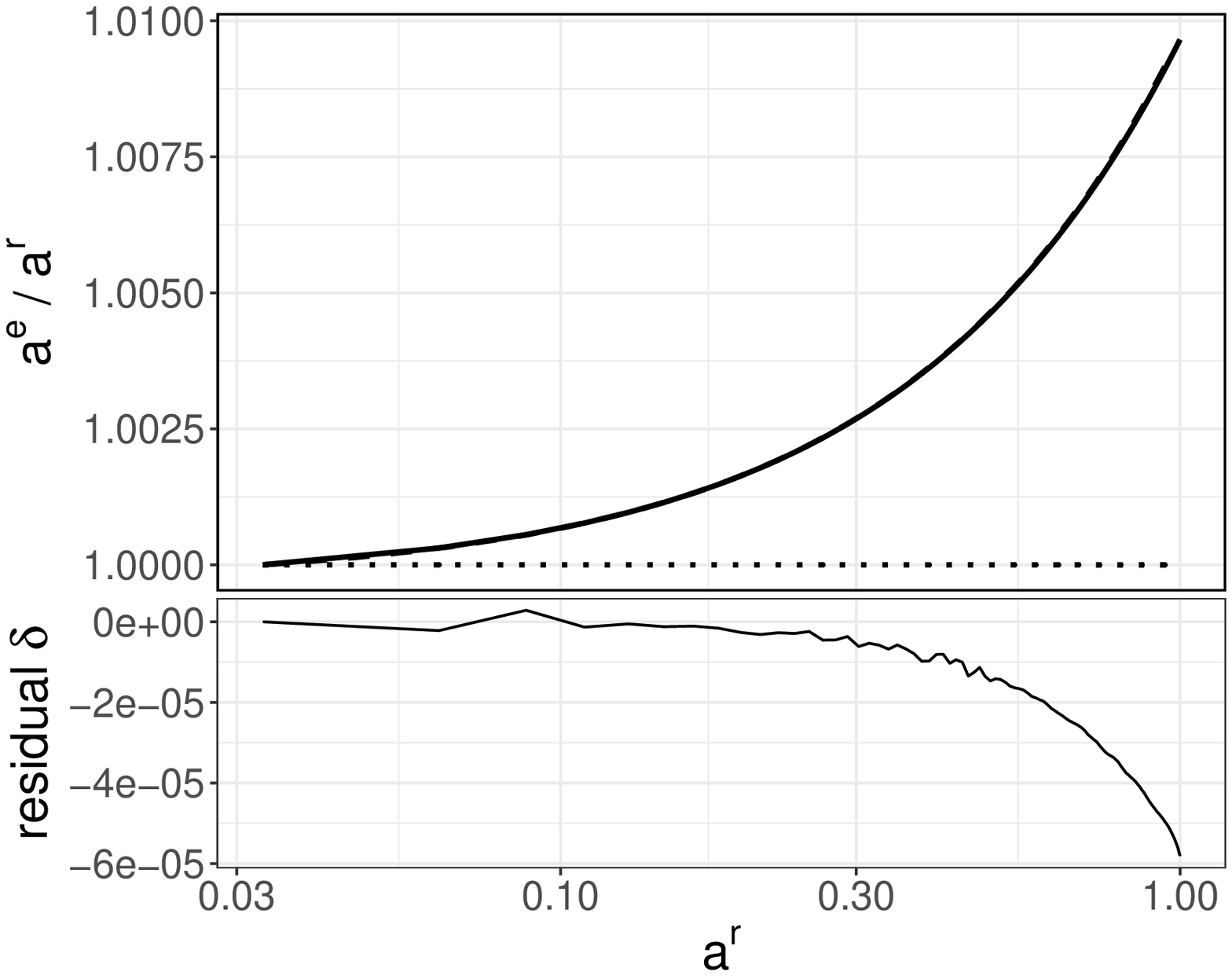}
  \includegraphics[width=0.6\columnwidth]{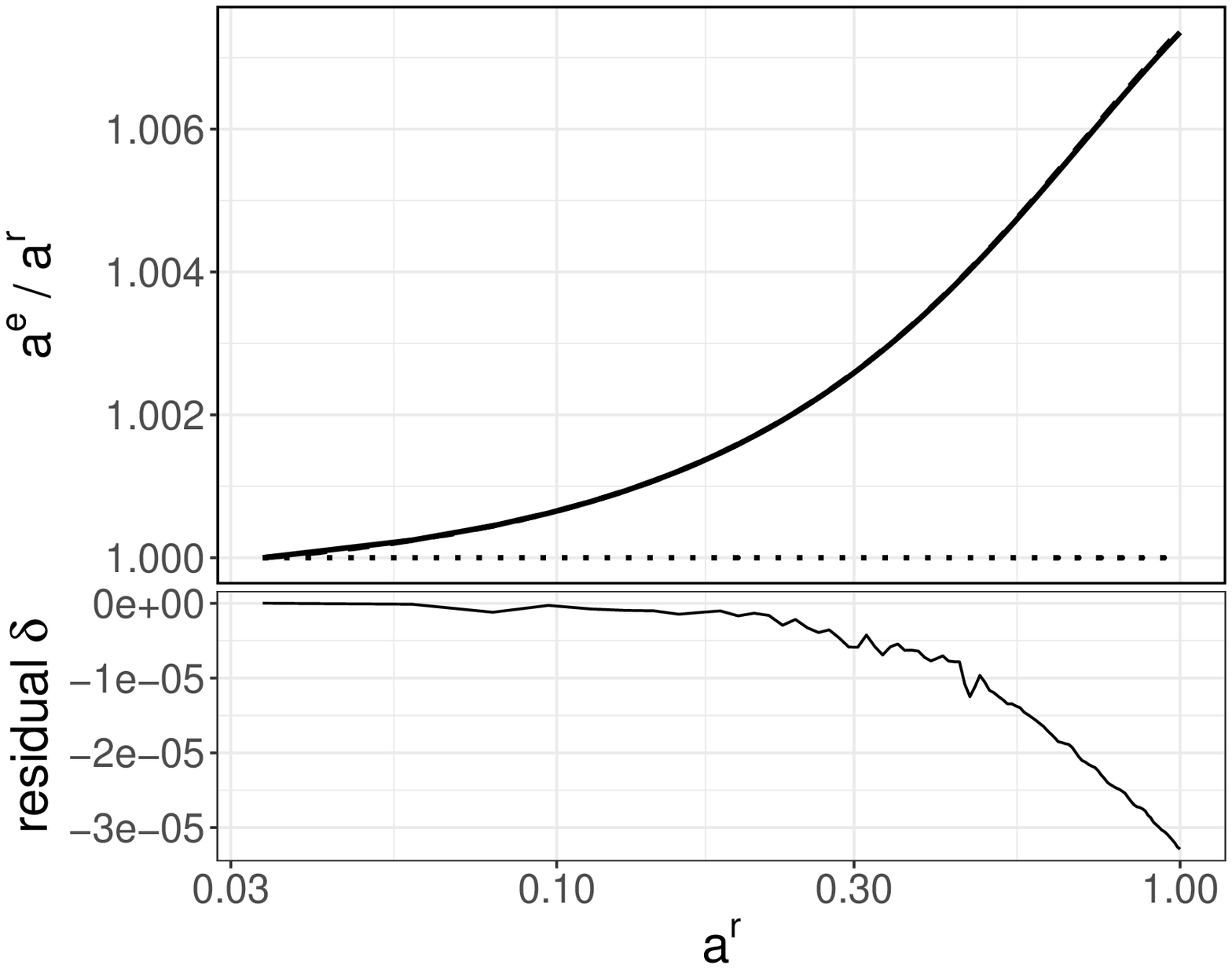}
  \caption{Emergent-to-reference model scale factor ratios
    $\effmodel{a}/\refmodel{a}$ for an {\inhomogname} simulation without addition of a perturbation (dotted line) and
    $\effmodel{a}/\refmodel{a}$ for an {\inhomogname} simulation with an initial invariant perturbation $\initial{\invI}=\InitPerturbInhomogPlusvalue$ (solid line).
    The expected FLRW evolution $\expected{\effmodel{a}}/\refmodel{a}$ is also plotted as a dashed line, but is visually indistinguishable from the numerically measured value.
    The residual $\delta$ (Eq.~\protect\eqref{e-resid-inhomog}) shows the numerical differences between these two curves.
    {\em Top:} EdS reference model; {\em bottom:} $\Lambda$CDM.
    Plain-text data for this figure and Fig.~\protect\ref{f-flat-FLRW-Iinv-test-minus} at \href{\projectzenodofilesbase/inhomog_scale_factors_EdS.dat}{\projectzenodoid/inhomog\_scale\_factors\_EdS.dat}, \href{\projectzenodofilesbase/inhomog_scale_factors_LCDM.dat}{\projectzenodoid/inhomog\_scale\_factors\_LCDM.dat}.
    \label{f-flat-FLRW-Iinv-test-plus}}
\end{figure}

\begin{figure}
  \centering
  \includegraphics[width=0.6\columnwidth]{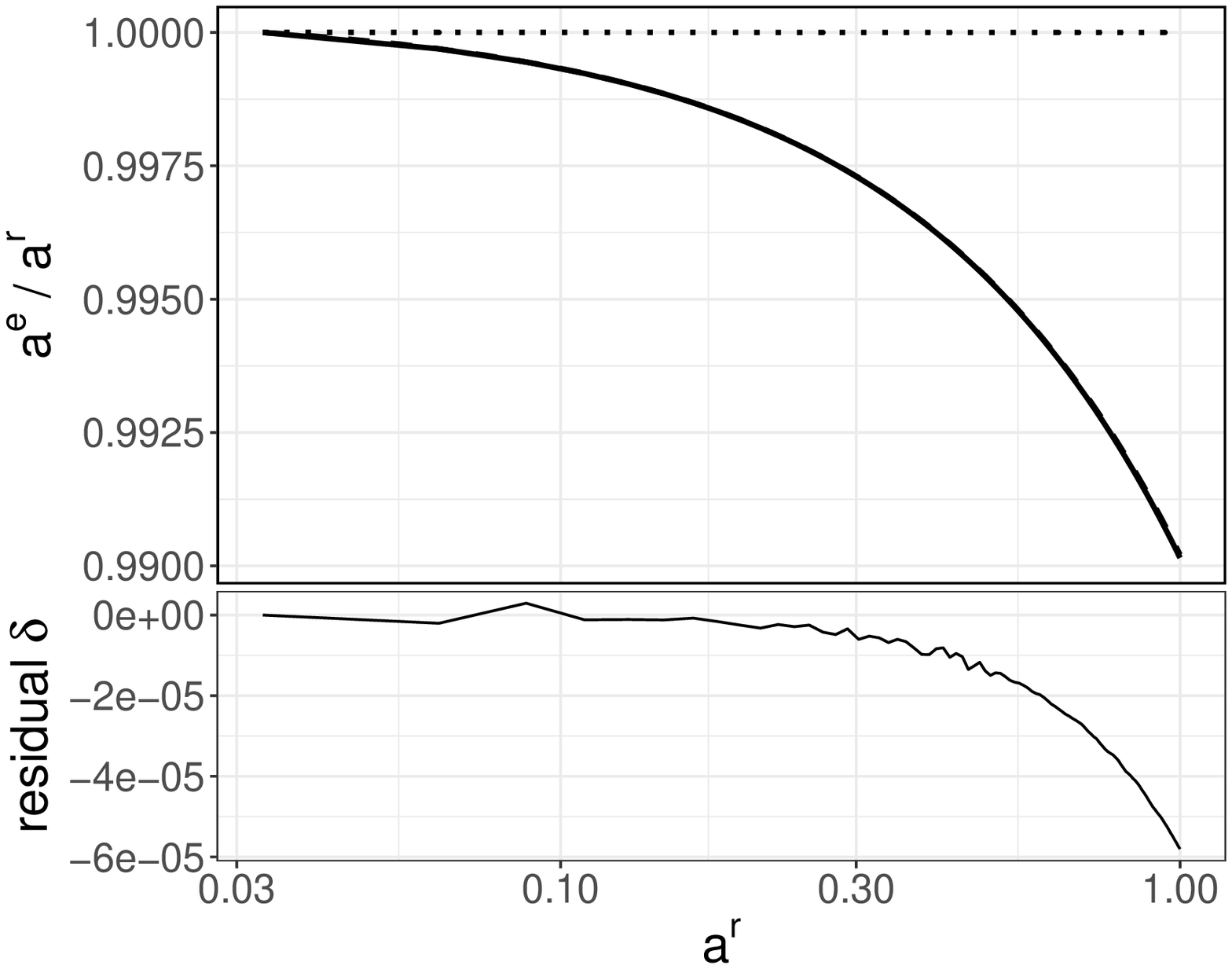}
  \includegraphics[width=0.6\columnwidth]{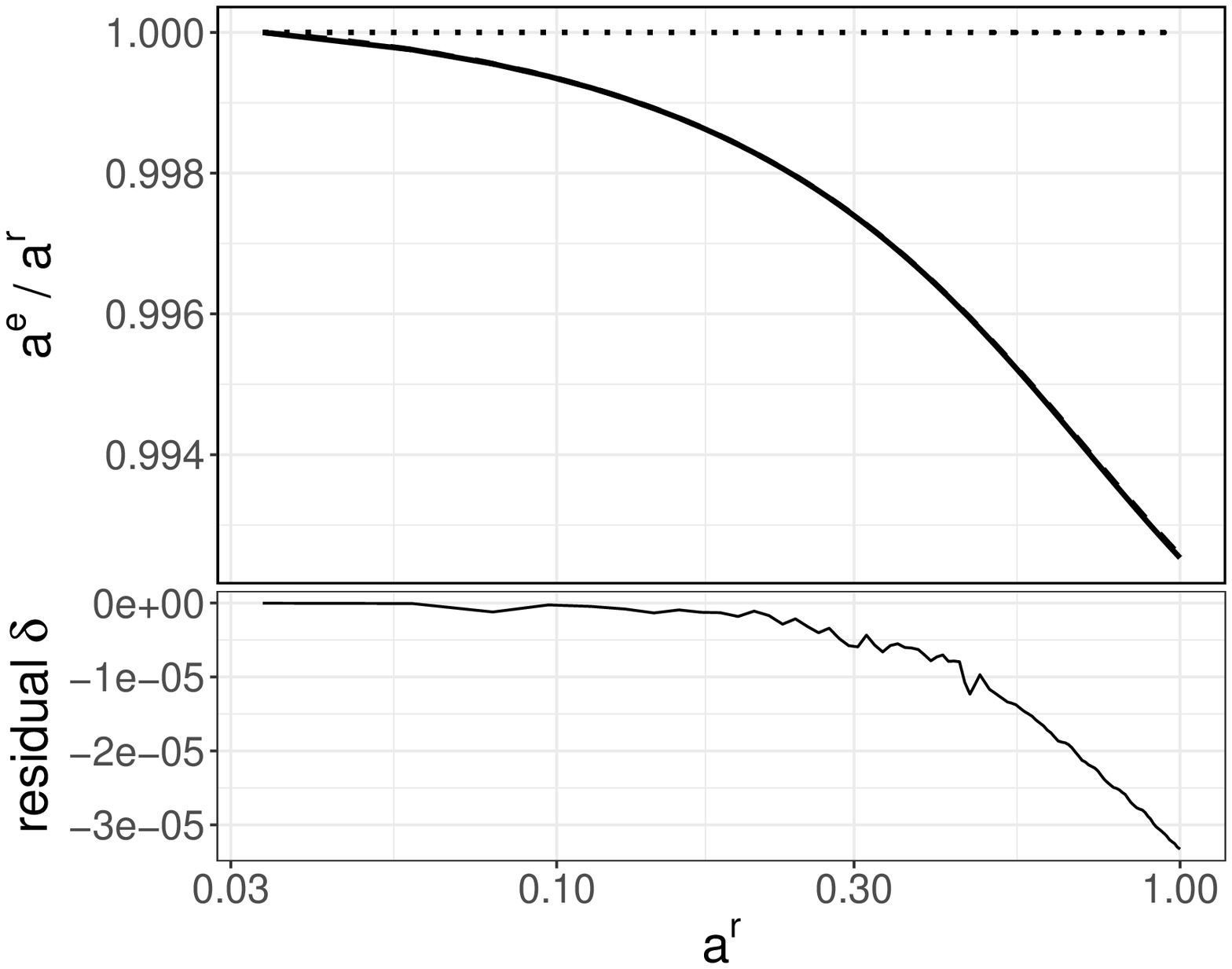}
  \caption{Emergent-to-reference model scale factor ratios, as for Fig.~\protect\ref{f-flat-FLRW-Iinv-test-plus}, for $\initial{\invI}=\InitPerturbInhomogMinusvalue$.
    {\em Top:} EdS reference model; {\em bottom:} $\Lambda$CDM.
    \label{f-flat-FLRW-Iinv-test-minus}}
\end{figure}

To evaluate the expansion rate $\effmodel{\dot{a}}$ in the emergent model, consider the thought experiment of a compact homogeneous domain $\CD$ within a global homogeneous spatial hypersurface, as presented above (\SSS\ref{s-thought-exper}).
We can make the conservative assumption that the line element (Eq.~\eqref{e-line-element}) correctly models both $\CD$ (where $\Phi \equiv \Psi \neq 0$) and the complement of $\CD$ (where $\Phi \equiv \Psi = 0$).
\postrefereechanges{The {\gevolutionname} model has no information about the different expansion rates $\dot{a}$ of $\CD$ and the complement of $\CD$ in this thought experiment; this information is indirectly contained only in the form of the scalar functions $\Phi$ (in the spatial directions) and $\Psi$ (in the time direction).
The time part of the line element in the effective model differs from that in the reference model by a factor of $\eeuler^\Psi$.
Returning to our reference and effective models, remembering that $\Phi \equiv \Psi$, differentiating Eq.~\eqref{e-ai-eff-gev} and dividing by the change in effective time gives}
\begin{align}
  \initial{\effmodel{\dot{a}}}
  &= \left( \initial{\refmodel{\dot{a}}} \, \eeuler^{-\Phi}  + \initial{\refmodel{a}}\left(-\eeuler^{-\Phi}\right) \dotPhi \right) \eeuler^{-\Phi}
  \nonumber \\
  &= \eeuler^{-2\Phi} \initial{\refmodel{\dot{a}}} \left( 1 - \frac{\initial{\refmodel{a}}}{\initial{\refmodel{\dot{a}}}} \dotPhi \right)
  \label{e-adoti-eff-gev}
  \\
  &= \eeuler^{-2\Phi} \initial{\refmodel{\dot{a}}} \left( 1 - \initial{\refmodel{a}} \initial{\left(\frac{\diffd \Phi}{\diffd \refmodel{a}}\right)} \right)
  \,.
  \label{e-adoti-eff-dphidaref-gev}
\end{align}
To first order in $\Phi$ and dropping terms in $\Phi \,{\diffd \Phi}/{\diffd \initial{\refmodel{a}}}$ or higher, Eq.~\eqref{e-adoti-eff-dphidaref-gev} reduces to
\begin{align}
  \initial{\effmodel{\dot{a}}} &\approx
     \initial{\refmodel{\dot{a}}}
     \left(1 - 2\Phi - \initial{\refmodel{a}}
     \initial{\left(\frac{\diffd \Phi}{\diffd \refmodel{a}}\right)} \right) \,,
\end{align}
as given in eq.~(5.11) of \citet{AdamekDDK16code}, where Eq.~\eqref{e-ai-eff-gev} corresponds to the second part of (5.11).

To evaluate Eq.~\eqref{e-adoti-eff-gev}, once a perturbation $\Phi$ is chosen, the Hamiltonian constraint in the {\gevolutionname} context, i.e. without allowing a mean per-mesh-cell curvature term, allows a flat perturbation that is either decaying or growing, as derived in \ref{app-phiprime} (the Raychaudhuri equation is derived in \ref{app-Raychaudhuri}).

Only one of these two solutions is allowed in {\gevolutionname}. This can be seen as follows.
\postrefereechanges{Both equation~(2.9) of \citet{AdamekDDK16code} -- for the older {\gevolutionname} form of the line element -- and eq.~(9) of \citet{Adamek2017radiation} -- for the form used here (Eq.~\eqref{e-line-element}) -- are linearised in $\Phi'$ (they are approximations that ignore some terms).
The non-linearised Einstein tensor time-time component $G^{00}$ is quadratic in $\Phi'$.
Thus, solving the Hamiltonian constraint for $\Phi'$ -- with the assumption of $\Phi$ being spatially homogeneous -- yields a pair of solutions.
Algebraically, these correspond to the choice of either a plus or a minus sign in solving a quadratic equation (see \ref{app-phiprime} for details).
Approximation by linearisation ignores one of the two solutions of the quadratic.
For example, if we write a general quadratic equation as $\alpha \xi^2 + \beta\xi + \gamma = 0$, then approximating $|\xi|^2 \ll |(\beta/\alpha)\xi| $ gives the solution $\xi \approx -\gamma/\beta$.
In reality, the quadratic equation has (for a positive discriminant) two real solutions.
In the case of {\gevolutionname}, the calculations in the code match the linearised solution (Eq.~(9) of \citet{Adamek2017radiation}) that omits terms in $(\Phi')^2$ or of higher order.
Thus, the linearisation effectively chooses the sign of the solution to the non-linearised equation (i.e.\/ the solution of Eq.~\eqref{e-Hamiltonian-constraint-quadratic} by linearisation is set to the negative branch of the square root in Eq.~\eqref{e-phiprime-solution-pair}).
This choice corresponds to the decaying mode of the Raychaudhuri equation, which relates $\Phi''$ to $\Phi'$ and $\Phi$ (Eq.~\eqref{e-Raychaudhuri-Phi}).
Without linearisation, the growing mode is not excluded.
We return to the physical meaning of the growing mode, and whether the approximation adopted by {\gevolutionname} could imply that the code fails to correctly model structure evolution, in \SSS\ref{s-discuss-gevolution} below.}

Thus, to check that {\gevolutionname} behaves as expected, using Eq.~\eqref{e-FLRW-initial-ref} for the reference model, the decaying branch in Eq.~\eqref{e-phiprime-solution-pair} (negative sign in $\pm$), Eqs~\eqref{e-ai-eff-gev} and \eqref{e-adoti-eff-gev} to obtain $\initial{\effmodel{a}}$ and $\initial{\effmodel{\dot{a}}}$, Eqs~\eqref{e-Omegami-eff-gev}, \eqref{e-OmegaLami-eff}, and \eqref{e-Omegaki-eff} for the effective initial $\Omega$ parameters, and Eq.~\eqref{e-FLRW-effective-currepoch} for the effective cosmological parameters at $\currepoch{\effmodel{t}}$, we obtain successive conversions from
$\refmodel{\{\currepoch{a}, \currepoch{\dot{a}}, \currepoch{\Omm}, \currepoch{\OmLam}, \currepoch{\Omk}, \currepoch{H}\}}$ to
$\refmodel{\{\initial{a}, \initial{\dot{a}}, \initial{\Omm}, \initial{\OmLam}, \initial{\Omk}, \initial{H}\}}$ to
$\effmodel{\{\initial{a}, \initial{\dot{a}}, \initial{\Omm}, \initial{\OmLam}, \initial{\Omk}, \initial{H}\}}$ to
$\effmodel{\{\currepoch{a}, \currepoch{\dot{a}}, \currepoch{\Omm}, \currepoch{\OmLam}, \currepoch{\Omk}, \currepoch{H}\}}$, i.e. we have the parameters of the expected FLRW scale factor solution.

\begin{figure}
  \centering
  \includegraphics[width=0.6\columnwidth]{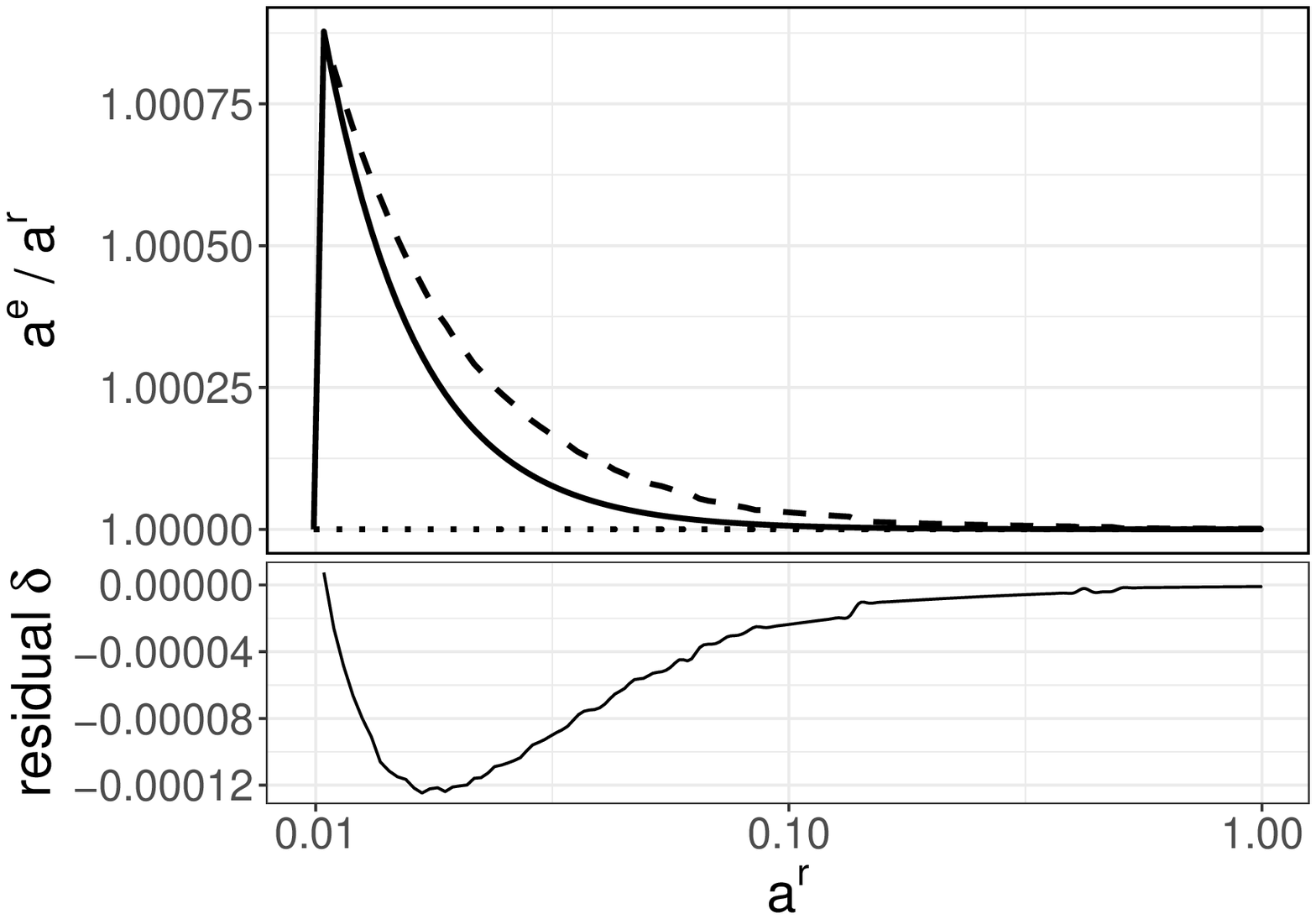}
  \includegraphics[width=0.6\columnwidth]{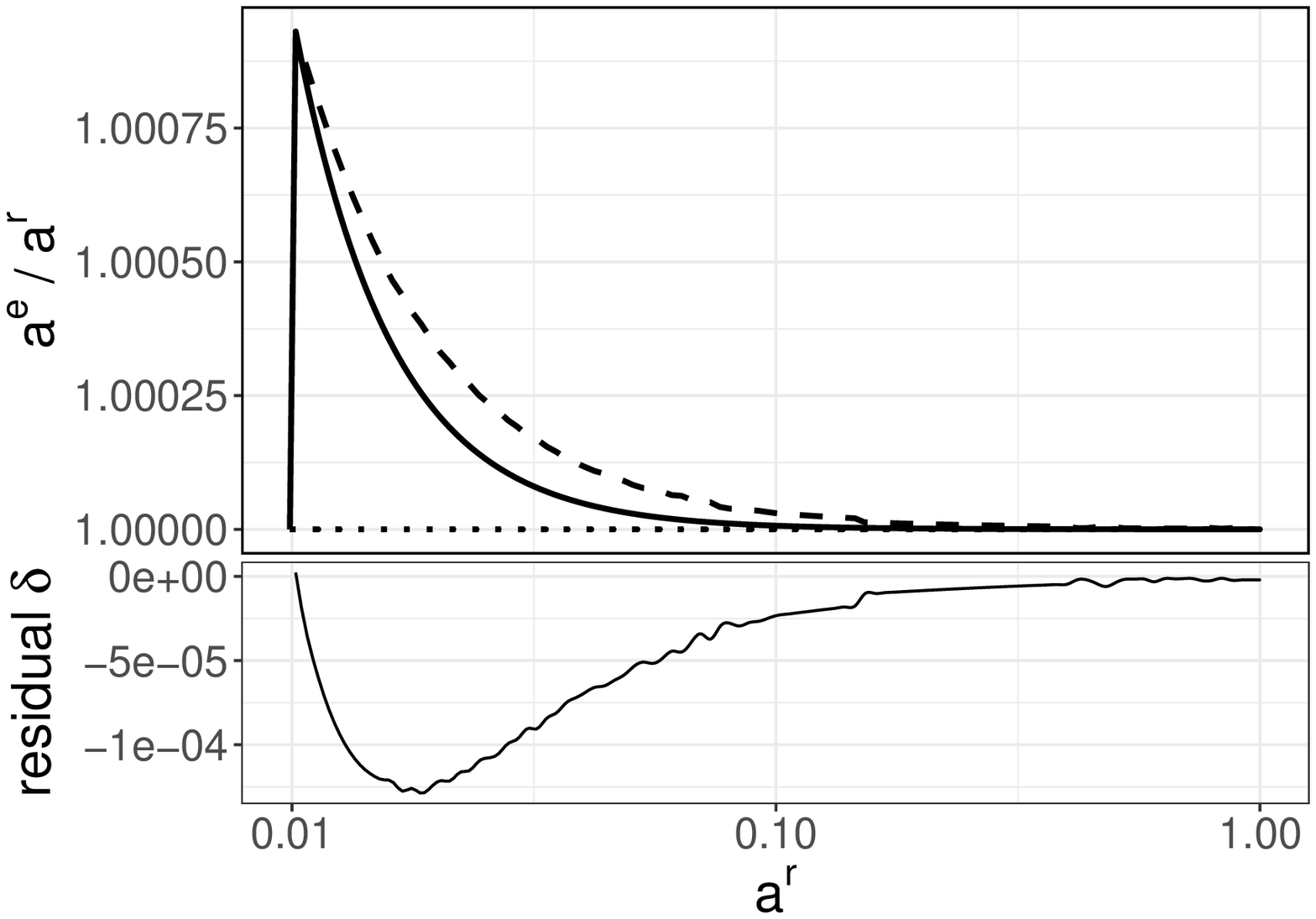}
  \caption{Scale factor ratios
    $\effmodel{a}_{\initial{\Phi}=0}/\refmodel{a}$ for a standard {\gevolutionname} simulation (dotted line);
    $\effmodel{a}_{\initial{\Phi}=\gevnparamZeromodeamplitudeEdSPlusHTwEightvalue}/\refmodel{a}$ for a {\gevolutionname} simulation with an initial potential perturbation $\initial{\Phi}=\gevnparamZeromodeamplitudeEdSPlusHTwEightvalue$ (solid line); and
    the expected FLRW evolution of $\effmodel{a}_{\mathrm{FLRW}}/\refmodel{a}$ (as defined in \SSS\ref{s-algebra-gev}; dashed line), for the decaying mode (negative square root in Eq.~\protect\ref{e-phiprime-solution-pair}).
    {\em Top:} EdS reference model; {\em bottom:} $\Lambda$CDM.
    The sudden increase in value for the perturbed {\gevolutionname} simulation (solid curve) at an early time step is the effect of the $\initial{\Phi}$ offset being inserted into the simulation, and this datapoint is removed from residuals to improve readability.
    The expected and actual curves match prior to this initial peak in the curve, by construction.
    Plain-text data for this figure and Fig.~\protect\ref{f-flat-FLRW-Phi-test-minus} at \href{\projectzenodofilesbase/gevolution_scale_factors_EdS.dat}{\projectzenodoid/gevolution\_scale\_factors\_EdS.dat}, \href{\projectzenodofilesbase/gevolution_scale_factors_LCDM.dat}{\projectzenodoid/gevolution\_scale\_factors\_LCDM.dat}.
    \label{f-flat-FLRW-Phi-test-plus}}
\end{figure}

\begin{figure}
  \centering
  \includegraphics[width=0.6\columnwidth]{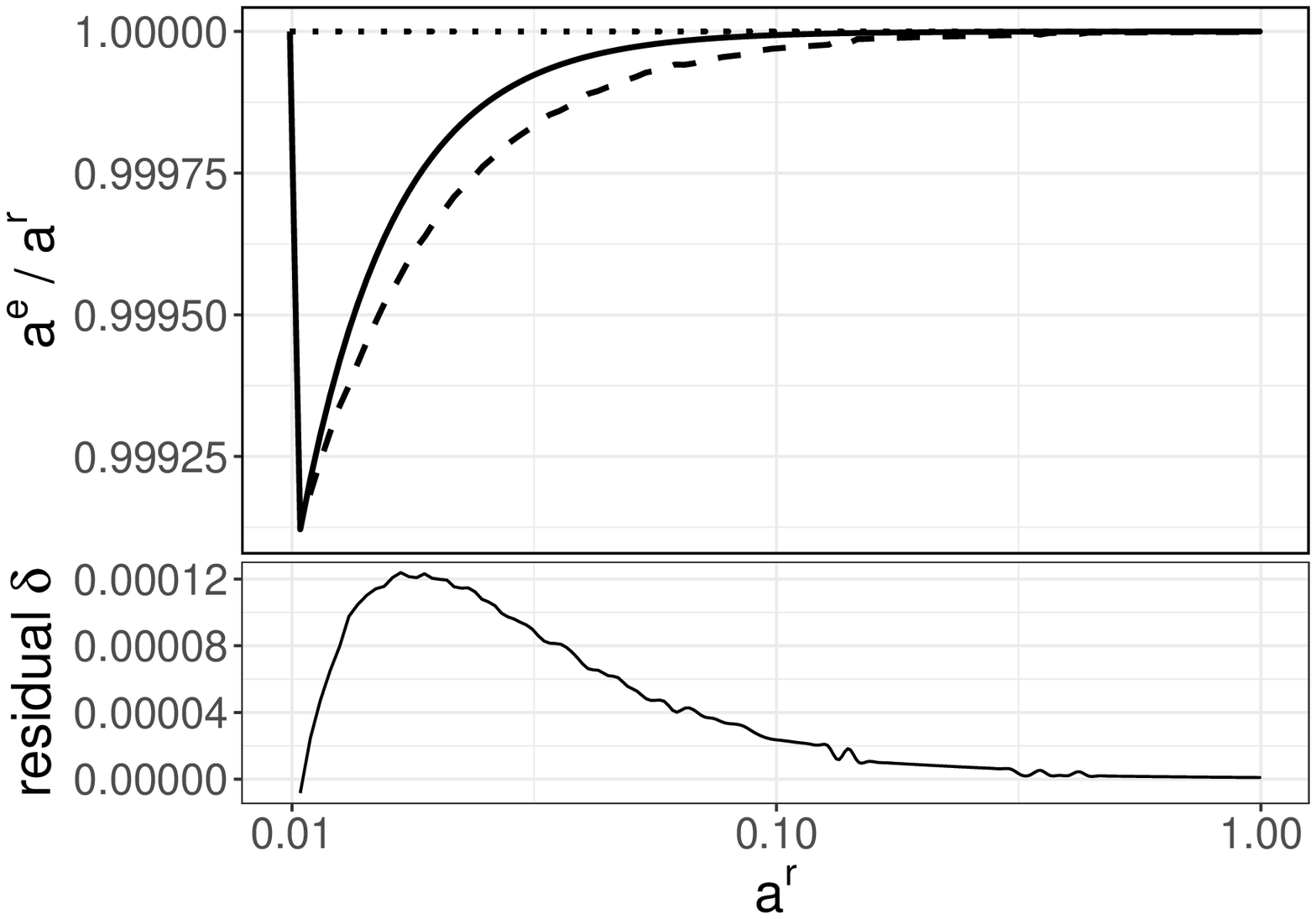}
  \includegraphics[width=0.6\columnwidth]{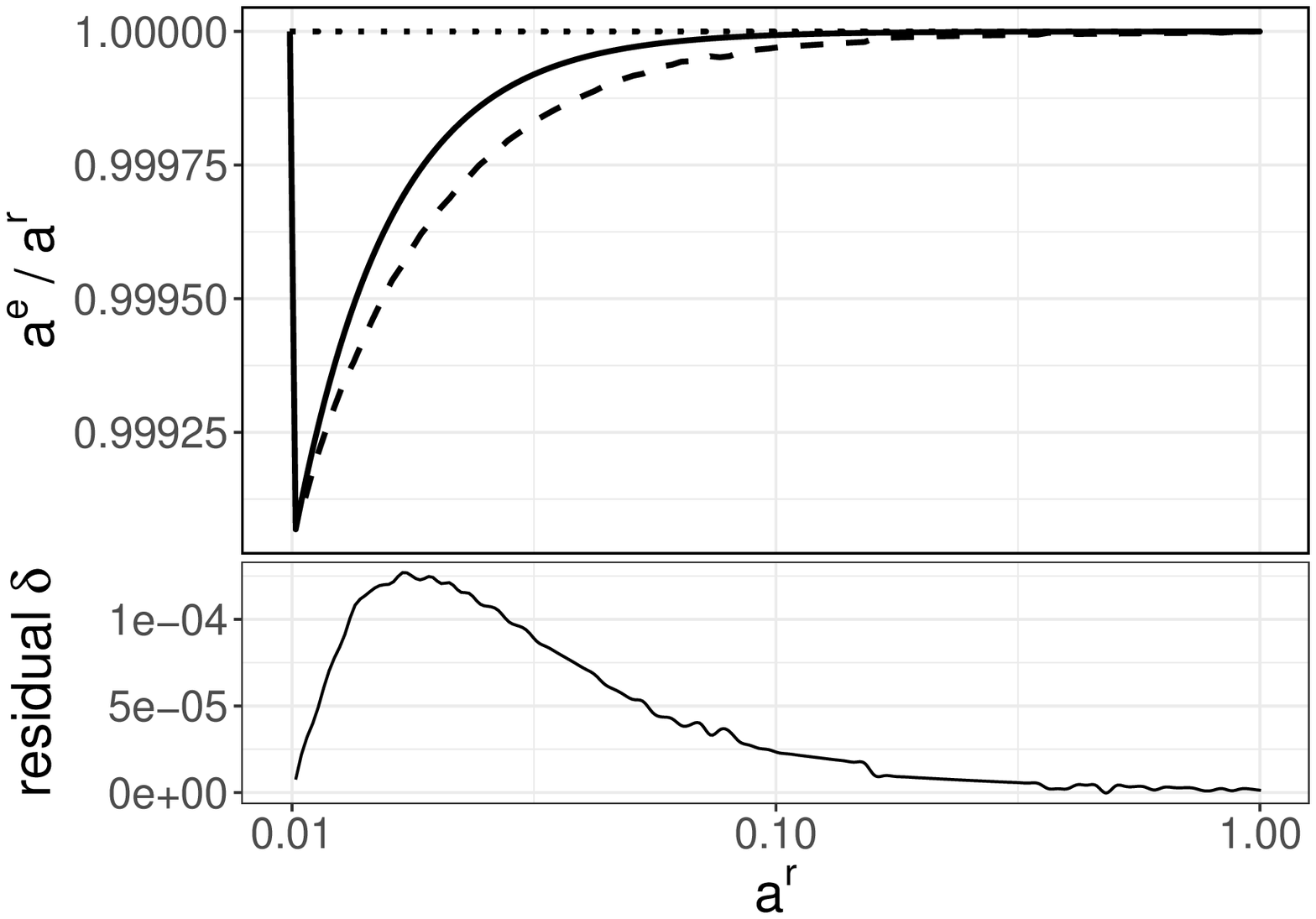}
  \caption{Scale factor ratios
    $\effmodel{a}_{\initial{\Phi}=0}/\refmodel{a}$ for a standard {\gevolutionname} simulation, as for Fig.~\protect\ref{f-flat-FLRW-Phi-test-minus}, but with a negative initial potential perturbation;
    $\effmodel{a}_{\initial{\Phi}=\gevnparamZeromodeamplitudeEdSMinusHTwEightvalue}/\refmodel{a}$ for a {\gevolutionname} simulation with an initial potential perturbation $\initial{\Phi}=\gevnparamZeromodeamplitudeEdSMinusHTwEightvalue$ (solid line); and
    the expected FLRW decaying mode evolution of $\effmodel{a}_{\mathrm{FLRW}}/\refmodel{a}$ (as defined in \SSS\ref{s-algebra-gev}; dashed line).
    {\em Top:} EdS reference model; {\em bottom:} $\Lambda$CDM.
    The behaviour mirrors that for the simulation with a positive $\initial{\Phi}$.
    Likewise, the initial point is removed from the residual plot for readability.
    \label{f-flat-FLRW-Phi-test-minus}}
\end{figure}

\subsection{\postrefereechanges{Introduction of perturbations in the simulation codes}}

\postrefereechanges{In this section, we provide the detailed method of how to apply the analytical methods described in section \SSS\ref{s-algebra-generic}.
For both codes, this was done with the aim of minimally disrupting the existing code when inserting the spatially uniform perturbation.}

\subsubsection{{\inhomogname}} \label{s-method-inhomog}

The {\inhomogname} package \citep{Roukema17silvir,RO19flatness} is a free-licensed C program and library that implements the relativistic Zel'dovich approximation (RZA) for the evolution of the kinematical backreaction $\CQ_\CD$ \citep{BuchRZA1,BuchRZA2}, i.e. the $\CQ_\CD$ Zel'dovich approximation (QZA) \citep{Roukema17silvir}.
The core functions of the library calculate the domain-averaged volume evolution represented by $a_{\CD}(t) := \scalaverageD{a}(t)$, using the RZA estimate for $\CQ_\CD(t)$, given an input of the initial values of the three invariants $\invI, \invII, \invIII$ of the extrinsic curvature in a comoving spatial domain in a flow-orthogonal time foliation.
This is used in the $N$-body simulation context by generating $N$-body simulation initial conditions from a power spectrum using {\mpgraficname} \citep{PrunetPichon08mpgrafic} and using the {\ramsesscalavname} extension of {\sc ramses} \citep{Teyssier02} as a front end.
The front end {\ramsesscalavname} reads the {\mpgraficname} realisation of initial conditions, calls {\dtfename} \citep{SchaapvdWeygaert00,vdWeygaertSchaap07,CW11,Kennel04kdtreeDTFE} to infer the Newtonian velocity gradients for non-overlapping subdomains that cover the full simulation volume, calculates approximations for $\initavinvI, \initavinvII, \initavinvIII$, and calls {\inhomogname} to calculate $a_{\CD}(t)$ for each of the subdomains.
The global mean volume evolution is calculated from the individual $a_{\CD}(t)$ evolution of the individual subdomains.
For the purposes of this paper, we avoid the issue of subdomain collapse and virialisation, since these are unnecessary for testing the match with FLRW solutions.

To see if a global perturbation inserted into {\inhomogname}-{\inhomogversion} yields the expected FLRW expansion rate while minimally intervening in the use of {\inhomogname} via the {\ramsesscalavname} front end, the routine {\tt amr\_dtfe\_interface} is modified to read the value of $\initavinvI$ from a plain-text file, and $\initavinvII$ and $\initavinvIII$ are set to zero.

\begin{table*}
  \centering
  \caption{Conversion between reference and expected FLRW models. Initial first invariant $\initial{\invI}$ ({\inhomogname}) and initial potential perturbation $\initial{\Phi}$ ({\gevolutionname}); initial reference model epoch $\initial{\refmodel{a}}$; reference model $\refmodel{}$ and expected FLRW model $\effmodel{}$ current-epoch cosmological parameters; expected ($\expected{\effmodel{a}}(\refmodel{a} = 1)$) and numerically measured ($\effmodel{a}(\refmodel{a} = 1)$) final scale factor.
    $\currepoch{\refmodel{H}}$ and $\currepoch{\effmodel{H}}$ are in units of km/s/Mpc; all other parameters are dimensionless.
    Plain-text version available at \href{\projectzenodofilesbase/flrw_ref_eff_constants.dat}{\projectzenodoid/flrw\_ref\_eff\_constants.dat}.
    \label{t-ref-expected-measured}}
  $\begin{array}{l r r  c c c c   c c r c  }
    \br \multicolumn{11}{c}{\mbox{\inhomogname}} \\
    \mbox{model}\rule{0ex}{2.7ex}
    & \multicolumn{1}{c}{\initial{\invI}}
    & \initial{\refmodel{a}}
    & \currepoch{\refmodel{\Omm}}
    & \currepoch{\refmodel{\OmLam}}
    & \currepoch{\refmodel{\Omk}}
    & \currepoch{\refmodel{H}}
    & \currepoch{\effmodel{\Omm}}
    & \currepoch{\effmodel{\OmLam}}
    & \currepoch{\effmodel{\Omk}}
    & \currepoch{\effmodel{H}}
\\
    \mr
    \mbox{EdS}\rule{0ex}{2.7ex}
    & \InitPerturbInhomogPlusvalue
    & \InhomogEdszZzoHtweightAInitRefvalue
    & \InhomogEdszZzoHtweightOmegaMzRefvalue
    & \InhomogEdszZzoHtweightOmegaLamzRefvalue
    & \InhomogEdszZzoHtweightOmegaKzRefvalue
    & \InhomogEdszZzoHtweightHzRefvalue
    & \InhomogEdszZzoHtweightOmegaMzEffvalue
    & \InhomogEdszZzoHtweightOmegaLamzEffvalue
    & \InhomogEdszZzoHtweightOmegaKzEffvalue
    & \InhomogEdszZzoHtweightADotZEffvalue
\\
    \mbox{EdS}
    & \InitPerturbInhomogMinusvalue
    & \InhomogEdsmzZzoHtweightAInitRefvalue
    & \InhomogEdsmzZzoHtweightOmegaMzRefvalue
    & \InhomogEdsmzZzoHtweightOmegaLamzRefvalue
    & \InhomogEdsmzZzoHtweightOmegaKzRefvalue
    & \InhomogEdsmzZzoHtweightHzRefvalue
    & \InhomogEdsmzZzoHtweightOmegaMzEffvalue
    & \InhomogEdsmzZzoHtweightOmegaLamzEffvalue
    & \InhomogEdsmzZzoHtweightOmegaKzEffvalue
    & \InhomogEdsmzZzoHtweightADotZEffvalue
\\
    \mbox{$\Lambda$CDM}
    & \InitPerturbInhomogPlusvalue
    & \InhomogLcdmzZzoHtweightAInitRefvalue
    & \InhomogLcdmzZzoHtweightOmegaMzRefvalue
    & \InhomogLcdmzZzoHtweightOmegaLamzRefvalue
    & \InhomogLcdmzZzoHtweightOmegaKzRefvalue
    & \InhomogLcdmzZzoHtweightHzRefvalue
    & \InhomogLcdmzZzoHtweightOmegaMzEffvalue
    & \InhomogLcdmzZzoHtweightOmegaLamzEffvalue
    & \InhomogLcdmzZzoHtweightOmegaKzEffvalue
    & \InhomogLcdmzZzoHtweightADotZEffvalue
\\
    \mbox{$\Lambda$CDM}
    & \InitPerturbInhomogMinusvalue
    & \InhomogLcdmmzZzoHtweightAInitRefvalue
    & \InhomogLcdmmzZzoHtweightOmegaMzRefvalue
    & \InhomogLcdmmzZzoHtweightOmegaLamzRefvalue
    & \InhomogLcdmmzZzoHtweightOmegaKzRefvalue
    & \InhomogLcdmmzZzoHtweightHzRefvalue
    & \InhomogLcdmmzZzoHtweightOmegaMzEffvalue
    & \InhomogLcdmmzZzoHtweightOmegaLamzEffvalue
    & \InhomogLcdmmzZzoHtweightOmegaKzEffvalue
    & \InhomogLcdmmzZzoHtweightADotZEffvalue
\\
    \mr
    \mbox{model}\rule{0ex}{2.7ex}
    & \multicolumn{1}{c}{\initial{\invI}}
    & \multicolumn{5}{c}{\expected{\effmodel{a}}(\refmodel{a} = 1)}
    & \multicolumn{4}{c}{\effmodel{a}(\refmodel{a} = 1)}
    \\
    \mr
    \mbox{EdS}\rule{0ex}{2.7ex}
    & \InitPerturbInhomogPlusvalue
    & \multicolumn{5}{c}{\InhomogExpectedAratioEdSzdzzouHTwEightvalue} & \multicolumn{4}{c}{\InhomogPerturbAratioEdSzdzzouHTwEightvalue}
    \\
    \mbox{EdS}
    & \InitPerturbInhomogMinusvalue
    & \multicolumn{5}{c}{\InhomogExpectedAratioEdSmzdzzouHTwEightvalue} & \multicolumn{4}{c}{\InhomogPerturbAratioEdSmzdzzouHTwEightvalue}
    \\
    \mbox{$\Lambda$CDM}
    & \InitPerturbInhomogPlusvalue
    & \multicolumn{5}{c}{\InhomogExpectedAratioLCDMzdzzouHTwEightvalue} & \multicolumn{4}{c}{\InhomogPerturbAratioLCDMzdzzouHTwEightvalue}
    \\
    \mbox{$\Lambda$CDM}
    & \InitPerturbInhomogMinusvalue
    & \multicolumn{5}{c}{\InhomogExpectedAratioLCDMmzdzzouHTwEightvalue} & \multicolumn{4}{c}{\InhomogPerturbAratioLCDMmzdzzouHTwEightvalue}
    \\
    \mr
    \multicolumn{11}{c}{\mbox{\gevolutionname}\rule{0ex}{2.7ex}} \\
    \mbox{model}\rule{0ex}{2.7ex}
    & \multicolumn{1}{c}{\initial{\Phi}}
    & \initial{\refmodel{a}}
    & \currepoch{\refmodel{\Omm}}
    & \currepoch{\refmodel{\OmLam}}
    & \currepoch{\refmodel{\Omk}}
    & \currepoch{\refmodel{H}}
    & \currepoch{\effmodel{\Omm}}
    & \currepoch{\effmodel{\OmLam}}
    & \currepoch{\effmodel{\Omk}}
    & \currepoch{\effmodel{H}}
\\
    \mr
    \mbox{EdS}\rule{0ex}{2.7ex}
    & \GevolutionEdszZzoHtweightPhiInitvalue
    & \GevolutionEdszZzoHtweightAInitRefvalue
    & \GevolutionEdszZzoHtweightOmegaMzRefvalue
    & \GevolutionEdszZzoHtweightOmegaLamzRefvalue
    & \GevolutionEdszZzoHtweightOmegaKzRefvalue
    & \GevolutionEdszZzoHtweightHzRefvalue
    & \GevolutionEdszZzoHtweightOmegaMzEffvalue
    & \GevolutionEdszZzoHtweightOmegaLamzEffvalue
    & \GevolutionEdszZzoHtweightOmegaKzEffvalue
    & \GevolutionEdszZzoHtweightADotZEffvalue
\\
    \mbox{EdS}
    & \GevolutionEdsmzZzoHtweightPhiInitvalue
    & \GevolutionEdsmzZzoHtweightAInitRefvalue
    & \GevolutionEdsmzZzoHtweightOmegaMzRefvalue
    & \GevolutionEdsmzZzoHtweightOmegaLamzRefvalue
    & \GevolutionEdsmzZzoHtweightOmegaKzRefvalue
    & \GevolutionEdsmzZzoHtweightHzRefvalue
    & \GevolutionEdsmzZzoHtweightOmegaMzEffvalue
    & \GevolutionEdsmzZzoHtweightOmegaLamzEffvalue
    & \GevolutionEdsmzZzoHtweightOmegaKzEffvalue
    & \GevolutionEdsmzZzoHtweightADotZEffvalue
\\
    \mbox{$\Lambda$CDM}
    & \GevolutionLcdmzZzoHtweightPhiInitvalue
    & \GevolutionLcdmzZzoHtweightAInitRefvalue
    & \GevolutionLcdmzZzoHtweightOmegaMzRefvalue
    & \GevolutionLcdmzZzoHtweightOmegaLamzRefvalue
    & \GevolutionLcdmzZzoHtweightOmegaKzRefvalue
    & \GevolutionLcdmzZzoHtweightHzRefvalue
    & \GevolutionLcdmzZzoHtweightOmegaMzEffvalue
    & \GevolutionLcdmzZzoHtweightOmegaLamzEffvalue
    & \GevolutionLcdmzZzoHtweightOmegaKzEffvalue
    & \GevolutionLcdmzZzoHtweightADotZEffvalue
\\
    \mbox{$\Lambda$CDM}
    & \GevolutionLcdmmzZzoHtweightPhiInitvalue
    & \GevolutionLcdmmzZzoHtweightAInitRefvalue
    & \GevolutionLcdmmzZzoHtweightOmegaMzRefvalue
    & \GevolutionLcdmmzZzoHtweightOmegaLamzRefvalue
    & \GevolutionLcdmmzZzoHtweightOmegaKzRefvalue
    & \GevolutionLcdmmzZzoHtweightHzRefvalue
    & \GevolutionLcdmmzZzoHtweightOmegaMzEffvalue
    & \GevolutionLcdmmzZzoHtweightOmegaLamzEffvalue
    & \GevolutionLcdmmzZzoHtweightOmegaKzEffvalue
    & \GevolutionLcdmmzZzoHtweightADotZEffvalue
\\
    \mr
    \mbox{model}\rule{0ex}{2.7ex}
    & \multicolumn{1}{c}{\initial{\Phi}}
    & \multicolumn{5}{c}{\expected{\effmodel{a}}(\refmodel{a} = 1)}
    & \multicolumn{4}{c}{\effmodel{a}(\refmodel{a} = 1)}
    \\
    \mr
    \mbox{EdS}\rule{0ex}{2.7ex}
    & \GevolutionEdszZzoHtweightPhiInitvalue
    & \multicolumn{5}{c}{\GevolutionExpectedAratioFinalEdSzdzzouHTwEightvalue} & \multicolumn{4}{c}{\GevolutionPhiAratioFinalEdSzdzzouHTwEightvalue}
    \\
    \mbox{EdS}
    & \GevolutionEdsmzZzoHtweightPhiInitvalue
    & \multicolumn{5}{c}{\GevolutionExpectedAratioFinalEdSmzdzzouHTwEightvalue} & \multicolumn{4}{c}{\GevolutionPhiAratioFinalEdSmzdzzouHTwEightvalue}
    \\
    \mbox{$\Lambda$CDM}
    & \GevolutionLcdmzZzoHtweightPhiInitvalue
    & \multicolumn{5}{c}{\GevolutionExpectedAratioFinalLCDMzdzzouHTwEightvalue} & \multicolumn{4}{c}{\GevolutionPhiAratioFinalLCDMzdzzouHTwEightvalue}
    \\
    \mbox{$\Lambda$CDM}
    & \GevolutionLcdmmzZzoHtweightPhiInitvalue
    & \multicolumn{5}{c}{\GevolutionExpectedAratioFinalLCDMmzdzzouHTwEightvalue} & \multicolumn{4}{c}{\GevolutionPhiAratioFinalLCDMmzdzzouHTwEightvalue}
    \\
    \br
  \end{array}$
\end{table*}

Our expectation is that this should yield an emergent FLRW solution for the expansion history different to that of the reference model.
The properties of the expected emergent solution are calculated using a shell script, using {\cosmdistname}-{\cosmdistversion} where appropriate, independently from the {\inhomogname} code and its {\ramsesscalavname} front end.
Equations~\eqref{e-Omegami-eff-inhomog}, \eqref{e-OmegaLami-eff}, and \eqref{e-Omegaki-eff} relate the reference and emergent matter density parameters, $\refmodel{\Omm}$ and $\effmodel{\Omm}$, at the early epoch when the intervention occurs.
The equations in \SSSS\ref{s-algebra-FLRW-rewritten}, \ref{s-algebra-ref-to-eff} and \ref{s-time-matching} then show how the emergent FLRW model constants $\effmodel{\{ \currepoch{\Omm}, \currepoch{\OmLam}, \currepoch{\Omk}, \currepoch{H}\}}$ are calculated from those of the reference model, $\refmodel{\{ \currepoch{\Omm}, \currepoch{\OmLam}, \currepoch{\Omk}, \currepoch{H}\}}$.

\subsubsection{{\gevolutionname}} \label{s-method-gev}

The {\gevolutionname} software is a free-licensed C++ cosmological simulation code developed by \citet{AdamekDDK15,AdamekDDK16code} aiming to be relativistically more accurate than traditional codes.
It has gained popularity as a tool for making FLRW observational predictions and for testing beyond-FLRW cosmological scenarios \citep[e.g.][]{Hassani19,Reverberi19}.
The input parameter file for the simulation includes physical parameters such as $\refmodel{\Omm}$ and $\refmodel{H_0}$, the comoving spatial side length of the 3-torus domain being simulated, the initial distribution of the particles prior to their perturbation by a power spectrum and starting and output redshifts.
The user selects desired outputs such as particle distributions at specified redshifts or power spectra of different perturbations in the Poisson gauge line element (Eq.~\eqref{e-line-element}).
The code is highly modular.
The numerical core of {\gevolutionname} is {\LATfieldtwo}, a library containing tools for solving equations of classical lattice field theory.
The code runs efficiently either on a personal computer of limited computational power or on a large computer cluster, depending primarily on the particle resolution.

The goal of testing {\gevolutionname} is to determine whether the code is currently ({\gevolutionname}-{\gevolutionversion}) able to handle a global perturbation while intervening only minimally in the internal code structure.
In order to test the effect of a spatially uniform global perturbation alone, we first provide an input power spectrum of perturbations with zero amplitude.
In the initial cycle of generating the particle distribution, we modify the function {\tt generateIC\_basic()} in the file {\tt ic\_basic.hpp}, adding a step that allows the insertion of a uniform perturbation in the scalar gravitational potential $\Phi$.
The optional insertion of the perturbation and the perturbation's amplitude are controlled in the initial settings file.
The {\gevolutionname} code outputs the reference model scale factors and other parameters.
A patch file with our exact set of modifications is provided in the reproducibility package of this paper and is archived at Software Heritage\footnote{\projectgevpatchfileSWHhref}.
When we refer to {\gevolutionname}-{\gevolutionversion}, strictly speaking we mean {\gevolutionname}-{\gevolutionversion} as modified using this patch file.

We again calculate the expected FLRW scale factor evolution using a shell script independent from the simulation code being tested, apart from the shared use of $\initial{\Phi}$ and $\initial{\dotPhi}$.
The reasoning is presented above in \SSS\ref{s-algebra-ref-to-eff-old}.
The conversion of initial scale factors to obtain $\initial{\effmodel{a}}$ is given in Eq.~\eqref{e-ai-eff-gev}.
The effective matter density $\initial{\effmodel{\Omm}}$ at the epoch of intervention is given in Eq.~\eqref{e-Omegami-eff-gev}.
The latter requires $\initial{\effmodel{\dot{a}}}$, which is provided by Eq.~\eqref{e-adoti-eff-gev}, provided that we know the value of $\dotPhi$ at this epoch.
The choice between growing and decaying modes for $\dotPhi$, corresponding to the positive and negative square roots in Eq.~\protect\ref{e-phiprime-solution-pair}, respectively (converted with Eq.~\ref{e-phi-t-tau-relation}), is set to the decaying mode, since {\gevolutionname} uses the linearised solution (see \SSS\ref{s-algebra-ref-to-eff-old}), which corresponds to the decaying perturbation.

As explained above (\SSS\ref{s-intro}, \SSS\ref{s-algebra-gev}), {\gevolutionname} and its backend {\LATfieldtwo} do not check the consistency between global spatial topology and curvature except in the sense that this is assumed to be implemented in the initial conditions.
Moreover, parameters representing the geometry per mesh cell are not stored.
Thus, the insertion of a spatially constant perturbation in $\Phi$ cannot correspond to offsetting the mean curvature in each mesh cell; instead it can only lead to a modified spatial section that is again of zero spatial curvature.
As shown in \ref{app-phiprime} and mentioned above, this allows two possible flat perturbation modes satisfying the Hamiltonian constraint (choice of $\Phi'$, or equivalently, $\dotPhi$).
We compare the numerical evolution of {\gevolutionname} with the decaying solution, which is clearly the one chosen by the code.
The relations in \SSSS\ref{s-algebra-FLRW-rewritten}, \ref{s-algebra-ref-to-eff} and \ref{s-time-matching} are then again used to obtain the emergent FLRW model constants, and {\cosmdistname} is used to generate the expected emergent scale factor evolution.

\section{Results} \label{s-results}

\subsection{{\inhomogname}} \label{s-results-inhomog}

We ran the relativistic Zel'dovich approximation evolution package {\inhomogname}-{\inhomogversion} from $N$-body initial conditions, with a resolution of $N=\NcrootInhomogHTwEightvalue^3$ particles, fundamental domain comoving (reference model) size {\Lboxvalue}~Mpc/$h$, and the {\sc ramses} resolution parameter {\ramsesparamlevelmaxEdSHTwEightname} $= \ramsesparamlevelmaxEdSHTwEightvalue$, with and without a perturbation as described in \SSS\ref{s-method-inhomog}.
We compared the numerically measured scale factor evolution with the expected FLRW model by defining a residual $\delta$ and a fractional deviation $\epsilon$, defined
\begin{align}
  \delta &:= \effmodel{a}/\refmodel{a} - \expected{\effmodel{a}}/\refmodel{a} \nonumber \\
  \epsilon &:= \frac{\effmodel{a}/\refmodel{a} - \expected{\effmodel{a}}/\refmodel{a}}{
     \expected{\effmodel{a}}/\refmodel{a}} \,.
  \label{e-resid-inhomog}
\end{align}
The fractional deviation $\epsilon$ describes the numerical deviation as a fraction of the excess expansion that should be generated by the perturbation.

Figure~\ref{f-flat-FLRW-Iinv-test-plus} (for an initial density perturbation $\initial{\invI} = \InitPerturbInhomogPlusvalue$) and Fig.~\ref{f-flat-FLRW-Iinv-test-minus} (for $\initial{\invI} = \InitPerturbInhomogMinusvalue$) show that the code behaves as expected to fair numerical accuracy, correctly taking into account the effect of spatial curvature and matching the emergent Friedmannian model.
For the EdS reference model, the deviation from the expected model increases in amplitude with time, with the fractional deviation at the final output time being \postrefereechanges{$\epsilon = \InhomogAccuracyPercentEdSzdzzouHTwEightvalue$\% for $\initial{\invI} = \InitPerturbInhomogPlusvalue$, and $\epsilon = \InhomogAccuracyPercentEdSmzdzzouHTwEightvalue$\%} for $\initial{\invI} = \InitPerturbInhomogMinusvalue$.
The corresponding $\Lambda$CDM reference model deviations are \postrefereechanges{$\epsilon = \InhomogAccuracyPercentLCDMzdzzouHTwEightvalue$\% and $\epsilon = \InhomogAccuracyPercentLCDMmzdzzouHTwEightvalue$\%,} respectively.
It is clear that {\sc inhomog} allows the emergence of scale factor evolution beyond that of the reference model, i.e. it allows the emergence of the homogeneous mode.

Table~\ref{t-ref-expected-measured} lists the scale factors at the final reference model epoch, and the reference and effective FLRW model current epoch constants.
In all eight cases, a dynamical non-null curvature parameter $\currepoch{\effmodel{\Omk}}$ appears in the expected FLRW model.
These are correctly handled by {\inhomogname}, while {\gevolutionname} instead returns to the reference model.

\subsection{{\gevolutionname}} \label{s-results-gevolution}

As in the {\inhomogname} case, we ran independent simulations for the EdS and $\Lambda$CDM models for {\gevolutionname}-{\gevolutionversion}, in each case with and without two cases of an initial perturbation: $\initial{\Phi}=\gevnparamZeromodeamplitudeEdSPlusHTwEightvalue$ and $\initial{\Phi}=\gevnparamZeromodeamplitudeEdSMinusHTwEightvalue$.
The simulations' particle resolution is $N={\NcrootGevolutionHTwEightvalue}^3$; the $\Lambda$CDM simulation was run with $\OmLamzeroref={\OmegaLLCDMvalue}$; and the fundamental domain comoving (reference model) size was {\Lboxvalue}~Mpc/$h$.

In the cases with $\initial{\Phi}=\gevnparamZeromodeamplitudeEdSPlusHTwEightvalue$, the perturbed spatial section is smaller than in the reference model, corresponding to a spatial section overdense in comparison to the background model.
However, this is insufficient to determine the expected spatial curvature, since density has to be normalised by the squared expansion rate in order to obtain the emergent density parameter, as shown in Eq.~\eqref{e-Omegami-eff-gev}.
Moreover, $\initial{\effmodel{\OmLam}}$ is also modified by the perturbation (Eq.~\eqref{e-OmegaLami-eff}), leaving $\initial{\effmodel{\Omk}}$ as the complement (Eq.~\eqref{e-Omegaki-eff}).
The evolution of the potential $\dotPhi$ enters into these relations via Eq.~\eqref{e-adoti-eff-gev}.
The linearisation in {\gevolutionname} implies the decaying mode of $\Phi$, which for $\Phi > 0$ yields $\dotPhi < 0$.
The formulae in \SSS\ref{s-algebra-generic} calculated using {\cosmdistname}-{\cosmdistversion} give a flat effective model ($\currepoch{\effmodel{\Omk}} = 0$) in these two cases, as shown in Table~\ref{t-ref-expected-measured}.
These formulae also give the expected behaviour of the $\effmodel{a}/\refmodel{a}$ ratio, which was calculated independently of {\gevolutionname}, using {\cosmdistname}.

Figures~\ref{f-flat-FLRW-Phi-test-plus} and \ref{f-flat-FLRW-Phi-test-minus} show the behaviour of {\gevolutionname} with a perturbation in comparison to the expected FLRW behaviour equivalent to that of the decaying mode.
We find that {\gevolutionname} provides a fair approximation to the expected FLRW behaviour, yielding decaying modes that return to the original FLRW reference solutions, though with a small difference between the numerical and analytical solutions.

Numerically, maximal {\gevolutionname}'s deviations $|\epsilon|$ from the expected behaviour for the EdS model are
\postrefereechanges{$\epsilon = \GevolutionAccuracyPercentMaxEdSzdzzouHTwEightvalue$\% for $\initial{\Phi} = \gevnparamZeromodeamplitudeEdSPlusHTwEightvalue$ and
$\epsilon = \GevolutionAccuracyPercentMaxEdSmzdzzouHTwEightvalue$\% for $\initial{\Phi} = \gevnparamZeromodeamplitudeEdSMinusHTwEightvalue$; and for the $\Lambda$CDM model the deviations are
$\epsilon = \GevolutionAccuracyPercentMaxLCDMzdzzouHTwEightvalue$\% for $\initial{\Phi} = \gevnparamZeromodeamplitudeLCDMPlusHTwEightvalue$ and
$\epsilon = \GevolutionAccuracyPercentMaxLCDMmzdzzouHTwEightvalue$\%} for $\initial{\Phi} = \gevnparamZeromodeamplitudeLCDMMinusHTwEightvalue$.
Those quantities exclude the initial spike seen in residual plots as an artefact of perturbation introduction.

In summary, we find that {\gevolutionname}-{\gevolutionversion} models the decaying mode well, but does not allow the growing mode of the quadratic equation in $\Phi'$ (we discuss this in \SSS\ref{s-discuss-gevolution}).

\section{Discussion} \label{s-discuss}
\subsection{{\inhomogname}} \label{s-discuss-inhomog}

As shown in Figs~\ref{f-flat-FLRW-Iinv-test-plus} and \ref{f-flat-FLRW-Iinv-test-minus}, {\inhomogname}-{\inhomogversion} is reasonably accurate in modelling the volume evolution implied by a density perturbation.
Nevertheless, the residual errors $\delta$ (Eq.~\eqref{e-resid-inhomog}) appear to grow systematically, especially at late times.
The residual errors at the current epoch of the reference model are of the order of the square of the change in scale factor, i.e. $\delta \sim \left[\effmodel{a}(\refmodel{a}=1)\right]^2$.
This would appear to suggest a second-order error in the QZA model.
Tightening of the integration accuracy parameters did not significantly affect this, although the visually identical nature of the patterns of the residuals for positive versus negative fluctuations for a fixed reference model was strengthened with improved numerical precision.
The pair of residual curves for a fixed reference model differ numerically, despite the visual resemblance.

Thus, although the QZA model is in several ways non-linear in the sense of traditional perturbation theory, and is exact in some special cases \citep[][\SSS{}V.A, V.B]{BuchRZA2}, it appears that for the evolution of average volume in a perturbed FLRW model, further improvements are likely to be needed, especially if the additional scale factor component is of the order of 16\% \citep{RMBO16Hbg1} rather than 1\%.

\subsection{{\gevolutionname}} \label{s-discuss-gevolution}

In \citet[][\SSS5.3]{AdamekDDK16code}, the possibility that a homogeneous mode could emerge is discussed, with suggestions of how it could be handled.
However, here we find that out of the two possible branches given in Eq.~\protect\eqref{e-phiprime-solution-pair}, i.e. the solution to the quadratic in $\Phi'$, the linearisation given in eqs~(2.9) of \citet{AdamekDDK16code} and eq.~(9) of \citet{Adamek2017radiation} restricts the evolution of a homogeneous mode to the decaying branch.
Equation~\eqref{e-Hamiltonian-constraint-quadratic} shows that this linearisation should be a good approximation when $\vert \Phi'\vert \ll \vert \refmodel{\dot{a}} \vert = \refmodel{\dot{a}}$, where the equality follows from only considering an expanding universe (not a contracting or stationary one).
For a decaying perturbation, if $\vert \Phi'\vert$ also decreases with increasing time, then an initial perturbation satisfying the condition is likely to lead to a solution in which the linearisation remains accurate.

The growing mode is the situation where an emergent model leads to
\begin{align}
  \vert \Phi' \vert > \refmodel{\dot{a}} &\Leftrightarrow \Phi' > \refmodel{\dot{a}}
  \Leftrightarrow \dotPhi > \refmodel{\dot{a}}/\refmodel{a} \,.
\end{align}
Equation~\eqref{e-adoti-eff-gev}, with the subscript $\initial{{}}$ reinserted, shows that for small $\vert\Phi\vert$, this implies that $\initial{\effmodel{\dot{a}}} < 0 $.
As a global effect, this would represent a universe starting to collapse, which cannot correspond to any EdS solution nor to the $\Lambda$CDM solution.
\postrefereechanges{The possible relevance of the growing mode could be that it is} an effective model for a locally collapsing spatial region, since average flatness in the scalar averaging sense does not strictly forbid gravitational turnaround from occurring in this way.
\postrefereechanges{In practice, in an EdS or $\Lambda$CDM reference model evolving from a cosmologically typical initial power spectrum of perturbations, it is rare for a region that is on average flat} to slow in its expansion and collapse past its turnaround epoch.
This was shown using the relativistic Zel'dovich approximation \citep{RO19flatness}.
\postrefereechanges{Thus, the relevance of the omitted growing mode in the context of our test for homogeneous modes is modest, since the mode is likely to be rare in practice.
  It would be more likely to be relevant in the more general case of an inhomogeneous perturbation, in which case local gravitational collapse is expected.
  In that case, the Hamiltonian constraint will still include squares of $\Phi'$, so the selection of the growing mode branch of the solutions would appear to be necessary for full modelling of structure formation within the {\gevolutionname} approach.
  An appropriate condition would have to be derived analytically and evaluated numerically to decide when to switch between the two modes.
  This would be particularly interesting for the formation of cosmic voids, whose role in effective volume expansion remains a key question of study.}

Nevertheless, \postrefereechanges{returning to homogeneous modes}, if the {\gevolutionname} convention of setting $\Phi = \Psi$ for the homogeneous mode were dropped, then more \postrefereechanges{possibilities} would be allowed.
An algebraic demonstration showing how both decaying and growing solutions are possible for uniform perturbations is given in App.~\ref{app-growing-decaying-modes} for the EdS case, for illustration.
Both the decaying and growing solutions are EdS solutions, i.e. they expand eternally.
Thus, allowing $\Phi \ne \Psi$ for the homogeneous mode could be expected to yield an extension of {\gevolutionname} that would enable the flat growing mode to be studied.

Given that {\gevolutionname} is free-licensed, it is likely that we, the original authors, or other cosmologists will extend the code to allow for the emergence of the growing homogeneous mode.
This would quite likely involve the inclusion of more second-order terms in the algebraic justification and in the coding itself, together with allowing $\Phi \ne \Psi$ for the homogeneous mode.
This future extension would not be completely trivial, since the code does not currently include unit tests or regression tests, and the built-in dependence on an implicitly fixed FLRW reference model of $a(t)$ could yield non-linearities that are numerically hard to control.
Nevertheless, this would be a promising followup of the current analysis.

Another option, but possibly unstable numerically, would be to use the formulae in \SSS\ref{s-algebra-generic} to calculate a new effective model at each time step and use this instead of the reference model in all appropriate formulae.
The source package for this paper includes an optional patch (not applied by default) that illustrates how this approach might be attempted.

One of the issues not considered here is the treatment of radiation and other particle species (eg. non-cold dark matter).
The {\gevolutionname} package has data structures for modelling these, but connecting these to the issue of global non-zero curvature significantly increases the complexity of the problem, so this is left for future work.

\subsection{\postrefereechanges{Role of simulation resolution}}

\begin{table*}
  \centering
  \caption{\postrefereechanges{Dependence of {\inhomogname} and {\gevolutionname} accuracy parameter $\epsilon$ (defined in Eq.~\protect\eqref{e-resid-inhomog}) as a function of simulation resolution $N^3$.
      Plain-text version available at \href{\projectzenodofilesbase/accuracy_parameters.dat}{\projectzenodoid/accuracy\_parameters.dat}.}
    \label{t-resolution-dependence}}
  \postrefereestart
  $\begin{array}{r rr rr}
    \br
    &
    \multicolumn{4}{c}{\mbox{{\inhomogname}}} \\
    &
    \multicolumn{2}{c}{\mbox{EdS}\rule{0ex}{2.7ex}} &
    \multicolumn{2}{c}{\mbox{$\Lambda$CDM}}
    \\
    N &
    \initial{\invI}=\InitPerturbInhomogPlusvalue &
    \initial{\invI}=\InitPerturbInhomogMinusvalue & 
    \initial{\invI}=\InitPerturbInhomogPlusvalue &
    \initial{\invI}=\InitPerturbInhomogMinusvalue  
    \\
    \mr
    32^3 &
    \InhomogAccuracyPercentEdSzdzzouThirtTwovalue\% &
    \InhomogAccuracyPercentEdSmzdzzouThirtTwovalue\% &
    \InhomogAccuracyPercentLCDMzdzzouThirtTwovalue\% &
    \InhomogAccuracyPercentLCDMmzdzzouThirtTwovalue\%
    \\
    64^3 &
    \InhomogAccuracyPercentEdSzdzzouSixtFourvalue\% &
    \InhomogAccuracyPercentEdSmzdzzouSixtFourvalue\% &
    \InhomogAccuracyPercentLCDMzdzzouSixtFourvalue\% &
    \InhomogAccuracyPercentLCDMmzdzzouSixtFourvalue\%
    \\
    \NcrootInhomogHTwEightvalue^3 &
    \InhomogAccuracyPercentEdSzdzzouHTwEightvalue\% &
    \InhomogAccuracyPercentEdSmzdzzouHTwEightvalue\% &
    \InhomogAccuracyPercentLCDMzdzzouHTwEightvalue\% &
    \InhomogAccuracyPercentLCDMmzdzzouHTwEightvalue\%
    \\
    \mr
    &
    \multicolumn{4}{c}{\mbox{{\gevolutionname}}} \\
    &
    \multicolumn{2}{c}{\mbox{EdS}\rule{0ex}{2.7ex}} &
    \multicolumn{2}{c}{\mbox{$\Lambda$CDM}}
    \\
    N &
\initial{\Phi}=\GevolutionEdszZzoHtweightPhiInitvalue &
    \initial{\Phi}=\GevolutionEdsmzZzoHtweightPhiInitvalue &
    \initial{\Phi}=\GevolutionLcdmzZzoHtweightPhiInitvalue &
    \initial{\Phi}=\GevolutionLcdmmzZzoHtweightPhiInitvalue
    \\
    \mr
    32^3 &
    \GevolutionAccuracyPercentMaxEdSzdzzouThirtTwovalue\% &
    \GevolutionAccuracyPercentMaxEdSmzdzzouThirtTwovalue\% &
    \GevolutionAccuracyPercentMaxLCDMzdzzouThirtTwovalue\% &
    \GevolutionAccuracyPercentMaxLCDMmzdzzouThirtTwovalue\%
    \\
    64^3 &
    \GevolutionAccuracyPercentMaxEdSzdzzouSixtFourvalue\% &
    \GevolutionAccuracyPercentMaxEdSmzdzzouSixtFourvalue\% &
    \GevolutionAccuracyPercentMaxLCDMzdzzouSixtFourvalue\% &
    \GevolutionAccuracyPercentMaxLCDMmzdzzouSixtFourvalue\%
    \\
    \NcrootGevolutionHTwEightvalue^3 &
    \GevolutionAccuracyPercentMaxEdSzdzzouHTwEightvalue\% &
    \GevolutionAccuracyPercentMaxEdSmzdzzouHTwEightvalue\% &
    \GevolutionAccuracyPercentMaxLCDMzdzzouHTwEightvalue\% &
    \GevolutionAccuracyPercentMaxLCDMmzdzzouHTwEightvalue\%
    \\
    \br
  \end{array}$
  \postrefereestop
\end{table*}

\postrefereechanges{In typical cosmological simulations, increasing the spatial resolution generally leads to numerical convergence for the calculation of quantities of interest.
  However, in the current work, our calibration test is that of a homogeneous perturbation.
  Increasing the resolution in the context of a homogeneous perturbation, for a fixed comoving reference volume and identical starting epoch, should lead to statistically equivalent numerical effects in each of a greater number of smaller regions, and is unlikely to show numerical convergence.

  We ran our full set of simulations for the resolutions listed in Table~\ref{t-resolution-dependence}, which shows the dependence of $\epsilon$ on the particle resolution $N$.
  The disagreement between the expected and numerically calculated scale factor evolution is worse for higher $N$ in the case of {\inhomogname}.
  This is consistent with the automatic choice of starting epoch set by {\mpgraficname}.
  For a fixed comoving reference domain size, {\Lboxvalue}~Mpc/$h$ in the current case, higher spatial resolution typically implies that a fixed amplitude of perturbations can be started in the linear regime at an earlier epoch.
  For example, the $N=32^3$ EdS simulations are started at $\initial{\refmodel{a}} = \InhomogEdszZzoThirttworesolutionAInitRefvalue$, while the $N=64^3$ and $N=\NcrootInhomogHTwEightvalue^3$ EdS simulations are started earlier, at $\initial{\refmodel{a}} = \InhomogEdszZzoSixtfourresolutionAInitRefvalue$ and $\initial{\refmodel{a}} = \InhomogEdszZzoHtweightresolutionAInitRefvalue$, respectively.
  The earlier starting epochs give longer evolution times, but start from the same initial homogeneous perturbation size.
  Since the expected and numerically calculated effective scale factors diverge as time increases (see Figs~\ref{f-flat-FLRW-Iinv-test-plus}, \ref{f-flat-FLRW-Iinv-test-minus}), the longer evolution time yields greater errors.

  In the case of {\gevolutionname}, which does not couple the starting epoch to the resolution, increasing $N$ has a negligible effect on $\epsilon$.}

\subsection{Numerical representation of geometrical information}
As explained above (\SSS\ref{s-method-projection}) in relation to {\gevolutionname}, the full representation of geometrical information in a numerical simulation is not trivial.
Most cosmological codes conceptually use three-dimensional meshes composed of polyhedral cells (typically cubes) as their numerical backbone, or particles that discretely represent a smooth mass distribution.
These do not completely represent the geometrical information present in the line element of the spacetime's metric, and instead use projections.
Given that not only spacetime curvature in general, but also spatial curvature in particular, is a crucial element of a general-relativistic spacetime, the issue of geometric projections cannot be avoided if the treatment of curvature in a code is to be correctly understood and analysed.
Codes claiming to be fully relativistic will need to correctly handle these geometrical issues in order to study their influence on the evolution of the Universe.

\subsection{Free-licensed cosmology software ecosystem}
The results of this paper illustrate how the use of free-licensed, modular codes written in C ({\cosmdistname} and {\inhomogname}) or C++ ({\gevolutionname}) can be easily and efficiently checked against each other.
These should be easy to use in checks or extensions of the Einstein Toolkit \citep{BentivegnaBruni15,Macpherson17}, part of a much bigger free-licensed relativistic code project, or of {\sc cosmograph} \citep{GiblinMS17}.

\section{Conclusions} \label{s-conclu}

The question of finding exact inhomogeneous cosmological solutions that are appropriate for the 3-torus spatial topology common to $N$-body and other cosmological simulations, in order to calibrate them, has long been a challenge.

The curved FLRW models do not provide a full solution to this challenge, but if we consider only the dynamical role of the curved models in volume evolution rather than their directly geometrical and topological role, as has been the tradition in the $N$-body simulation community, then it is clear that they can provide a robust calibration for relativistic cosmological simulations.
The QZA code {\inhomogname} passes this calibration test to first order in the change in the scale factor, and the higher order deviations hint at possibilities for improvements.

The flat FLRW models, appropriately recalibrated to match the numerical code, have shown their utility in illustrating the risk of linearisation.
In the case of {\gevolutionname}, we found that out of the two solutions of a quadratic solution, the choice of solution is hardwired.
The choice made by {\gevolutionname} is the decaying mode that is implied by linearisation in the first conformal-time derivative of the potential, $\Phi_{,\tau}$.
Thus, inhomogeneous initial conditions in a cosmological simulation that lead to an emergent homogeneous mode that is growing rather than decaying are likely to be excluded by construction in {\gevolutionname}.
Work extending {\gevolutionname} to allow the growing mode might provide a useful extension of the code.

The checks presented here have been done aiming at long-term reproducibility; the git commit hash of the source package of this paper, for those wishing to confirm that attempted reproductions are using the same version of the source package, is \postrefereechanges{\projectversion}.

\section*{Data Availability Statement} \label{s-data-availability}
The full reproducibility package for this paper, including the choice of simulation parameters (\enquote*{input data}), is available at {\postrefereechanges{\projectzenodoid}}\footnote{{\projectzenodohrefShowURL}} and in live\footnote{\projectgitrepository} and archived\footnote{\projectgitrepositoryarchived} {\sc git} repositories.
The specific version of the source package used to produce this paper can be identified by its {\sc git} commit hash \postrefereechanges{\projectversion}.

\section*{Acknowledgments}
The authors wish to thank Julian Adamek for very detailed and rapid feedback, and Marius Peper, Krzysztof Bolejko, Jan Ostrowski, Pierre Mourier, Asta Heinesen, Quentin Vigneron \postrefereechanges{and our two referees} for many useful suggestions.

Work on this paper has been supported by the ``A next-generation worldwide quantum sensor network with optical atomic clocks'' project of the TEAM IV programme of the Foundation for Polish Science co-financed by the European Union under the European Regional Development Fund.
Part of this work has been supported by the Polish MNiSW grant DIR/WK/2018/12.
Part of this work has been supported by the Pozna\'n Supercomputing and Networking Center (PSNC) computational grant 314.
Part of this work was supported by Universitas Copernicana Thoruniensis in Futuro under NCBR grant POWR.03.05.00-00-Z302/17.

{} 
We gratefully acknowledge the use of the following free-software packages and libraries. Boost 1.73.0, Bzip2 1.0.6, Cairo 1.16.0, CGAL 5.0.2, CMake 3.18.1, cosmdist 0.3.12, cURL 7.71.1, Dash 0.5.10.2, Discoteq flock 0.2.3, dtfe 1.1.1.Q-8efe489, Expat 2.2.9, fftw2 2.1.5-4.2, FFTW 3.3.8 \citep{fftw}, File 5.39, Fontconfig 2.13.1, FreeType 2.10.2, gevolution 1.2-0404c0b, Git 2.28.0, GNU Autoconf 2.69.200-babc, GNU Automake 1.16.2, GNU AWK 5.1.0, GNU Bash 5.0.18, GNU Binutils 2.35, GNU Compiler Collection (GCC) 11.2.0, GNU Coreutils 9.0, GNU Diffutils 3.7, GNU Findutils 4.7.0, GNU gettext 0.21, GNU gperf 3.1, GNU Grep 3.4, GNU Gzip 1.10, GNU Integer Set Library 0.18, GNU libiconv 1.16, GNU Libtool 2.4.6, GNU libunistring 0.9.10, GNU M4 1.4.18-patched, GNU Make 4.3, GNU Multiple Precision Arithmetic Library 6.2.0, GNU Multiple Precision Complex library, GNU Multiple Precision Floating-Point Reliably 4.0.2, GNU Nano 5.2, GNU NCURSES 6.2, GNU Patch 2.7.6, GNU Readline 8.0, GNU Scientific Library 2.6, GNU Sed 4.8, GNU Tar 1.32, GNU Texinfo 6.7, GNU Wget 1.20.3, GNU Which 2.21, GPL Ghostscript 9.52, HDF5 library 1.10.5, icu release-67-1, inhomog 0.1.10-1da3bed, Less 563, Libbsd 0.10.0, Libffi 3.2.1, libICE 1.0.10, Libidn 1.36, Libjpeg v9b, Libpaper 1.1.28, Libpng 1.6.37, libpthread-stubs (Xorg) 0.4, libSM 1.2.3, Libtiff 4.0.10, libXau (Xorg) 1.0.9, libxcb (Xorg) 1.14, libXdmcp (Xorg) 1.1.3, libXext 1.3.4, Libxml2 2.9.9, libXt 1.2.0, Lzip 1.22-rc2, Metastore (forked) 1.1.2-23-fa9170b, mpgrafic 0.3.19-4b78328, Open MPI 4.0.4, OpenSSL 1.1.1a, PatchELF 0.10, Perl 5.32.0, Perl Compatible Regular Expressions 8.44, Pixman 0.38.0, pkg-config 0.29.2, Python 3.8.5, ramses-scalav 0.0-482f90f, r-cran 4.0.2, Unzip 6.0, util-Linux 2.35, util-macros (Xorg) 1.19.2, Valgrind 3.17.0, X11 library 1.6.9, XCB-proto (Xorg) 1.14, xorgproto 2020.1, xtrans (Xorg) 1.4.0, XZ Utils 5.2.5, Zip 3.0 and Zlib 1.2.11. 
\LaTeX packages used for producing the pdf file include the following. alegreya 54512 (revision), biber 2.18, biblatex 3.18, bitset 1.3, caption 62757 (revision), courier 61719 (revision), csquotes 5.2l, datetime 2.60, ec 1.0, enumitem 3.9, environ 0.3, epstopdf 2.28, etoolbox 2.5k, fancyhdr 4.0.3, fmtcount 3.07, fontaxes 1.0e, fontspec 2.8a, footmisc 6.0d, fp 2.1d, kastrup 15878 (revision), lastpage 1.2n, latexpand 1.6, letltxmacro 1.6, listings 1.8d, logreq 1.0, mnras 3.1, mweights 53520 (revision), newtx 1.71, pdfescape 1.15, pdftexcmds 0.33, pgf 3.1.9a, pgfplots 1.18.1, preprint 2011, pstricks 3.13, pst-tools 0.12, setspace 6.7a, tcolorbox 5.1.1, tex 3.141592653, texgyre 2.501, times 61719 (revision), titlesec 2.14, trimspaces 1.1, txfonts 15878 (revision), ulem 53365 (revision), xcolor 2.14, xkeyval 2.9 and xstring 1.84. 
{}

\begin{appendix}
\section{Evolution of the perturbation: Hamiltonian constraint} \label{app-phiprime}

In order to properly calculate the scale factor change in Eq.~\eqref{e-adoti-eff-gev}, a correct value of $\dot\Phi$ is needed -- whether as numerical data output by the code or calculated analytically.
As kindly pointed out by J.\/ Adamek, an analytical equation for the evolution of the homogeneous perturbation can be straightforwardly derived.
We start from the pointwise (rather than volume-averaged) Hamiltonian constraint in the $3+1$ decomposition of the Einstein equation \citep{Gourg07lecture}
\begin{align}
  {}^3{\CR} - K_{ij}K^{ij} + K^2 &= 16\pi G \mathbf{T}(\vec{n}, \vec{n}) = 16 \pi G T^{00}= 16 \pi G \rho\,,
  \label{e-Hamiltonian-constraint}
\end{align}
where the (pointwise) quantities are the spatial curvature ${}^3{\CR}$, the extrinsic curvature at the same spacetime location $K_{ij}$, its trace $K$, the Newtonian gravitational constant $G$, the momentum--energy tensor $\mathbf{T}$ and its time--time component, which in this case is the matter density, $T^{00} = \rho$, and $\vec{n}$ is the normal to the spatial hypersurface at that point.
This equation needs to be expressed using the Poisson gauge formalism used in {\gevolutionname}.
The spatial part of the Ricci curvature can be calculated from the metric expressed as a line element, which in Poisson gauge (Eq.~\eqref{e-line-element}) and the case $B_{i} =0 = h_{ij}$ reduces to
\begin{align}
  {}^3{\CR} = 2 \left(\refmodel{a}\right)^{-2} e^{2\Phi} \delta^{ij} (2 \Phi_{,ij} - \Phi_{,i}\Phi_{,j}) \,,
  \label{e-3Ricci-Poisson}
\end{align}
where $\Phi_{,i}$ denotes a spatial derivative in the direction $i$.
The extrinsic curvature $K_{ij}$ can be calculated from the spatial line element and the lapse and shift functions.
Since in our example we assume that vector and tensor perturbations are zero, our shift covector $B_i$ is zero; the lapse function is
\begin{align}
  N = \refmodel{a}{\mathrm e}^{\Psi}\,.
  \label{e-lapse-function-Poisson}
\end{align}
Using $K_{ij} = -1/(2N)\,\partial/\partial \tau \ \gamma_{ij}$ [e.g., \citet[][(4.63); $\gamma_{ij}$ is the spatial part of the line element]{Gourg07lecture}], and Eq.~\eqref{e-phi-t-tau-relation}, gives
\begin{align}
  K_{ij} = \refmodel{a} \left(\Phi' - \refmodel{\dot{a}} \right) \, \mathrm{e}^{-2\Phi-\Psi} \delta_{ij} \,,
  \label{e-extrinsic-curvature}
\end{align}
making the switch between proper to coordinate time derivatives for convenience ($\refmodel{a}' = \refmodel{a}\refmodel{\dot{a}}$). The mixed, contravariant, and scalar extrinsic curvature are
\newcommand\detailedExCurvOne{\textbf{\em (A few intermediate details are given in red for extra checking only; these will be toggled off in the submitted version.)}
  \begin{align*}\gamma^{ij} &= a^{-2} e^{2\Phi} \delta^{ij}
  \end{align*}}\iftoggle{showDetailedDerivations}{{\color{red}\detailedExCurvOne}}{}\begin{align}
  K^i_{\ j} &= \gamma^{ik} K_{kj} = \left(\refmodel{a}\right)^{-1} (\Phi' - \refmodel{\dot{a}}) \mathrm{e}^{-\Psi} \delta^i_{\ j} \,,\nonumber\\
  K^{ij} &= \gamma^{ik} K^k_{\ j} = \left(\refmodel{a}\right)^{-3} (\Phi' - \refmodel{\dot{a}}) e^{2\Phi - \Psi} \delta^{ij} \,,\quad\mathrm{and}
  \nonumber\\ K &= K^i_{\ i} = 3 \left(\refmodel{a}\right)^{-1} (\Phi' - \refmodel{\dot{a}}) e^{-\Psi} \,,
  \label{e-extrinsic-curvature-scalar}
\end{align}
respectively. \newcommand\detailedExCurvTwo{
  \begin{align*}
    K^{ij} K_{ij} = 3 a^{-2} (\Phi' - \refmodel{\dot{a}})^2 e^{-2\Psi}
  \end{align*}
}\iftoggle{showDetailedDerivations}{{\color{red}\detailedExCurvTwo}}{}Thus, using Eqs~\eqref{e-extrinsic-curvature} and \eqref{e-extrinsic-curvature-scalar} in Eq.~\eqref{e-Hamiltonian-constraint}, we obtain
\begin{align}
  &2 \left(\refmodel{a}\right)^{-2} e^{2\Phi} \delta^{ij} (2 \Phi_{,ij} - \Phi_{,i}\Phi_{,j}) \nonumber\\
  &\phantom{2\ } - 3 a^{-2} (\Phi' - \refmodel{\dot{a}})^2 e^{-2\Psi} \nonumber\\
  &\phantom{2\ }\, +  9 a^{-2} (\Phi' - \refmodel{\dot{a}})^2 e^{-2\Psi} = 16\pi G \rho \,.
  \label{e-Hamiltonian-constraint-expanded}
\end{align}
On the right-hand side of Eq.~\eqref{e-Hamiltonian-constraint-expanded} we have the density:
\begin{align}
  G\,T^{00} = G\,\rho &= \frac{3G \currepoch{\refmodel{H}}^2 }{8\pi G\left(\refmodel{a}\right)^2}
    \left( \currepoch{\refmodel{\Omm}}\, \left({\refmodel{a}}\right)^{-1} \mathrm{e}^{3\Phi} +
    \currepoch{\refmodel{\OmLam}}\, \left({\refmodel{a}}\right)^2 \right)
    \,,
  \label{e-rho-in-Omegas}
\end{align}
from the reference model (FLRW) Hamiltonian constraint, which is $\left(\refmodel{\dot{a}}\right)^2 = \currepoch{\refmodel{H}}^2 \left( \currepoch{\refmodel{\Omm}} \left({\refmodel{a}}\right)^{-1}  +  \currepoch{\refmodel{\OmLam}} \left({\refmodel{a}}\right)^2 \right)$, together with a correction by a factor of $\mathrm{e}^{3\Phi}$ from the perturbed line element (Eq.~\eqref{e-line-element}).
Since we have a flat perturbation field with $\Phi_{,i} = 0 = \Phi_{,ij} \,\forall\, i,j$,  Eq.~\eqref{e-Hamiltonian-constraint-expanded} simplifies to
\begin{align}
  \postrefereechanges{\mathrm{e}^{-2\Psi}} \left(\refmodel{\dot{a}} - \Phi'\right)^2
  &= \frac{8\pi G}{3} \left(\refmodel{a}\right)^2 \rho
  \nonumber \\
  &= \currepoch{\refmodel{H}}^2 \left( \currepoch{\refmodel{\Omm}}\, \left({\refmodel{a}}\right)^{-1} \mathrm{e}^{3\Phi} +
  \currepoch{\refmodel{\OmLam}}\, \left({\refmodel{a}}\right)^2 \right)
  \,,
  \label{e-Hamiltonian-constraint-quadratic}
\end{align}
Subtracting the reference model Hamiltonian constraint gives
\begin{align}
  \Phi' &= \refmodel{\dot{a}} \pm \mathrm{e}^{\Psi} \,
                           \sqrt{\left({\refmodel{\dot{a}}}\right)^2 +
                                         \frac{\currepoch{\refmodel{H}}^2\, \currepoch{\refmodel{\Omm}}}
                                                  {\refmodel{a}}
                                                  \left(\mathrm{e}^{3\Phi} - 1\right)}
                           \,.
  \label{e-phiprime-solution-pair}
\end{align}

The choice of the sign before the square root and the setting of $\Psi = \Phi$ may constrain the behaviour of the solution.
The negative square root together with $\Psi=\Phi$ give a decaying mode, in which over- and under-densities weaken with time.
The positive square root, possibly together with $\Psi \ne \Phi$, may give a growing mode.
\postrefereechanges{The possible physical relevance of this case is discussed in \SSS\ref{s-discuss-gevolution}}.
A solution that linearises Eq.~\eqref{e-Hamiltonian-constraint-quadratic} in terms of $\Phi'$ (ignores the $\Phi'{}^2$ term) has only the decaying solution.
Examples of exact decaying and growing modes in the EdS case are given in \SSS\ref{app-growing-decaying-modes}.

\section{Evolution of the perturbation: Raychaudhuri equation} \label{app-Raychaudhuri}

As stated above, the shift function in our case is zero and the lapse has no spatial dependence, so
\begin{align}
  S &= 0 \nonumber \\
  S_{ij} &= 0 \nonumber \\
  \mathrm{D}_i \mathrm{D}_j N &= 0 \,.
\end{align}
Thus, the Raychaudhuri equation \citep[][(4.64), cf (4.79 for the $N=1$ case)]{Gourg07lecture} becomes
\begin{align}
  \frac{\partial}{\partial\tau} K_{ij}
  &=
  N \left( 0+ K K_{ij} -2 K_{ik} K^k_j -4\pi GT^{00} \gamma_{ij} \right)\,,
\end{align}
i.e., using Eqs~\eqref{e-lapse-function-Poisson}, \eqref{e-extrinsic-curvature}, \eqref{e-extrinsic-curvature-scalar}, and \eqref{e-rho-in-Omegas},
\newcommand\detailedRaychaudhuri{
\begin{align*}
  &\left[ \refmodel{a}\refmodel{\dot{a}}  \mathrm{e}^{-\Psi-2\Phi}
    \left(\Phi' - \refmodel{\dot{a}} \right)
    \right.
    \nonumber\\
    &\phantom{[} + \refmodel{a} \left(-\Psi'-2\Phi'\right) \mathrm{e}^{-\Psi-2\Phi}
    \left(\Phi' - \refmodel{\dot{a}} \right)
    \nonumber\\
    &\phantom{[} + \left.
    \refmodel{a} \left(-\Psi'-2\Phi'\right)
    \left(\Phi'' - \refmodel{a}\refmodel{\dot{a}} \right)
    \vphantom{e^{-\Psi}} \right] \delta_{ij}
  \nonumber\\
  &\phantom{[+} = \refmodel{a} \mathrm{e}^{\Psi}
    \left[
      \vphantom{\frac{3\currepoch{\refmodel{H}}^2}{8\pi \left(\refmodel{a}\right)^2}} 3 \left(\refmodel{a}\right)^{-1} \mathrm{e}^{-\Psi}
    \left(\Phi' - \refmodel{\dot{a}} \right) \,
    \refmodel{a} \mathrm{e}^{-\Psi-2\Phi}
    \left(\Phi' - \refmodel{\dot{a}} \right)
    \right.
    \nonumber\\
    &\phantom{[+=} -2 \refmodel{a} \mathrm{e}^{-\Psi-2\Phi}
    \left(\Phi' - \refmodel{\dot{a}} \right) \,
    \left(\refmodel{a}\right)^{-1} \mathrm{e}^{-\Psi}
    \left(\Phi' - \refmodel{\dot{a}} \right)
    \nonumber\\
    &\phantom{[+=}\left.
      -4\pi \left(\refmodel{a}\right)^2 \mathrm{e}^{-2\Phi}
      \frac{3\currepoch{\refmodel{H}}^2}{8\pi \left(\refmodel{a}\right)^2}
      \left( \currepoch{\refmodel{\Omm}}\, \left({\refmodel{a}}\right)^{-1} \mathrm{e}^{3\Phi} +
      \currepoch{\refmodel{\OmLam}}\, \left({\refmodel{a}}\right)^2 \right)
    \right] \delta_{ij}
    \nonumber\\
    \\
  & \mathrm{e}^{-\Psi-2\Phi}
  \left[
    \left(\Phi' - \refmodel{\dot{a}} \right)
    \left(\refmodel{\dot{a}} -2\Psi' - \Phi' \right)
    + \Phi'' - \refmodel{a} \refmodel{\ddot{a}}\right]
  \nonumber\\
  &\phantom{e} =
  \mathrm{e}^\Psi \left[ \vphantom{\frac{3}{2}}
    \mathrm{e}^{-2\Psi-2\Phi}
    \left(\Phi' - \refmodel{\dot{a}} \right)^2
    \right.
    \nonumber\\
    &\phantom{e= \mathrm{e}^\Psi [}
    \left.
    - \frac{3}{2}\, \mathrm{e}^{-2\Phi}
    \currepoch{\refmodel{H}}^2
      \left( \currepoch{\refmodel{\Omm}}\, \left({\refmodel{a}}\right)^{-1} \mathrm{e}^{3\Phi} +
      \currepoch{\refmodel{\OmLam}}\, \left({\refmodel{a}}\right)^2 \right)
      \right]
    \nonumber\\
    \\
    & (\Phi'-\refmodel{\dot{a}}) \left(\refmodel{\dot{a}} -\Psi' -2\Phi'\right)
    + \Phi'' - \refmodel{a}\, \refmodel{\ddot{a}}
    \nonumber\\
    &\phantom{xx} =
    (\Phi'-\refmodel{\dot{a}})^2
    - \frac{3}{2} \mathrm{e}^{2\Psi} H_0^2
    \left(\Ommzero\, \refmodel{a}^{-1} \mathrm{e}^{3\Phi}+ \OmLamzero\, \left(\refmodel{a}\right)^{2} \right) \,.
  \end{align*}}
\iftoggle{showDetailedDerivations}{{\color{red}\detailedRaychaudhuri}}{}
\begin{align}
  \Phi'' &= \refmodel{a}\, \refmodel{\ddot{a}}
            - (\Phi'-\refmodel{\dot{a}}) \left(\refmodel{\dot{a}} -\Psi' -2\Phi'\right)
            + (\Phi'-\refmodel{\dot{a}})^2
            \nonumber\\
            &\phantom{=\ }- \frac{3}{2} \mathrm{e}^{2\Psi} H_0^2
            \left(\Ommzero\, \refmodel{a}^{-1} \mathrm{e}^{3\Phi}+ \OmLamzero\, \left(\refmodel{a}\right)^{2} \right) \,.
            \label{e-Raychaudhuri-Phi}
\end{align}

\section{Decaying and growing perturbations in effective solutions} \label{app-growing-decaying-modes}

When finding the FLRW effective scale factor history that corresponds to a reference model combined with a uniform perturbation, the EdS solutions illustrate how both decaying and growing modes are possible.
Although the EdS model for a fixed initial singularity has only one free parameter (either the Hubble--Lema\^{\i}tre constant or, equivalently, the current universe age), the effective model is not required to have its pre-switching evolution match that of the reference model, since the effective model is only meaningful after the switching point.
In other words, the implied big bang moment of the effective model is a virtual value rather than a physical constraint.
Thus, a second free parameter is provided by the time offset (see Eq.~\eqref{e-match-t}).
The reference and effective models are
\begin{align}
  \refmodel{a} &= \left(\frac{\refmodel{t}}{\currepoch{\refmodel{t}}}\right)^{2/3} \,,\nonumber\\
  \effmodel{a} &= \left(\frac{\effmodel{t}}{\currepoch{\effmodel{t}}}\right)^{2/3}
  = \left(\frac{\refmodel{t} - \initial{\refmodel{t}} + \initial{\effmodel{t}}}{\currepoch{\effmodel{t}}}\right)^{2/3}
  \,.
  \label{e-EdS-illustration-2-free-params}
\end{align}

After setting $\currepoch{\refmodel{t}}$ for the reference model and $\initial{\refmodel{t}}$ for the insertion of the perturbation, the two parameters $\currepoch{\effmodel{t}}$ and $\initial{\effmodel{t}}$ both need to be chosen to provide the effective model.
A decaying perturbation occurs when $\currepoch{\refmodel{t}} = \currepoch{\effmodel{t}}$, since, as $t \rightarrow \infty$,
\begin{align}
  \frac{\vert \effmodel{a} - \refmodel{a} \vert}{\refmodel{a}}
  &\rightarrow
  \left(\frac{\initial{\effmodel{t}} - \initial{\refmodel{t}}}{\refmodel{t}} \right)^{2/3}
  \rightarrow 0 \,,
\end{align}
which will only be valid if $\initial{\effmodel{t}} - \initial{\refmodel{t}} > 0$.
When $\currepoch{\refmodel{t}} \neq \currepoch{\effmodel{t}}$, as $t \rightarrow \infty$,
\begin{align}
  \frac{\vert \effmodel{a} - \refmodel{a} \vert}{\refmodel{a}}
  &\rightarrow
  \left| \left(\frac{\currepoch{\refmodel{t}}}{\currepoch{\effmodel{t}}}\right)^{2/3} -1 \right|
  \rightarrow \infty \ \mathrm{or}\ 1 \,,
\end{align}
giving a perturbation that grows in amplitude.

\end{appendix}

\end{document}